\documentclass[showpacs,preprintnumbers,amssymb]{revtex4}

\usepackage{graphicx}
\usepackage{epsfig}
\usepackage{dcolumn}
\usepackage{bm}
\usepackage{amsmath}
\usepackage{amssymb}

\def\be{\begin{equation}}
\def\ee{\end{equation}}
\def\bea{\begin{eqnarray}}
\def\eea{\end{eqnarray}}
\def\nnb{\nonumber}
\def\bra {\langle}
\def\ket {\rangle}

\newcommand{\f}{\frac}
\newcommand{\al}{\alpha_s}

\def\lsim{\:\raisebox{-0.5ex}{$\stackrel{\textstyle<}{\sim}$}\:}
\def\gsim{\:\raisebox{-0.5ex}{$\stackrel{\textstyle>}{\sim}$}\:}

\def\gsim{\lower0.5ex\hbox{$\:\buildrel >\over\sim\:$}}
\def\lsim{\lower0.5ex\hbox{$\:\buildrel <\over\sim\:$}}

\def\missET {{\not\!\! E_T}}

\begin{document}

\preprint{Udem-GPP-TH-11-199}

\title{Two Higgs doublets with 4th generation fermions - models for TeV-scale compositeness}
\author{Shaouly Bar-Shalom}
\email{shaouly@physics.technion.ac.il}
\affiliation{Physics Department, Technion-Institute of Technology, Haifa 32000, Israel}
\author{Soumitra Nandi}
\email{soumitra.nandi@gmail.com}
\affiliation{Physique des Particules, Universit\'e de Montr\'eal, C.P. 6128, succ.\ centre-ville, Montr\'eal, QC, Canada H3C 3J7}
\author{Amarjit Soni}
\email{soni@bnl.gov}
\affiliation{Theory Group, Brookhaven National Laboratory, Upton, NY 11973, USA}

\date{\today}

\begin{abstract}
We construct a class of two Higgs doublets models with a 4th sequential generation of fermions that
may effectively accommodate the low energy characteristics and phenomenology
of a dynamical electroweak
symmetry breaking scenario which is triggered by the condensates of the 4th family fermions.
In particular, we single out the heavy quarks by coupling the ``heavier" Higgs doublet
($\Phi_h$) which possesses a much larger VEV only to them while the
``lighter'' doublet ($\Phi_\ell$) couples
only to the light fermions.
We study the constraints on these models from precision electroweak data
as well as from flavor data. We also discuss some distinct new features
that have direct consequences on the production and decays of the 4th family
quarks and leptons in high energy colliders; in particular the conventional search
strategies for $t^\prime$ and $b^\prime$ may need to be significantly revised.
\end{abstract}


\maketitle

\section{Introduction}

One of the most studied, yet unresolved theoretical puzzles in modern
particle physics
is the origin of ElectroWeak symmetry breaking (EWSB).
Indeed, it is widely anticipated that the LHC will provide us
with crucial answers regarding the underlying nature of EWSB: is the Higgs
a fundamental scalar needing protection from SUSY or is it a composite object.
In the Standard Model (SM), EWSB is triggered by the Higgs mechanism, which assumes
a single fundamental scalar, the Higgs, with a mass at the EW-scale.
This leads to the long standing
difficulty known as the hierarchy problem:
the presence of a fundamental EW-scale seems unnatural
since there is a problem of stabilizing the Higgs mass
against radiative corrections without introducing a cutoff to the
theory at the nearby TeV scale. The hierarchy problem, which is
usually being interpreted as evidence for new TeV-scale physics,
has fueled much scientific effort
in the past several decades, both in theory and in experiment.

Furthermore, recent flavor physics studies have revealed some degree of tension in
the CKM fits for the SM with 3 generations \cite{sl07,sl08,lenz1,bona,slprl10}.
For example, there are indications that the ``predicted'' value of $\sin2\beta$ is larger than the
value measured directly via the ``gold-plated'' $\psi K_s$ mode by as much as $\sim 3.3\sigma$ \cite{sl10}.
On the other hand, the announced CDF and DO results on the CP asymmetry
$S_{\psi\phi}$ in $B_s\to \psi\phi$ (at a higher
luminosity around 6/fb) are larger than the SM prediction by about $1\sigma$ \cite{hfag10}, and
at the same time, they find a surprisingly large CP-asymmetry in the same-sign dimuons signal,
which they attribute primarily to $a^s_{sl}$ - the semileptonic asymmetry in $B_s\to X_s \mu\nu$ \cite{d0dimuonprd,d0dimuonprl}.

Interestingly, perhaps the simplest variant of the SM, known as the SM4, in which only a 4th sequential
generation of fermion doublets is added to the theory (for reviews see \cite{sher0,hou2009-rev,SM4proc})
can address some of the theoretical challenges associated with the hierarchy problem
\cite{DEWSB,holdom-new,hung-new}
and can readily account for the CKM anomalies mentioned above \cite{SAGMN08,SAGMN10,ajb10B,buras_charm,gh10,NS3,lenz_fourth11,lenz_fourth12,Alok:2010zj}.
In particular,
as was suggested over two decades ago, a heavy 4th generation
fermion may trigger dynamical EWSB \cite{DEWSB}. The picture that arises in
this scenario is of new heavy fermions which have large Yukawa couplings
that are driven to a Landau pole or a fixed point
(which acts as a cutoff), possibly at the nearby TeV scale \cite{holdom-new,hung-new}.
Consequently, some form of strong dynamics
and/or compositeness may occur and
the Higgs particles are viewed as composites
primarily of the 4th generation fermions (see e.g., \cite{gustavo1,gustavo2,hashimoto1}), with condensates
$<Q_{L}^\prime t_R^\prime> \neq 0$, $<Q_{L}^\prime b_R^\prime> \neq 0$ (and possibly
also $<L_L^\prime \nu_R^\prime> \neq 0$, $<L_L^\prime \tau_R^\prime> \neq 0$), which induce
EWSB and generate a dynamical mass for the condensing fermions.
As for the CKM anomalies, the two extra phases that the SM4 possesses can give rise to a host
of non-standard CP asymmetries \cite{gh08,jarlskog} and, in addition, can significantly ameliorate the difficulties with regard to baryogenesis that the SM has \cite{gh08,fok,gh2011}.

Indeed, recent searches for 4th generation heavy quarks by the CDF Collaboration have found
that $m_{t'},m_{b'} \gsim 350$ GeV \cite{limits} -
in support of the compositeness scenario.
Thus, any theory that contains these heavy fermionic states
is inevitably cutoff at the near by TeV-scale where compositeness
is expected to occur.
As was realized already 20 years ago by Luty \cite{luty},
the compositeness picture which emerges in this case may be
more naturally embedded at low energies in multi-Higgs
theories, since one should expect
several composite scalars to emerge as manifestations of the
different possible bound states of the fundamental heavy fermions.
This idea was further studied
recently in \cite{sher1,hung,hashimoto1,hashimoto2,wise1}.
Moreover, as will be shown in this paper,
the addition of more scalar doublets relaxes
the constraints
from precision EW data (PEWD) (see also \cite{hashimoto2}),
and allows for interesting new dynamics associated
with the 4th generation fermions that can be tested
at high energy colliders.

Adopting this viewpoint, in section II we construct a class of models with two scalar
doublets and four generations
of fermions that can serve as effective low energy frameworks and
capture the key ingredients of the TeV-scale compositeness picture,
by giving a special status to the heavy fermionic
states. We then analyze in section III the constraints on these models from PEWD and from flavor physics
in b-quark systems, and in section IV we discuss some of the new distinct phenomenological consequences
of our multi-Higgs setup for collider searches of the 4th generation fermions.
Finally, in section V we summarize our findings.

\section{\label{sec:t2hdm} Two Higgs doublet models for the 4th generation fermions - 4G2HDMs}

Recall that in a type II 2HDM (see \cite{HHG}) one Higgs doublet couples only to the
the up-quarks while the 2nd Higgs doublet couples to the down-quarks. It is straight
forward to extend such a setup to the case of a 4th generation fermion doublet - this
was considered in \cite{hashimoto2,bernr,sher2HDM} and within a SUSY framework in \cite{fok,shersusy,dawson,rizzo}.

Our aim here is to construct a new class of two Higgs doublet models (2HDMs) that can
serve as a viable low energy effective framework for models of 4th generation condensation. Thus,
in analogy with the 2HDM setup proposed in \cite{Das}, we construct our 4G2HDMs using different
Yukawa textures than the ``standard" 2HDM of type II. In particular, in our 4G2HDMs
one of the Higgs fields (call it the ``heavier'' field) couples only to heavy fermionic states,
while the second Higgs field (the ``lighter'' field)
is responsible for the mass generation of all other (lighter) fermions.
In this way, the heavier field may be viewed as a $\bar q^\prime q^\prime$ composite
with a condensate $<q^\prime q^\prime> \neq 0$.

The Higgs potential is a general 2HDM one
\cite{HHG} and the Yukawa interaction Lagrangian of the quark sector is defined as:

\begin{widetext}
\begin{eqnarray}
\mathcal{L}_{Y}= -\bar{Q}_{L}
\left( \Phi_{\ell}F \cdot \left( I-{\cal I}_d^{\alpha_d \beta_d} \right) +
\Phi_{h}F \cdot {\cal I}_d^{\alpha_d \beta_d} \right) d_{R}
-\bar{Q}_{L}
\left( \tilde\Phi_{\ell} G \cdot \left( I - {\cal I}_u^{\alpha_u \beta_u} \right) +
\Phi_{h} G \cdot {\cal I}_u^{\alpha_u \beta_u} \right)
u_{R} + h.c.\mbox{ ,}
\label{eq:LY}
\end{eqnarray}
\end{widetext}
where $f_{L(R)}$
are left(right)-handed fermion fields, $Q_{L}$ is the left-handed
$SU(2)$ quark doublet and $F,G$ are general $4\times4$
Yukawa matrices in flavor space. Also, $\Phi_{\ell,h}$ are the Higgs doublets:
\begin{eqnarray*}
\Phi_i & =\left(\begin{array}{c}
\phi^{+}_i\\
\frac{v_i+\phi^{0}_i}{\sqrt{2}}\end{array}\right),\quad\tilde{\Phi_i}=\left(\begin{array}{c}
\frac{v_i^{*}+\phi^{0*}_i}{\sqrt{2}}\\
-\phi^{-}_i\end{array}\right) ~,
\end{eqnarray*}
$I$ is the identity matrix and ${\cal I}_q^{\alpha_q \beta_q}$ ($q=d,u$)
are diagonal $4\times4$ matrices defined by
${\cal I}_q^{\alpha_q \beta_q} \equiv {\rm diag}\left(0,0,\alpha_q,\beta_q\right)$.

The Yukawa texture of (\ref{eq:LY}) can be realized in terms of a $Z_2$-symmetry under which the fields transform as follows:
\begin{eqnarray}
 && \Phi_{\ell}\to-\Phi_{\ell},~ \Phi_{h}\to+\Phi_{h},~ Q_{L}\to+Q_{L}, \nonumber \\
 && d_{R}\to-d_{R}\;(d=d,s),~ u_{R}\to-u_{R}\;(u=u,c),\nonumber \\
&& b_{R}\to (-1)^{1+\alpha_d} b_{R},~ b^\prime_{R}\to (-1)^{1+\beta_d} b^\prime_{R},\nonumber \\
&& t_{R}\to (-1)^{1+\alpha_u} t_{R},~ t^\prime_{R}\to (-1)^{1+\beta_u} t^\prime_{R}.
\label{eq:z2}
 \end{eqnarray}

One can thus construct several models in which
the Yukawa interactions of the heavy fermionic states have a non-trivial
structure, possibly associated with the compositeness scenario. Three particularly
interesting models which we will study in this paper are:

\begin{itemize}
\item {\bf 4G2HDM-I}: $\left(\alpha_d,\beta_d,\alpha_u,\beta_u\right)=\left(0,1,0,1\right)$.
In this case $\Phi_h$ gives masses only to $t^\prime$ and $b^\prime$, while $\Phi_\ell$
generates masses for all other quarks (including the top-quark).
\item {\bf 4G2HDM-II}: $\left(\alpha_d,\beta_d,\alpha_u,\beta_u\right)=\left(1,1,1,1\right)$.
In this case the heavy condensate $\Phi_h$ is responsible for the mass generation of the heavy
quarks states of both the 3rd and 4th generation quarks, whereas $\Phi_\ell$
generates masses for the light quarks of the 1st and 2nd generations.
\item {\bf 4G2HDM-III}: $\left(\alpha_d,\beta_d,\alpha_u,\beta_u\right)=\left(0,1,1,1\right)$.
In this case $m_t,m_{b^\prime},m_{t^\prime} \propto v_h$, so that only quarks with
masses at the EW-scale are coupled to the heavy doublet $\Phi_h$.
\end{itemize}

The above 3 models represent, in our view, the minimal
set of multi-Higgs frameworks that capture the compositeness
scenarios associated with the heavy 4th generation fermions.
Defining $\tan\beta \equiv v_h/v_\ell$, in the 4G2HDM-I we expect
$\tan\beta \sim m_{q^\prime}/m_t \sim {\cal O}(1)$
($q^\prime = t^\prime$ or $b^\prime$), while
for the 4G2HDM-II and 4G2HDM-III models,
$\tan\beta \gg 1$ seems to be a more natural choice.

As mentioned earlier, the construction of our 4G2HDM models
was inspired in part by the 2HDM ``for the top-quark", which was
introduced by Das and Kao in \cite{Das} and which
was designed to give an effective explanation for the large top-quark mass
via $v_h \gg v_\ell$.
However, there is a fundamental difference between
our 4G2HDMs and the Das and Kao 2HDM: the Das and Kao model which was constructed
with three fermion generations
has no new heavy fermions (the heavier Higgs doublet, $\Phi_h$, couples only to the top-quark). Thus,
without the new heavy fermionic degrees of freedom,
the top-quark Yukawa coupling remains perturbative
up to the Planck scale, so that their 2HDM does not
have a natural low-energy cutoff as one would expect for the condensation picture.
On the other hand, in our 4G2HDMs the strong Yukawa couplings of
the heavier Higgs field
to the new heavy 4th family fermions reaches a Landau pole at the near by TeV-scale, thus signaling
new physics - possibly in the form of compositeness. Alternatively, our framework might be more naturally
embedded into weakly coupled theories in 5 dimensions, see e.g., \cite{gustavo1,gustavo3}.

From the point of view of the leptonic sector, the type-I 4G2HDM
is the more natural underlying setup that can effectively accommodate the
heavy masses of the 4th generation neutrino $\nu^\prime$. In particular, recall that
the current bounds on $m_{\nu^\prime}$ \cite{PDG} indicate that $\nu^\prime$ should have
a mass at least at the EW-scale. The main glaring problem for the SM4 is the fact that
it does not address the origin of such a heavy mass for $\nu^\prime$ \cite{king}. On the other hand,
within our 4G2HDM-I the heaviness of the 4th generation leptons
(with respect to the lighter three generations) is effectively accommodated by coupling
them to the heavy Higgs doublet. This setup for the leptonic sector might also be
an effective underlying description of more elaborate construction in models of warped extra
dimensions, see e.g., \cite{gustavo3}.

The physical Higgs fields $H^\pm$ and $h,H,A$ ($h$ and $H$ are the lighter
and heavier CP-even neutral states, respectively, and $A$ is the neutral CP-odd state)
are obtained by diagonalizing
the neutral and charged Higgs mass matrices:
\begin{eqnarray}
\Phi_\ell^+ &=& c_\beta G^+ - s_\beta H^+  ~, \nonumber \\
\Phi_h^- &=& s_\beta G^+ + c_\beta H^+  ~, \nonumber \\
\Phi_\ell^0 &=& c_\alpha H - s_\alpha h + i \left(c_\beta G^0 - s_\beta A \right) ~, \nonumber \\
\Phi_h^0 &=& s_\alpha H + c_\alpha h + i \left(s_\beta G^0 + c_\beta A \right) ~, \nonumber \\
\end{eqnarray}
where $G^+,G^0$ are the goldstone bosons, $c_\beta,~s_\beta \equiv \cos\beta,~\sin\beta$,
$c_\alpha,s_\alpha \equiv \cos\alpha,\sin\alpha$ and $\alpha$ is the mixing
angle in the CP-even neutral Higgs sector.

The Yukawa interactions between the physical Higgs bosons and quark states are then given by:
\begin{widetext}
\begin{eqnarray}
{\cal L}(h q_i q_j) &=& \frac{g}{2 m_W} \bar q_i \left\{ m_{q_i} \frac{s_\alpha}{c_\beta} \delta_{ij}
- \left( \frac{c_\alpha}{s_\beta} + \frac{s_\alpha}{c_\beta} \right) \cdot
\left[ m_{q_i} \Sigma_{ij}^q R + m_{q_j} \Sigma_{ji}^{q \star} L \right] \right\} q_j h \label{Sff1}~, \\
{\cal L}(H q_i q_j) &=& \frac{g}{2 m_W} \bar q_i \left\{ -m_{q_i} \frac{c_\alpha}{c_\beta} \delta_{ij}
+ \left( \frac{c_\alpha}{c_\beta} - \frac{s_\alpha}{s_\beta} \right) \cdot
\left[ m_{q_i} \Sigma_{ij}^q R + m_{q_j} \Sigma_{ji}^{q \star} L \right] \right\} q_j H ~, \\
{\cal L}(A q_i q_j) &=& - i I_q \frac{g}{m_W} \bar q_i \left\{ m_{q_i} \tan\beta \gamma_5 \delta_{ij}
- \left( \tan\beta + \cot\beta \right) \cdot
\left[ m_{q_i} \Sigma_{ij}^q R - m_{q_j} \Sigma_{ji}^{q \star} L \right] \right\} q_j A ~, \\
{\cal L}(H^+ u_i d_j) &=& \frac{g}{\sqrt{2} m_W} \bar u_i \left\{
\left[ m_{d_j} \tan\beta \cdot V_{u_id_j} - m_{d_k} \left( \tan\beta + \cot\beta \right) \cdot
V_{ik} \Sigma^{d}_{kj} \right] R \right. \nonumber \\
&& \left. + \left[ -m_{u_i} \tan\beta \cdot V_{u_id_j} + m_{u_k} \left( \tan\beta + \cot\beta \right) \cdot
\Sigma^{u \star}_{ki} V_{kj} \right] L
 \right\} d_j H^+ \label{Sff2}~,
\end{eqnarray}
\end{widetext}
where $q=d$ or $u$ for down or up-quarks with
weak Isospin $I_d=-\frac{1}{2}$
and $I_u=+\frac{1}{2}$, respectively,
and $R(L)=\frac{1}{2}\left(1+(-)\gamma_5\right)$. Also, $V$ is the $4 \times 4$ CKM matrix
and $\Sigma^d(\Sigma^u)$ are new mixing matrices in the down(up)-quark sectors, obtained
after diagonalizing the quarks mass matrices:
\begin{widetext}
\begin{eqnarray}
\Sigma_{ij}^d &=& \Sigma_{ij}^d(\alpha_d,\beta_d,D_R) = \alpha_d D_{R,3i}^\star D_{R,3j} + \beta_d D_{R,4i}^\star D_{R,4j}~, \nonumber \\
\Sigma_{ij}^u &=& \Sigma_{ij}^u(\alpha_u,\beta_u,U_R) = \alpha_u U_{R,3i}^\star U_{R,3j} + \beta_u U_{R,4i}^\star U_{R,4j} ~, \label{sigma}
\end{eqnarray}
\end{widetext}
where $D_R,U_R$ are the rotation (unitary) matrices of the right-handed
down and up-quarks, respectively. Notice that $\Sigma^u$ and $\Sigma^d$ depend
only on the elements of the 3rd and 4th rows of $U_R$ and $D_R$, respectively,
and on whether $\alpha_q$ and/or $\beta_q$ are ``turned on". For example,
in model 4G2HDM-I, for which $\left(\alpha_d,\beta_d,\alpha_u,\beta_u\right)=\left(0,1,0,1\right)$,
only the 4th row elements of $U_R$ and $D_R$ are relevant.

Recall that in standard frameworks such as the single-Higgs SM4 or 2HDMs of types
I and II \cite{HHG}, the right-handed mixing matrices $U_R$ and $D_R$ are non-physical
in the sense that they are ``rotated away" in the diagonalization procedure of the quark masses.
On the other hand, in our 4G2HDMs some elements of these matrices
can, in principle, be measured in Higgs-fermion systems, as we will later show.
One can, thus, treat
these matrices as unknowns, by expressing physical observables
in terms of the elements of the 3rd and 4th rows
of $U_R$ and $D_R$, or
study there properties under some theoretically motivated parameterization. In particular,
inspired by the working assumption of our 4G2HDMs
and by the observed flavor pattern in the up and down-quark sectors,
we may assume the following
structure (see also \cite{Das} for the $3 \times 3$ case):
\begin{widetext}
\begin{eqnarray}
D_R = \left(\begin{array}{cccc}
\cos\theta_{ds} & -\sin\theta_{ds} & \sin\theta_{ds} \cos\theta_{bb'} \epsilon_s^\star & - \cos\theta_{ds} \cos\theta_{bb'} \epsilon_s^\star \\
\sin\theta_{ds} & \cos\theta_{ds} & -\sin\theta_{ds} \sin\theta_{bb'} \epsilon_s^\star e^{-i \delta_b} &  \cos\theta_{ds} \sin\theta_{bb'} \epsilon_s^\star e^{-i \delta_b}\\
0 & \epsilon_s & \cos\theta_{bb'} & -\sin\theta_{bb'} e^{-i \delta_b} \\
0 & 0 & \sin\theta_{bb'} e^{i \delta_b} & \cos\theta_{bb'}
\end{array}\right),\label{eq:DR}
\end{eqnarray}
\begin{eqnarray}
U_R = \left(\begin{array}{cccc}
\cos\theta_{uc} & -\sin\theta_{uc} & \sin\theta_{uc} \cos\theta_{tt'} \epsilon_c^\star & - \cos\theta_{uc} \cos\theta_{tt'} \epsilon_c^\star \\
\sin\theta_{uc} & \cos\theta_{uc} & -\sin\theta_{uc} \sin\theta_{tt'} \epsilon_c^\star e^{-i \delta_t} &  \cos\theta_{uc} \sin\theta_{tt'} \epsilon_c^\star e^{-i \delta_t}\\
0 & \epsilon_c & \cos\theta_{tt'} & -\sin\theta_{tt'} e^{-i \delta_t} \\
0 & 0 & \sin\theta_{tt'} e^{i \delta_t} & \cos\theta_{tt'}
\end{array}\right),\label{eq:UR}
\end{eqnarray}
\end{widetext}
where $\epsilon_{s} = \frac{m_s}{m_b} e^{i \delta_s}$ and $\epsilon_{c} = \frac{m_c}{m_t} e^{i \delta_c}$,
so that unitarity of $D_R$ and $U_R$ is restored at 1st order
in $\epsilon_{s}$ and $\epsilon_c$, respectively. In the limit
$\sin\theta_{uc} \sim m_u/m_{c} <<1 $ and
$\sin\theta_{ds} \sim m_d/m_{s} <<1$,$^{\footnotemark[1]}$\footnotetext[1]{The mixing angles
$\theta_{uc}$ and
$\theta_{ds}$ have no effect in our models as they enter only in the 1st and 2nd rows
of $U_R$ and $D_R$ which have no physical outcome.}
$U_R$ and $D_R$ simplify to (similar textures can be found in 
Randall-Sundrum warped models of flavor \cite{aps2005,neubert}):
\begin{widetext}
\begin{eqnarray}
D_R = \left(\begin{array}{cccc}
 1 & 0 & 0 & - \epsilon_s^\star \left( 1- \frac{|\epsilon_b|^2}{2} \right) \\
0 & 1 & 0 &  \epsilon_s^\star \epsilon_b^\star \\
0 & \epsilon_s & \left( 1- \frac{|\epsilon_b|^2}{2} \right) & -\epsilon_b^\star \\
0 & 0 & \epsilon_b & \left( 1- \frac{|\epsilon_b|^2}{2} \right)
\end{array}\right), ~
U_R = \left(\begin{array}{cccc}
 1 & 0 & 0 & - \epsilon_c^\star \left( 1- \frac{|\epsilon_t|^2}{2} \right) \\
0 & 1 & 0 &  \epsilon_c^\star \epsilon_t^\star \\
0 & \epsilon_c & \left( 1- \frac{|\epsilon_t|^2}{2} \right) & -\epsilon_t^\star \\
0 & 0 & \epsilon_t & \left( 1- \frac{|\epsilon_t|^2}{2} \right)
\end{array}\right),
\end{eqnarray}
\end{widetext}
where we have further defined
\begin{eqnarray}
\epsilon_{b} = \sin\theta_{bb'} e^{i \delta_b}  ~, \epsilon_{t} = \sin\theta_{tt'} e^{i \delta_t}
~.
\end{eqnarray}

We thus obtain for the $\Sigma$ mixing matrices in Eq.~\ref{sigma}
(in each element keeping only the leading terms in $\epsilon_{q}$, $q=s,c,b,t$):
\begin{widetext}
\begin{eqnarray}
\Sigma^d &=& \left(\begin{array}{cccc}
0 & 0 & 0 & 0 \\
0 & \alpha_d |\epsilon_s|^2 & \alpha_d \epsilon_s^\star \left( 1- \frac{|\epsilon_b|^2}{2} \right) & - \alpha_d \epsilon_s^\star \epsilon_b^\star \\
0 & \alpha_d \epsilon_s \left( 1- \frac{|\epsilon_b|^2}{2} \right) & \alpha_d \left( 1- \frac{|\epsilon_b|^2}{2} \right) + \beta_d |\epsilon_b|^2 & (\beta_d -\alpha_d) \epsilon_b^\star \left( 1- \frac{|\epsilon_b|^2}{2} \right) \\
0 & - \alpha_d \epsilon_s \epsilon_b &  (\beta_d -\alpha_d) \epsilon_b \left( 1- \frac{|\epsilon_b|^2}{2} \right) & \alpha_d |\epsilon_b|^2 + \beta_d \left( 1- \frac{|\epsilon_b|^2}{2} \right)
\end{array}\right), \label{sigsimple}
\end{eqnarray}
\end{widetext}
and similarly for $\Sigma^u$ by replacing $\alpha_d,\beta_d \to \alpha_u,\beta_u$ and
$\epsilon_s,\epsilon_b \to \epsilon_c,\epsilon_t$.

A natural choice which we will adopt
in some instances below is: $|\epsilon_t| = \sin\theta_{tt'} \sim m_t/m_{t^\prime}$ and
$|\epsilon_b| = \sin\theta_{bb'} \sim m_b/m_{b^\prime}$.

\section{\label{sec:constraints} Constraints on the 4G2HDMs}

We now consider constraints from PEWD and from flavor
physics in b-quark systems; namely ${\bar B} \to X_s \gamma$ and $B_q - {\bar B_q}$ ($q = d,s$) mixing.
The PEWD constraints can be divided into the effects
of the heavy new physics which does and does not couple directly
to the SM ordinary fermions. For the former we consider
constraints from $Z \to b \bar b$, which is mainly sensitive to the
$H^+ t^\prime b$ and $W^+ t^\prime b$ couplings in our models. The effects
which do not involve direct couplings to the ordinary fermions,
are analyzed by the quantum oblique corrections to the gauge-bosons 2-point functions,
which can be parameterized in terms of the oblique parameters S,T and U
\cite{peskin}. It should be noted that, as far as the oblique parameters are
concerned, the contribution from our 4G2HDMs is identical
at the 1-loop level to that of any 2HDM,
since the new $Hff$ Yukawa interactions in our models
do not contribute at 1-loop to the gauge-bosons self energies.

\subsection{${\bar B}\to X_s\gamma$ and $B_q$-${\bar B_q}$ mixing}

\noindent\underline{1. $\bar{B} \to X_s \gamma$}
\bigskip

The inclusive radiative decays of the $B$ meson are known to be a
very sensitive probe of new physics. Strong constraints on new physics from $\bar{B} \to X_s \gamma$
\cite{CDGG98,DGG00,MPR98} crucially depend on theoretical
uncertainties in the SM prediction for this decay.
At the parton level, the decay process $B\to X_s \gamma$ is induced by the
flavor changing (FC) decay of the b-quark into a strange quark.

The current experimental world average is given by \cite{hfag10},
\be
{\rm BR}[ \bar{B} \to X_s \gamma] = (3.55 \pm 0.24 \pm 0.09)\times 10^{-4} ~.
\ee

In the SM, the calculation of the decay rate is most conveniently performed after
decoupling the electroweak bosons and the top quark. In the resulting
effective theory, the relevant FC weak interactions are given by
a linear combination of dimension-five and -six operators \cite{Chetyrkin:1996vx}
\bea
O_{1,2} &=& (\bar{s} \Gamma_i c)(\bar{c} \Gamma'_i b), \hspace{1.6cm}
\begin{array}{l} \mbox{\footnotesize (current-current} \\[-1mm]
                 \mbox{\footnotesize ~operators)} \end{array}\nnb\\
O_{3,4,5,6} &=&  (\bar{s} \Gamma_i b) {\textstyle \sum_q} (\bar{q} \Gamma'_i q), \hspace{1cm}
\begin{array}{l} \mbox{\footnotesize (four-quark} \\[-1mm]
                 \mbox{\footnotesize ~penguin operators)} \end{array}\nnb\\
O_7 &=& \f{e m_b}{16 \pi^2}\, \bar{s}_L \sigma^{\mu \nu} b_R F_{\mu \nu}, \hspace{6.5mm}
\begin{array}{l} \mbox{\footnotesize (photonic dipole} \\[-1mm]
                 \mbox{\footnotesize ~operator)} \end{array}\nnb\\
O_8 &=& \f{g m_b}{16 \pi^2}\, \bar{s}_L \sigma^{\mu \nu} T^a b_R G^a_{\mu\nu}. \hspace{2mm}
\begin{array}{l} \mbox{\footnotesize (gluonic dipole} \\[-1mm]
                 \mbox{\footnotesize ~operator)} \end{array} \label{ops}.
\eea
The Wilson coefficients, $C_i$, of these operators are perturbatively calculable at the renormalization scale $\mu_0 \sim (m_W, m_t)$ and
the Renormalization Group Equations (RGE) can be used to evaluate $C_i$
at the scale $\mu_b \sim m_b/2$. Finally, the operator on-shell matrix
elements are calculated at $\mu_b$.
At present, all the relevant Wilson coefficients $C_i(\mu_b)$ are known at the
Next-to-Next-to-Leading-Order (NNLO) \cite{GHW96,CMM97,AG95,P96,MM95,match2,Czakon:2006ss,Bobeth:1999mk}.
However, the matrix elements of
the operators $O_i$ consists of perturbative and non-perturbative corrections. As far as
the perturbative corrections are concerned, they are reduced dramatically after the completion of
Next-to-Leading-Order (NLO) and
NNLO QCD calculations. A further improvement comes from electroweak corrections \cite{KN99,CM98,BM00,GH00}.
On the other hand, no satisfactory quantitative estimates of all the non-perturbative effects are available,
but they are believed to be $\approx 5\%$ \cite{misiak08}.

In the SM within the leading log approximation, the ${\bar B} \to X_s \gamma$
amplitude is proportional to the (effective) Wilson coefficient of the
operator $O_7$. The well-known \cite{BMMP94} expression for this
coefficient reads
%
\begin{eqnarray}      \label{c7eff0}
C^{(0)\rm eff}_7(\mu_b) = \eta^{\f{16}{23}} C^{(0)}_7(\mu_0) +
\f{8}{3} \left( \eta^{\f{14}{23}} - \eta^{\f{16}{23}}
\right) C^{(0)}_8(\mu_0)
+ \sum_{i=1}^8 h_i \eta^{a_i},
\end{eqnarray}
%
where $\eta = \al(\mu_0)/\al(\mu_b)$ and
\begin{widetext}
\begin{eqnarray}
h_i = \left( \begin{array}{cccccccc}
\f{626126}{272277} & -\f{56281}{51730} & -\f{3}{7} & -\f{1}{14} &
-0.6494 & -0.0380 & -0.0185 & -0.0057 \end{array} \right) ~.
\end{eqnarray}
\end{widetext}

Separating the charm and top contributions, and neglecting the CKM-suppressed
$u$-quark contribution, eq.~(\ref{c7eff0}) can be written
as \cite{Gambino:2001ew}
\be
C^{(0)\rm eff}_7(\mu_b) = X_c + X_t,
\ee
where the charm contribution, given by $X_c$, is obtained from eq.~(\ref{c7eff0}) by the replacement:
$C^{(0)}_7(\mu_0) \to -\f{23}{36}$ and $C^{(0)}_8(\mu_0) \to -\f{1}{3}$,
\bea      \label{Xc}
X_c &=& -\f{23}{36} \eta^{\f{16}{23}}
-\f{8}{9} \left( \eta^{\f{14}{23}} - \eta^{\f{16}{23}}\right)
+ \sum_{i=1}^8 h_i \eta^{a_i},
\eea
which is equivalent to including only charm contributions to the matching
conditions for the corresponding operators. Analogously, only the top-loop contributes to $X_t$ and the expression is given
by
\bea
X_t &=& -\f{1}{2} A_0^t\left( x_t \right) \eta^{\f{16}{23}}
-\f{4}{3} F_0^t \left( x_t \right)
\left(\eta^{\f{14}{23}} - \eta^{\f{16}{23}}\right), \nonumber \\
\eea
with $x_t \equiv (m_t(\mu_0)/m_W)^2$ and
\be \label{A0F0}
\begin{array}{rcl}
A^t_0(x) &=& \f{-3x^3 + 2x^2}{2(x-1)^4} \ln x
+\f{-22x^3+153x^2-159x+46}{36(x-1)^3}, \\[3mm]
F^t_0(x) &=& \f{3x^2}{2(x-1)^4} \ln x
+\f{-5x^3+9x^2-30x+8}{12(x-1)^3}.\\[-3mm] \nonumber
\end{array}
\ee

Including the perturbative, electroweak and the available
non-perturbative corrections,
the branching ratio of $\bar{B} \to X_s \gamma$, with an energy cut--off
$E_0$ in the $\bar{B}$-meson rest frame, can be written as follows \cite{Gambino:2001ew}:
\begin{widetext}
\mathindent0cm
\be \label{main}
{\rm BR}[\bar{B} \to X_s \gamma]^{{\rm subtracted~} \psi,\;\psi'
}_{E_{\gamma} > E_0}
= {\rm BR}[\bar{B} \to X_c e \bar{\nu}]_{\rm exp}
\left| \f{ V^*_{ts} V_{tb}}{V_{cb}} \right|^2
\f{6 \alpha_{\rm em}}{\pi\;C}
\left[ P(E_0) + N(E_0) \right],
\ee
\mathindent1cm
\end{widetext}
where $\alpha_{\rm em} = \alpha_{\rm em}^{\rm on~shell}$ \cite{CM98},
$N(E_0)$ denotes the non-perturbative correction
and $P(E_0)$ is given by the perturbative ratio
\be \label{pert.ratio}
\f{\Gamma[ b \to X_s \gamma]_{E_{\gamma} > E_0}}{
|V_{cb}/V_{ub}|^2 \; \Gamma[ b \to X_u e \bar{\nu}]} =
\left| \f{ V^*_{ts} V_{tb}}{V_{cb}} \right|^2
\f{6 \alpha_{\rm em}}{\pi} \; P(E_0) ~.
\ee
In their approach (see \cite{Gambino:2001ew}) the {\em charmless}
semileptonic rate has been chosen as the
normalization factor in eq.~(\ref{pert.ratio}), whereas $C$ in eq.~(\ref{main})
is given by
\be \label{phase1}
C = \left| \f{V_{ub}}{V_{cb}} \right|^2
\f{\Gamma[\bar{B} \to X_c e \bar{\nu}]}{\Gamma[\bar{B} \to X_u e \bar{\nu}]} ~.
\ee

Furthermore, the perturbative quantity $P(E_0)$ can be written as \cite{Gambino:2001ew}:
\begin{eqnarray} \label{Pdel}
P(E_0) = \left| K_c +
\left( 1 + \f{\al(\mu_0)}{\pi} \ln \f{\mu_0^2}{m_t^2} \right)
r(\mu_0) K_t + \varepsilon_{\rm ew} \right|^2
 + B(E_0),
\end{eqnarray}
where $K_t$ contains the top-quark contribution to the $b \to s \gamma$
amplitude and $K_c$ contains the remaining contributions, among which the
charm loops are by far the dominant one.
Also, the electroweak correction to the
$b \to s \gamma$ amplitude is denoted in Eq.~\ref{Pdel}
by $\varepsilon_{\rm ew}$ and $B(E_0)$ is
the bremsstrahlung function which contains the effects of $b \to s
\gamma g$ and $b \to s \gamma q \bar{q}$ \linebreak $(q=u,d,s)$
transitions and which is the only $E_0$-dependent part in $P(E_0)$.

The NLO expression for $K_t$ is given by~\cite{Gambino:2001ew}
\begin{widetext}
\bea
K_t &=& \left[ 1 -\f{2}{9} \al(m_b)^2 +\f{\al(\mu_0)}{\pi}
\ln \f{\mu_0}{m_t} \;\; 4 x \f{\partial}{\partial x} \right]
\left[ -\f{1}{2} \eta^{\f{4}{23}} A_0(x_t)
+ \f{4}{3} \left( \eta^{\f{4}{23}} - \eta^{\f{2}{23}} \right) F_0(x_t) \right]
\nonumber\\&& \hspace{6mm}
+ \f{\al(\mu_b)}{4\pi} \left\{ E_0(x_t) \sum_{k=1}^8 e_k \eta^{\left(a_k+\f{11}{23}\right)}
\right. \nonumber\\ && \left. \hspace{-5mm}
+ \eta^{\f{4}{23}} \left[ -\f{1}{2} \eta A_1(x_t)
+ \left( \f{12523}{3174} ~~-\f{7411}{4761} \eta ~~~~-\f{2}{9} \pi^2
      ~~-\f{4}{3} ~ \left( \ln \f{m_b}{\mu_b} + \eta \ln \f{\mu_0}{m_t} \right) \right) A_0(x_t)
\right. \right. \nonumber\\ &&
\left. \left. \hspace{6mm}
                           +\f{4}{3} \eta F_1(x_t)
+ \left( -\f{50092}{4761} +\f{1110842}{357075} \eta +\f{16}{27} \pi^2
         +\f{32}{9} \left( \ln \f{m_b}{\mu_b} + \eta \ln \f{\mu_0}{m_t} \right) \right) F_0(x_t)
\right]
\right. \nonumber\\ && \left. \hspace{-5mm}
+ \eta^{\f{2}{23}} \left[ -\f{4}{3} \eta F_1(x_t)
+ \left( \f{2745458}{357075} -\f{38890}{14283}  \eta -\f{4}{9} \pi (\pi+i)
         -\f{16}{9} \left( \ln \f{m_b}{\mu_b} + \eta \ln \f{\mu_0}{m_t} \right) \right) F_0(x_t)
\right] \right\} ~,
\label{Kt}
\eea
\end{widetext}
where the functions $A^t_1(x)$ and $F^t_1(x)$ and the expression for $K_c$ are given in Ref.~\cite{Gambino:2001ew}.

For the electroweak ($\varepsilon_{\rm ew}$) and non-perturbative ($N(E_0)$)
corrections in eq.~(\ref{main})
we consider the following values \cite{Gambino:2001ew},
\bea
\varepsilon_{\rm ew} &\approx& 0.0035 + 0.0012 + 0.0028 = 0.0075  \nonumber \\
N(E_0) &=& 0.0036 \pm 0.0006 \,.   \label{ewnp}
\eea

Other required inputs which we take from \cite{Gambino:2001ew} are,
\bea
r(\mu_0 = m_t) &=& 0.578 \pm 0.002_{\mu_b} \pm (\mbox{parametric errors}) \\
C &=& 0.575 \; ( 1 \pm 0.01 \pm 0.02 \pm  0.02 )\\
a(z) &=& (0.97 \pm 0.25) \;+\; i ( 1.01 \pm 0.15 )\\
b(z) &=& (-0.04 \pm 0.01) \;+\; i ( 0.09 \pm 0.02 ) ~,
\eea
where $a(z)$ and $b(z)$ are the $z$-dependent terms in $K_c$ 
($z=(m_c/m_b)^2$, see Eq.~3.7 in \cite{Gambino:2001ew}).

With these inputs the NLO prediction for the branching fraction of $B\to X_s\gamma$
is \cite{Gambino:2001ew}
\be
{\rm BR}[\bar{B} \to X_s \gamma]_{E_{\gamma} > 1.6~{\rm GeV}} = (3.60 \pm 0.30) \times 10^{-4}.
\ee

\begin{figure*}[t]
\vspace{5mm}
\hspace*{-10mm}  $\gamma$ \hspace{36.5mm} $\gamma$
\hspace{42.0mm} $\gamma$ \hspace{36.5mm} $\gamma$ \\[7mm]
\hspace*{-0mm} $u_i$ \hspace{11mm} $u_i$ \hspace{18mm}
            $W^{\pm}$ \hspace{11mm}   $W^{\pm}$ \hspace{21mm}
    $u_i$ \hspace{11mm}  $u_i$ \hspace{18mm} $H^{\pm}$ \hspace{11mm} $H^{\pm}$ \\[-18mm]
\includegraphics[width=75mm,angle=0]{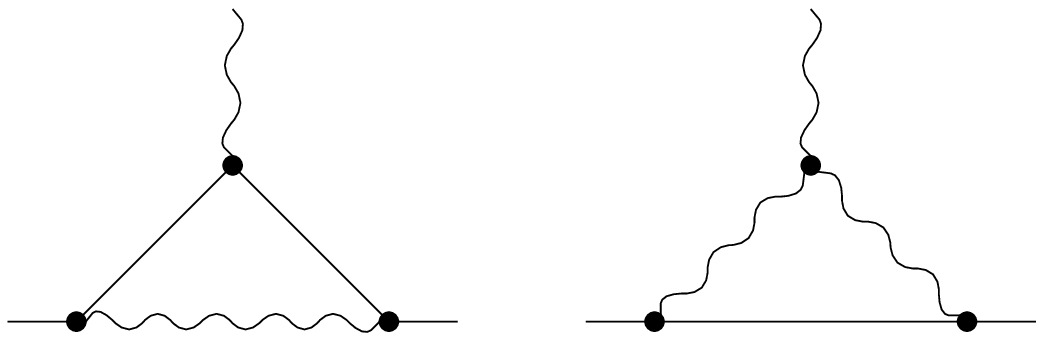}
\hspace{1cm}
\includegraphics[width=75mm,angle=0]{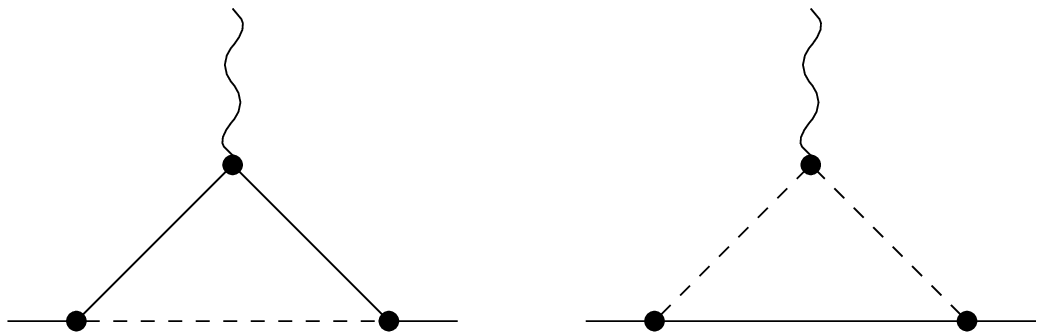}\\[-1mm]
\hspace*{-2mm}
$b$ \hspace{0.5cm}  $W^{\pm}$  \hspace{8mm} $s$ \hspace{12mm}
$b$ \hspace{9mm}   $u_i$   \hspace{7mm} $s$ \hspace{14.5mm}
$b$ \hspace{1cm} $H^{\pm}$ \hspace{8mm} $s$ \hspace{9mm}
$b$ \hspace{12mm}   $u_i$   \hspace{7mm} $s$ \\[-1cm]
\begin{center}
\caption{\sf Examples of one-loop 1PI diagrams that contribute to $b \to s \gamma$ in the 4G2HDM, with $W$-bosons, charged Higgs and 4th generation quarks exchanges ($u_i=u,c,t,t^\prime$).}
\label{fig:one-loop}
\end{center}
\vspace{-1cm}
\end{figure*}

In the SM4 there are no new operators other than the ones present in the SM. However, there are extra contributions
to the Wilson coefficients corresponding to the operators $O_7$ and $O_8$ from $t'$-loop~\cite{SAGMN08,SAGMN10,ajb10B,buras_charm}.
In our 4G2HDMs the new ingredient with respect to the SM4 is the
presence of the charged Higgs which gives new contributions
to the Wilson coefficients of the effective theory.
Examples of the 1-loop diagrams that contribute to $b \to s \gamma$ in our
4G2HDMs are given in Fig.~\ref{fig:one-loop}.$^{\footnotemark[2]}$\footnotetext[2]{We are considering only the charged Higgs contributions to $b \to s \gamma$ and neglecting the flavor changing neutral Higgs 1-loop exchanges, which are much
smaller in our
models due to the very small $b -s$ and $b^\prime - s$ transitions
as embedded in $\Sigma^d$ (see Eq.~\ref{sigsimple}).}

In order to include the charged-Higgs effect we need to compute the new
Wilson coefficients at the matching scale $\mu_0$ (The new
$H^+ u_i d_j$ Yukawa interactions in our models are given in Eq.~\ref{Sff2}).
At the LO, the charged-Higgs contributions, with the $t$-quark in the loops
are given by (see also \cite{Hou:1987kf}),
\be
\delta C_{i}^{(0)eff}(\mu0) =0 ~~~~~i=1,...,6 \label{c1-6}
\ee
\be
\delta C_{7,8}^{(0_t)eff}(\mu0) = \frac{A_{U_t}}{3}
F_{7,8}^{(1)}(y_t)+A_{D_t} F_{7,8}^{(2)}(y_t), \label{c78}
\ee
and that of $t'$ in the loops are given by
\be
\delta C_{7,8}^{(0)'eff}(\mu0) = \frac{A_{U_{t'}}}{3}
F_{7,8}^{(1)}(y_{t'})+A_{D_{t'}} F_{7,8}^{(2)}(y_{t'})~, \label{c78-2}
\ee
where $y_i=\frac{\bar m_i^2(\mu_0 )}{m_{H^+}^2}$, and the functions $F_{7,8}^{(1,2)}(y_i)$ are given by \cite{Hou:1987kf,CDGG98,MPR98}
\bea
F_7^{(1)}(y_i)&=&\frac{y_i(7-5y_i-8y^2_i)}{24(y_i-1)^3}+\frac{y^2_i(3y_i-2)}{4(y_i-1)^4}\ln y_i, \nonumber \\
F_8^{(1)}(y_i)&=&\frac{y_i(2+5y_i-y^2_i)}{8(y_i-1)^3}-\frac{3y^2_i}{4(y_i-1)^4}\ln y_i,  \nonumber \\
F_7^{(2)}(y_i)&=&\frac{y_i(3-5y_i)}{12(y_i-1)^2}+\frac{y_i(3y_i-2)}{6(y_i-1)^3}\ln y_i, \nonumber \\
F_8^{(2)}(y_i)&=&\frac{y_i(3-y_i)}{4(y_i-1)^2}-\frac{y_i}{2(y_i-1)^3}\ln y_i ~.
\eea

Dropping terms proportional to $m_s$ (the strange-quark mass)
and also neglecting the terms proportional to $\Sigma_{bb} \propto |\epsilon_b|^2$ (which is expected to be small
compared to the leading terms),
the factors $A_{U_{t/t'}}$ and $A_{D_{t/t'}}$ in Eqs.~\ref{c78} and \ref{c78-2}
are given by

\begin{widetext}
\begin{eqnarray}
A_{U_t} &=& (A_{u_1}- A_{u_2} \Sigma_{tt})^2 + \sqrt{\frac{y_{t'}}{y_t}} (\frac{V^{\ast}_{t's}}{V^{\ast}_{ts}}+\frac{V_{t'b}}{V_{tb}})\Sigma_{t't}
(A_{u_2}^2 \Sigma_{tt}  -A_{u_1} A_{u_2} ) + \frac{y_{t'}}{y_t} \frac{\lambda^{t'}_{sb}}{\lambda^t_{sb}} A_{u_2}^2 \Sigma_{t't}^2, \nonumber \\
A_{D_{t}} &=& - A_{d_1} A_{u_1} + A_{d_1} A_{u_2} \Sigma_{tt}+\frac{m_{b'}}{m_b} \frac{V_{tb'}}{V_{tb}} (A_{d_2} A_{u_1} -
A_{d_2} A_{u_2} \Sigma_{tt}) \Sigma_{b'b} \nonumber \\
&{}& -\sqrt{\frac{y_{t'}}{y_{t}}} \frac{m_{b'} \lambda^{t'}_{bs}}{m_b \lambda^{t}_{bs}} A_{u_2} A_{d_2} \Sigma_{t't} \Sigma_{b'b} +
\sqrt{\frac{y_{t'}}{y_{t}}} \frac{V^{\ast}_{t's}}{V^{\ast}_{ts}} A_{d_1} A_{u_2} \Sigma_{t't}, \nonumber \\
A_{U_{t'}} &=& (A_{u_1}- A_{u_2} \Sigma_{t't'})^2 + \sqrt{\frac{y_{t}}{y_{t'}}} (\frac{V^{\ast}_{ts}}{V^{\ast}_{t's}}+\frac{V_{tb}}{V_{t'b}})\Sigma_{tt'}
(A_{u_2}^2 \Sigma_{t't'}  -A_{u_1} A_{u_2} ) + \frac{y_{t}}{y_{t'}} \frac{\lambda^{t}_{sb}}{\lambda^{t'}_{sb}} A_{u_2}^2 \Sigma_{tt'}^2, \nonumber \\
A_{D_{t'}} &=& - A_{d_1} A_{u_1} + A_{d_1} A_{u_2}\Sigma_{t't'} + \frac{m_{b'}}{m_b} \frac{V_{t'b'}}{V_{t'b}} (A_{d_2} A_{u_1}-A_{d_2} A_{u_2}
\Sigma_{t't'}) \Sigma_{b'b}\nonumber \\
&{}& - \sqrt{\frac{y_t}{y_{t'}}} \frac{m_{b'} \lambda^{t}_{bs}}{m_b \lambda^{t'}_{bs}} \frac{V_{t'b'}}{V_{tb}} A_{u_2} A_{d_2} \Sigma_{t't} \Sigma_{b'b}
+ \sqrt{\frac{y_t}{y_{t'}}} \frac{V^{\ast}_{ts}}{V^{\ast}_{t's}} A_{d_1} A_{u_2} \Sigma_{tt'} .
\label{eq36}
\end{eqnarray}
\end{widetext}
where for later convenience we have defined
\begin{eqnarray}
A_{u_1} = A_{d_1} = \tan\beta ~, ~ A_{u_2} = A_{d_2} = \tan\beta + \cot\beta ~.
\label{defaf}
\end{eqnarray}

In all the cases where the new physics contributions
do not involve new operators
(and in which \linebreak $C_k^{\rm new}(\mu_0) = 0$~ for
~$k=1,2,3,5,6$ as in our case - see Eq.~\ref{c1-6}),
it is straightforward to incorporate the extra
terms to the NLO formulae. In particular, these
contributions effectively modify $K_t$ given
in Eq.~\ref{Pdel}, which in our 4G2HDMs should be replaced by

\be
K_t\to K_t^{W} + \frac{V_{t'b}V^{\ast}_{t's}}{V_{tb}V^{\ast}_{ts}} K_{t'}^{W} + K_t^{H} + \frac{V_{t'b}V^{\ast}_{t's}}{V_{tb}V^{\ast}_{ts}} K_{t'}^{H}~,
\label{kt4hdm}
\ee
where $K_t^{W}$, $K_{t'}^{W}$, $K_t^{H}$ and $K_{t'}^H$ represent the $W$ and charged-Higgs contributions to the
$b\to s\gamma$ amplitudes from $t$ and $t'$ loops (see Fig.~\ref{fig:one-loop}).
In particular, $K_{t'}^W$ can be obtained simply by replacing (neglecting $ln(\frac{\mu_0}{m_t})$)

\begin{eqnarray}
E_0(x_t) &\to& E_0(x_{t'}), \nonumber \\
A_0(x_t) &\to& A_0(x_{t'}), \nonumber \\
A_1(x_t) &\to& A_1(x_{t'}), \nonumber \\
F_0(x_t) &\to& F_0(x_{t'}), \nonumber \\
F_1(x_t) &\to& F_1(x_{t'}),
\end{eqnarray}
in eq.~\ref{Kt}. On the other hand, $K_t^{H}$ and $K_{t'}^{H}$, which represent
the net contributions to the $b\to s\gamma$ amplitude from charged-Higgs exchanges (with $t$ and $t'$ as the internal quark, respectively),
can be obtained from Eq.~\ref{Kt} by
calculating the functions $E_0(y_i),~A_0(y_i),~A_1(y_i),~F_0(y_i)$ and $F_1(y_i)$
($i = t$ or $i=t'$). The LO functions
$A_0(y_t)$ and $A_0(y_{t'})$ are given by

\begin{widetext}
\bea
A_0(y_{t}) &=& - 2~\delta C_{7}^{(0)eff}(\mu_0), \hskip 10pt F_0(y_{t}) = - 2~\delta C_{8}^{(0)eff}(\mu_0), \nonumber \\
A_0(y_{t'}) &=& - 2~\delta C_{7}^{(0)'eff}(\mu_0), \hskip 10pt F_0(y_{t'}) = - 2~\delta C_{8}^{(0)'eff}(\mu_0).
\label{AFld}
\eea
and the NLO functions
$A_1(y_t)$ and $A_1(y_{t'})$ by
\bea
A_1(y_i) &=& - 2~\delta C_{7}^{(1)}(\mu_0), \hskip 10pt F_1(y_i) = - 2~\delta C_{8}^{(1)}(\mu_0).
\eea
\end{widetext}

The NLO contributions to the Wilson coefficients (in our 4G2HDMs)
are given by$^{\footnotemark[3]}$\footnotetext[3]{The NLO results for the Wilson coefficients in
 a 2HDM can be found in \cite{CDGG98,Bobeth:1999ww}.}

\begin{widetext}
\be
\delta C_{i}^{(1)eff}(\mu_0) =0 ~~~~~i=1,2,3,5,6
\ee
\be
E_0(y_i) = \delta C_4^{(1)}(\mu_0) =  A_{U_i} \left[
         \f{3y_i^2-2y_i}{6(1-y_i)^4} \ln y_i \;+\; \f{-7y_i^3+29y_i^2-16y_i}{36(1-y_i)^3} \right],
\ee

and

\bea
\delta C_7^{(1)}(\mu_0) &=&  A_{U_i} \left\{
      \f{16y_i^4-74y_i^3+36y_i^2}{9(1-y_i)^4} Li_2\left(1-\f{1}{y_i}\right) \right.
\;+\;  \f{-63y_i^4+807y_i^3-463y_i^2+7y_i}{81(1-y_i)^5} \ln y_i \nonumber \\
 &+& \f{-1202y_i^4+7569y_i^3-5436y_i^2+797y_i}{486(1-y_i)^4}
\;+\; \left[ \f{6y_i^4+46y_i^3-28y_i^2}{9(1-y_i)^5} \ln y_i \right. \nonumber \\
 &+&  \left.\left. \f{-14y_i^4+135y_i^3-18y_i^2-31y_i}{27(1-y_i)^4} \right]  \ln \f{\mu_0^2}{m_i^2} \right\}
\;+\; A_{D_i} \left\{\f{-32y_i^3+112y_i^2-48y_i}{9(1-y_i)^3} Li_2\left(1-\f{1}{y_i}\right) \right. \nonumber \\
 &+&  \f{14y_i^3-128y_i^2+66y_i}{9(1-y_i)^4} \ln y_i
\;+\; \f{8y_i^3-52y_i^2+28y_i}{3(1-y_i)^3}
\;+\;  \left[ \f{-12y_i^3-56y_i^2+32y_i}{9(1-y_i)^4} \ln y_i \right. \nonumber \\
&+& \left.\left. \f{16y_i^3-94y_i^2+42y_i}{9(1-y_i)^3} \right]  \ln \f{\mu_0^2}{m_i^2} \right\},
\eea
\bea
\delta C_8^{(1)}(\mu_0) &=& A_{U_i} \left\{
      \f{13y_i^4-17y_i^3+30y_i^2}{6(1-y_i)^4} Li_2\left(1-\f{1}{y_i}\right) \right.
\;+\;   \f{-468y_i^4+321y_i^3-2155y_i^2-2y_i}{216(1-y_i)^5} \ln y_i \nonumber \\
 &+& \f{-4451y_i^4+7650y_i^3-18153y_i^2+1130y_i}{1296(1-y_i)^4}
\;+\; \left[ \f{-17y_i^3-31y_i^2}{6(1-y_i)^5} \ln y_i  \right. \nonumber \\
&+& \left. \left. \f{-7y_i^4+18y_i^3-261y_i^2-38y_i}{36(1-y_i)^4}  \right]  \ln \f{\mu_0^2}{m_i^2}\right\}
\;+\; A_{D_i} \left\{ \f{-17y_i^3+25y_i^2-36y_i}{6(1-y_i)^3} Li_2\left(1-\f{1}{y_i}\right) \right. \nonumber \\
&+& \f{34y_i^3-7y_i^2+165y_i}{12(1-y_i)^4} \ln y_i
\;+\; \f{29y_i^3-44y_i^2+143y_i}{8(1-y_i)^3}
\;+\;  \left[ \f{17y_i^2+19y_i}{3(1-y_i)^4} \ln y_i \right. \nonumber \\
&+& \left. \left. \f{7y_i^3-16y_i^2+81y_i}{6(1-y_i)^3} \right]  \ln \f{\mu_0^2}{m_i^2} \right\}.
\eea
\end{widetext}

The electroweak and non-perturbative corrections are retained to their SM predictions as given in
\cite{Gambino:2001ew} (see also eq.~\ref{ewnp}), i.e.,
we do not take into account the effect of our 4G2HDM on these corrections.

\pagebreak
\noindent\underline{2. $B_{q}-\bar B_{q}$ mixing}
\bigskip

 In the SM, $B_{q}-\bar B_{q}$ mixing ($q=d,s$)
 proceeds to an excellent approximation only
 through the box diagrams with internal top quark exchanges. On the other hand, in our
 4G2HDMs there are additional contributions to $B_{q}-\bar B_{q}$ mixing
coming from the loop exchanges of the $t'$ and charged-Higgs.

In the 4G2HDM, the mass difference $\Delta M_q = 2|M_{{12}_q}|$ is given
at LO by$^{\footnotemark[4]}$\footnotetext[4]{The LO
results for $B_{q}-\bar B_{q}$ mixing in a ``standard" 2HDM of type II with three generations
of fermion doublets are given in \cite{buras_2hdm}.}
\be
M_{{12}_q} = \frac{G_F^2}{12\pi^2} M_W^2 f_{B_q}^2 B_q M_{B_q} \left[M_{WW} + M_{HH} + M_{HW} \right],
\ee
where we have used
\bea
\bra B_q|({\bar s}b)_{(V-A)}({\bar s}b)_{(V-A)}| B_q \ket &=& \f{8}{3} f_{B_q}^2 B_q M_{B_q}^2, \\
\bra B_q|({\bar s}b)_{(S+P)}({\bar s}b)_{(S+P)}| B_q \ket &=& - \f{5}{3} f_{B_q}^2 B_q M_{B_q}^2.
\eea
and
\begin{widetext}
\bea
M_{WW} &=& {\lambda^t_{bq}}^2 \eta_{tt} S_{WW}(x_t) + {\lambda^{t'}_{bq}}^2 \eta_{t't'} S_{WW}(x_{t'}) +
2~ \lambda^t_{bq}  \lambda^{t'}_{bq} \eta_{tt'} S_{WW}(x_t,x_{t'}), \nonumber \\
M_{HH} &=& {\lambda^t_{bq}}^2 S_{HH}(y_t) + {\lambda^{t'}_{bq}}^2 S_{HH}(y_{t'}) +
2~ \lambda^t_{bq}  \lambda^{t'}_{bq} S_{HH}(y_t,y_{t'}), \nonumber \\
M_{HW} &=& {\lambda^t_{bq}}^2 S_{HW}(x_t,z) + {\lambda^{t'}_{bq}}^2 S_{HW}(x_{t'},z) +
2~ \lambda^t_{bq}  \lambda^{t'}_{bq} S_{HW}(x_t,x_{t'},z),
\eea
\end{widetext}
 with $z = \frac{m_{H^+}^2}{m_W^2}$, $x_i = \frac{m_i^2}{m_W^2}$, $y_i = \frac{m_i^2}{m_{H^+}^2}$ ($i = t$ or $t'$) and
 $\lambda^u_{d_i d_j} \equiv V_{u d_i}^\star V_{u d_j}$.

The contributions from $W$-exchange diagrams with
$q_i$ and $q_j$ ($i,j$ are generation indices)
as the internal quarks are given by,
\begin{widetext}
\bea
S_{WW}(x_i,x_j) &=& x_i x_j \left\{\left[\f{1}{4}+\f{3}{2}\f{1}{(1-x_j)}-\f{3}{4}\f{1}{(1-x_j)^2}\right] \f{\ln x_j}{(x_j-x_i)} +
(x_j\to x_i) - \f{3}{4}\f{1}{(1-x_i)(1-x_j)}\right\} ~,
\label{ilww}
\eea
\end{widetext}
 and $S_{WW}(x_i) \equiv S_{WW}(x_i,x_i)$ can be obtained from Eq.~\ref{ilww} by taking the limit $x_j\to x_i$.

The contributions from the $H^+$-exchange diagrams are given by
\begin{widetext}
\bea
S_{HH}(y_t,y_{t'}) &=& z S_{L_1} S_{L_2} \left[\f{S^{k2}_{HH}(y_t,y_{t'})}{4} B_{L_1} B_{L_2}- \f{5}{8} x_b S^{m}_{HH}(y_t,y_{t'})
B_{R_1} B_{R_2}\right],\nonumber \\
S_{HH}(y_t) &=& z S_{L_2}^2 \left[\f{S^{k2}_{HH}(y_t)}{4} B_{L_2}^2- \f{5}{8} x_b S^{m}_{HH}(y_t) B_{R_2}^2 \right],\nonumber \\
S_{HH}(y_{t'}) &=& z S_{L_1}^2 \left[ \f{S^{k2}_{HH}(y_{t'})}{4} B_{L_1}^2 - \f{5}{8} x_b S^{m}_{HH}(y_{t'}) B_{R_1}^2 \right],
\eea
\end{widetext}
where $x_b = \f{m_b^2}{m_{H^+}^2} \f{M_{B_q}^2}{{m_b(m_b)}^2}$,
\begin{widetext}
\bea
S^{k2}_{HH}(y_i,y_j) &=& y_i y_j \left\{\f{1}{(y_i-y_j)}\left(\f{y_i^2 \ln y_i}{(1-y_i)^2}-\f{y_j^2 \ln y_j}{(1-y_j)^2}\right)
+ \f{1}{(1-y_i)(1-y_j)} \right\}, \\
S^{k2}_{HH}(y_i) &=& S^{k2}_{HH}(y_i,y_j)_{y_j\to y_i}, \\
S^{m}_{HH}(y_i,y_j) &=& y_i y_j \left\{\f{1}{(y_i-y_j)}\left(\f{y_i \ln y_i}{(1-y_i)^2}-\f{y_j \ln y_j}{(1-y_j)^2}\right)
+ \f{1}{(1-y_i)(1-y_j)} \right\}, \\
S^{m}_{HH}(y_i) &=& S^{m}_{HH}(y_i,y_j)_{y_j\to y_i}.
\eea
\end{widetext}
and the terms
\bea
B_{L_1} &=& - A_{u_1} + A_{u_2} \Sigma_{t't'} + A_{u_2} \f{m_t}{m_{t'}}\f{V_{tb}}{V_{t'b}} \Sigma_{tt'}, \nonumber \\
B_{L_2} &=& - A_{u_1} + A_{u_2} \Sigma_{tt} +  A_{u_2} \f{m_{t'}}{m_t} \f{V_{t'b}}{V_{tb}} \Sigma_{t't}, \nonumber \\
S_{L_1} &=& - A_{u_1} + A_{u_2} \Sigma_{t't'} + A_{u_2} \f{m_t}{m_{t'}} \f{V^{\ast}_{ts}}{V^{\ast}_{t's}} \Sigma_{tt'}, \nonumber \\
S_{L_2} &=& - A_{u_1} + A_{u_2} \Sigma_{tt} +  A_{u_2} \f{m_{t'}}{m_t} \f{V^{\ast}_{t's}}{V^{\ast}_{ts}} \Sigma_{t't}, \nonumber \\
B_{R_1} &=&  A_{d_1} - A_{d_2} \Sigma_{bb} - A_{d_2} \f{m_{b'}}{m_b} \f{V_{t'b'}}{V_{t'b}} \Sigma_{b'b}, \nonumber \\
B_{R_2} &=&  A_{d_1} - A_{d_2} \Sigma_{bb} - A_{d_2} \f{m_{b'}}{m_b} \f{V_{tb'}}{V_{tb}} \Sigma_{b'b},
\label{eq56}
\eea
are obtained from the $b\to t,t'$ and $t,t' \to s$ vertices in the
box diagrams.

The functions $S_{HW}(x_i,x_j,z)$ obtained from diagrams with
both $W$ and $H^+$-exchanges are given by
\begin{widetext}
\bea
S_{HW}(x_t,x_{t'},z) &=& 2 \,x_t \, x_{t'} (S_{L_1}B_{L_2} + S_{L_2} B_{L_1})\left[\f{S_1(x_t,x_{t'},z)}{4} + S_2(x_t,x_{t'},z) \right],  \\
S_{HW}(x_t,z) &=& 2 \,x_t^2\, S_{L_2} B_{L_2}\left[\f{S_1(x_t,z)}{4} + S_2(x_t,z) \right],  \\
S_{HW}(x_{t'},z) &=& 2\, x_{t'}^2\, S_{L_1} B_{L_1}\left[\f{S_1(x_{t'},z)}{4} + S_2(x_{t'},z) \right],
\eea
\end{widetext}
where
\begin{widetext}
\bea
S_1(x_i,x_j,z) &=& \f{z \ln z}{(1-z)(z-x_i)(z-x_j)} + \f{x_i \ln x_i}{(1-x_i)(x_i-z)(x_i-x_j)}
+ \f{x_j \ln x_j}{(1-x_j)(x_j-z)(x_j-x_i)}, \nonumber \\
S_2(x_i,x_j,z) &=& - \f{z^2 \ln z}{(1-z)(z-x_i)(z-x_j)} - \f{x_i^2 \ln x_i}{(1-x_i)(x_i-z)(x_i-x_j)}
- \f{x_j^2 \ln x_j}{(1-x_j)(x_j-z)(x_j-x_i)} ~,
\eea
\end{widetext}
and the functions $S_{1}(x_i,z)$ and $S_{2}(x_i,z)$ can be derived from the expressions for $S_{1}(x_i,x_j,z)$ and $S_{2}(x_i,x_j,z)$, respectively, by taking the limit $x_j \to x_i$.

\bigskip
\noindent\underline{3. Combined constraints}
\bigskip

Using the analysis above, we derive below the constraints on our 4G2HDMs that come from
$Br(B\to X_s \gamma)$ and $\Delta M_q$ $(q =d,s)$.
For the B-physics parameters we use the inputs given in
Table \ref{tab1}. As an illustration, the 4th generation quark masses are fixed
to $m_{t'}=500$ GeV and $m_{b'}=450$ GeV, consistent
with the direct limits from the Tevatron \cite{limits} and the
perturbative unitarity upper bounds \cite{PU,sher1}.$^{\footnotemark[5]}$\footnotetext[5]{There is a very weak dependence
of $B \to X_s \gamma$ and $B-\bar B$-mixing on the $b'$-mass,
since it enters only in the the $H^+ u d$ Yukawa couplings
with no dynamical and/or kinematical dependence.}
We vary the charged Higgs mass in the range
$200~ {\rm GeV} < m_{H^+} < 1~ {\rm TeV}$ and study
the dependence on $\epsilon_t$ in the range
$0 < \epsilon_t < 1$, while fixing $\epsilon_b = m_b/m_{b'} (\sim 0.01)$.
We also vary the $4\times4$ CKM element $V_{t' b}$
in the range $0 < |V_{t' b}| < 0.2$ (see also next section), keeping
$|\lambda^{t'}_{sb}| \leq 0.02$
and varying $\tan\beta$ in the range, $1 < \tan\beta < 30$.
We made a scan over the entire parameter space by a flat random number
generator and obtained bounds and correlations among the various parameters mentioned above.

\begin{widetext}
\begin{table*}[htbp]
\begin{center}
\begin{tabular}{|c|c|}
\hline
$f_{bd}\sqrt{B_{bd}} = 0.224 \pm 0.015$\, {GeV} \cite{Gamiz:2009ku,gamizp}& $|V_{ub}| = (32.8 \pm 2.6)\times 10^{-4}$
\footnote{It is the weighted average of
$V_{ub}^{inl}=(40.1\pm 2.7 \pm 4.0)\times 10^{-4}$  and
$V_{ub}^{exl}=(29.7 \pm 3.1)\times 10^{-4}$. In our numerical analysis,
we increase the error on $V_{ub}$ by 50\% and take the total error to be
around 12\% due to the appreciable
disagreement between the two determinations.} \\
$\xi = 1.232 \pm 0.042$ \cite{Gamiz:2009ku,gamizp} & $|V_{cb}| = (40.86 \pm 1.0)\times 10^{-3} $ \\
$\eta_t = 0.5765\pm 0.0065$ \cite{buras1}  & $\gamma = (73.0 \pm 13.0)^{\circ} $\\
$\Delta{M_s} = (17.77 \pm 0.12) ps^{-1}$    & ${\cal{BR}}(B\to X_s \gamma) = (3.55 \pm 0.25)\times 10^{-4}$\\
$\Delta{M_d} = (0.507 \pm 0.005) ps^{-1}$    & $m_b(m_b) = 4.23 \, GeV$ \\
$f_B = (0.208 \pm 0.008)$  GeV  & $\alpha_s(M_Z) = 0.11$ \\
$m_t^{pole} = (170 \pm 4) $ GeV  & $\tau_{B^+} = 1.63\,ps$\\
 & $m_{\tau} = 1.77$  GeV  \\
\hline
\end{tabular}
\caption{Inputs used in order to constrain the 4G2HDM parameter space. When not
explicitly stated, we take the inputs from Particle Data Group \cite{PDG}.}
\label{tab1}
\end{center}
\end{table*}
\end{widetext}

Let us first consider the case $V_{t' b} \to 0$,
corresponding to the ``3+1" scenario, in which
the 4th generation quarks do not mix with the quarks of
the 1st three generations (we assume that
$|V_{t' b}| >> |V_{t' s}|, |V_{t' d}|$).
In this case, the top-quark loops become dominant, since
contributions to the amplitudes of $B \to X_s \gamma$ and $B_q$-${\bar B_q}$ mixing from $t'$-loops
are mostly suppressed apart from the terms which are proportional to
$(m_{b^\prime}/m_b) \cdot \lambda^t_{bs}$ (see Eqs.~\ref{eq36} and \ref{eq56}).

\begin{widetext}
\begin{figure}[htb]
\begin{center}
\epsfig{file=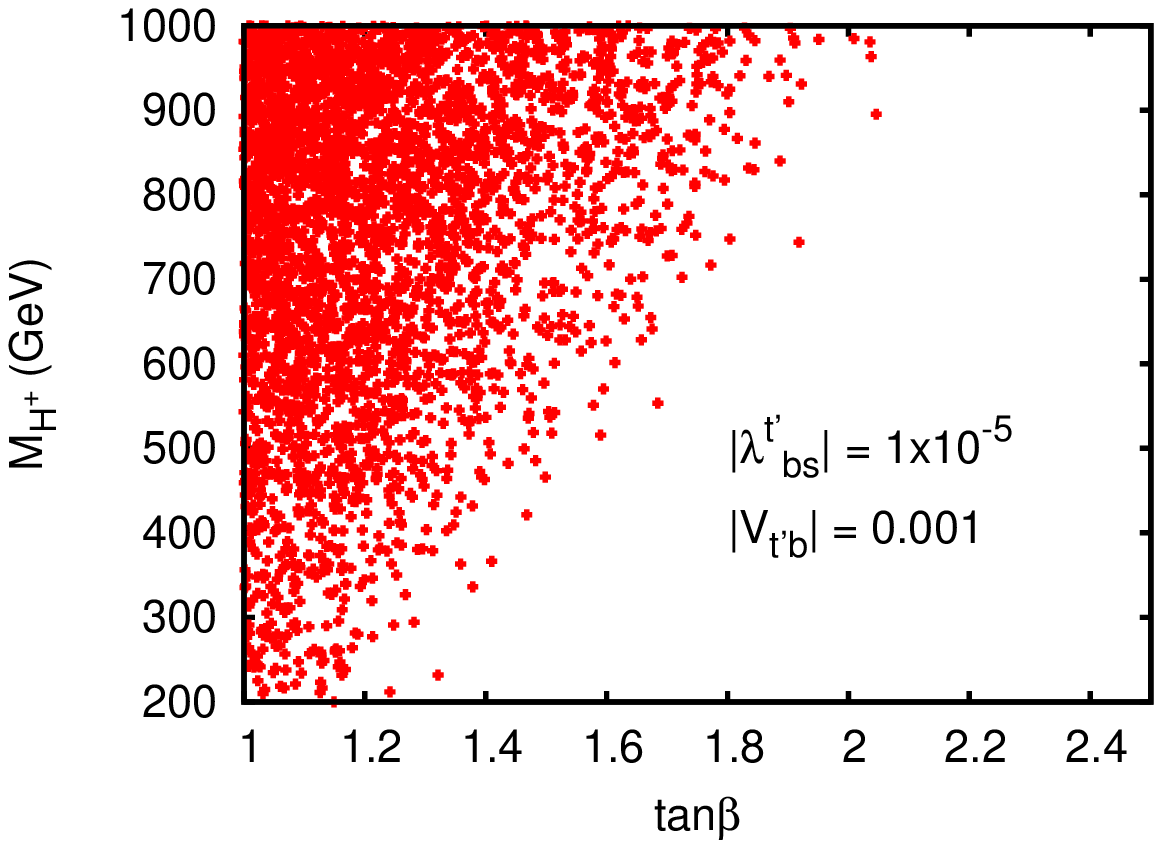,height=5cm,width=5cm,angle=0}
\epsfig{file=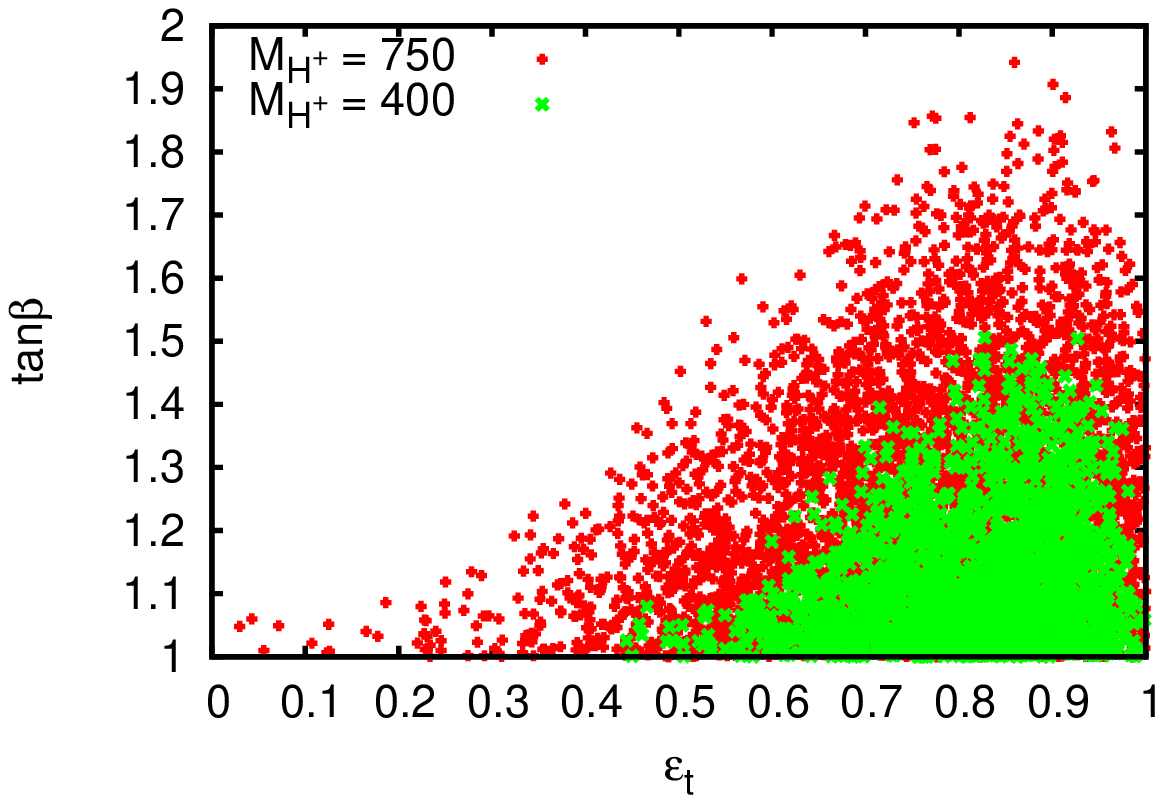,height=5cm,width=5cm,angle=0}
\caption{\emph{The allowed parameter space in the
$m_{H^+}-\tan\beta$ and $\tan\beta - \epsilon_t$ planes, following
constraints from $B \to X_s \gamma$ and $B_q$-${\bar B_q}$ mixing,
in the 4G2HDM-I, for
$V_{t' b} = 0.001$, $m_{t'}=500$ GeV, $m_{b'}=450$ GeV and
$\epsilon_b = m_b/m_{b'}$.}}
\label{figI-vtb0}
\end{center}
\end{figure}
\end{widetext}

\begin{widetext}
\begin{figure}[htb]
\begin{center}
\epsfig{file=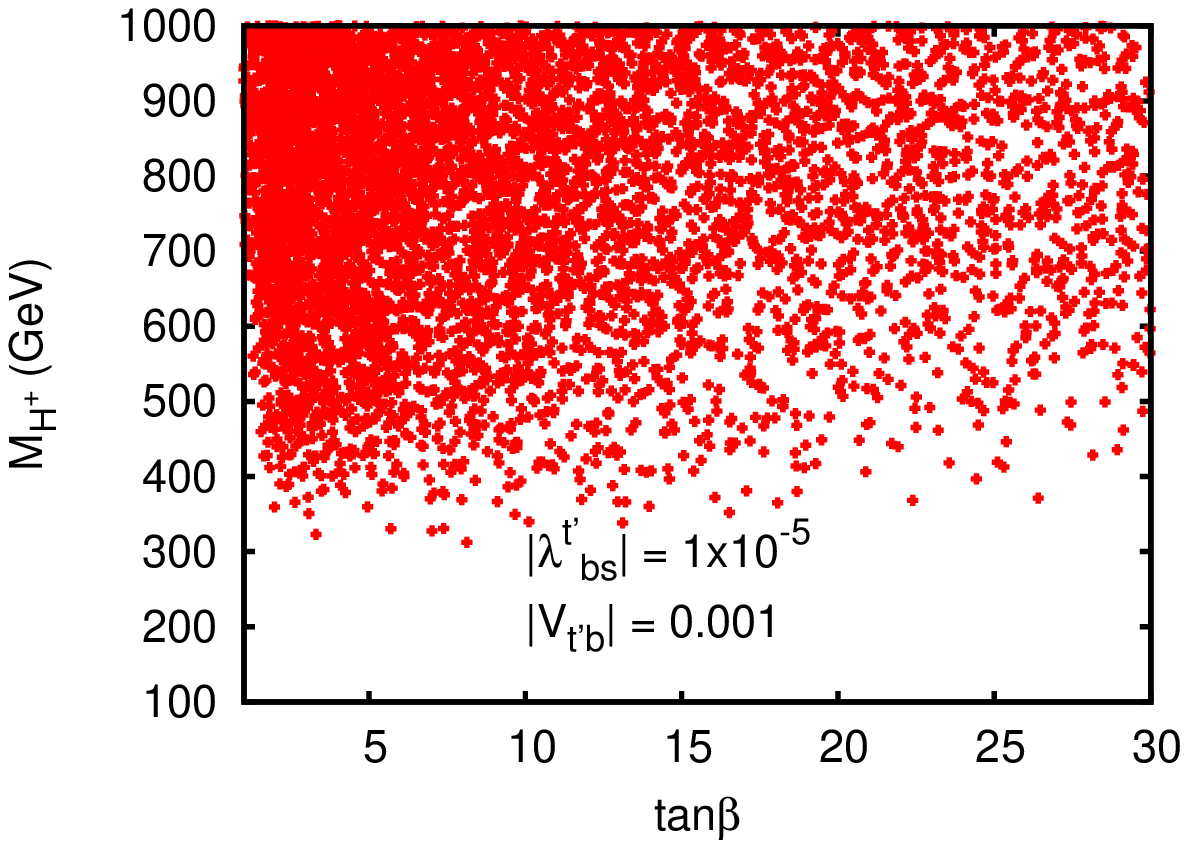,height=5cm,width=5cm,angle=0}
\epsfig{file=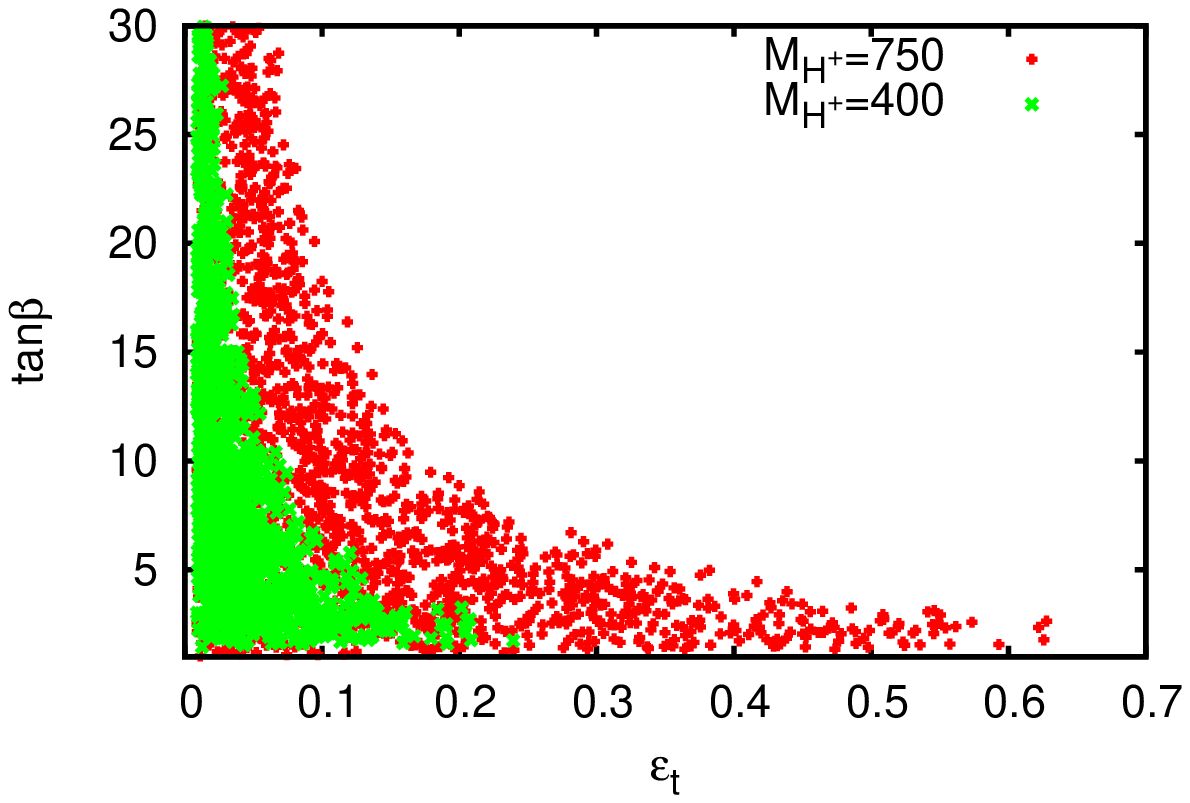,height=5cm,width=5cm,angle=0}
\caption{\emph{Same as Fig.~\ref{figI-vtb0} for the 4G2HDM-II.}}
\label{figII-vtb0}
\end{center}
\end{figure}
\end{widetext}

\begin{widetext}
\begin{figure}[htb]
\begin{center}
\epsfig{file=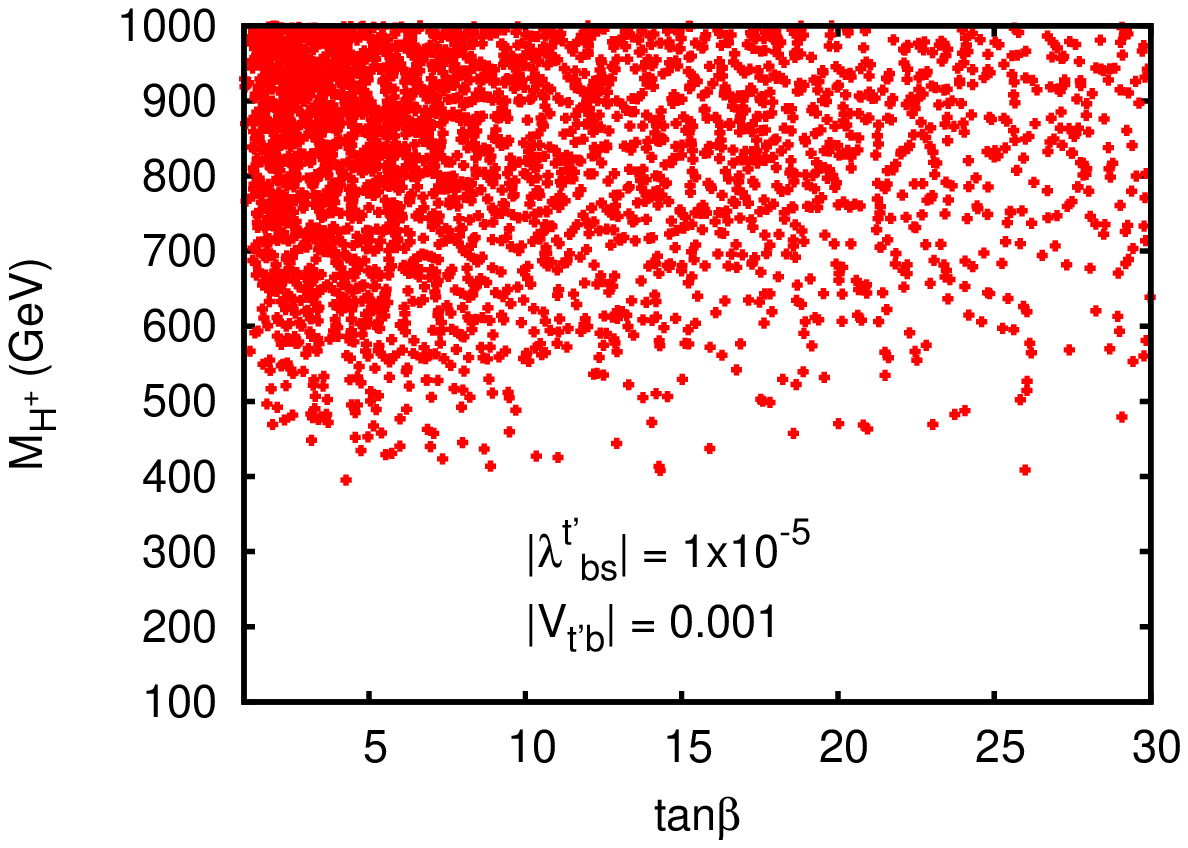,height=5cm,width=5cm,angle=0}
\epsfig{file=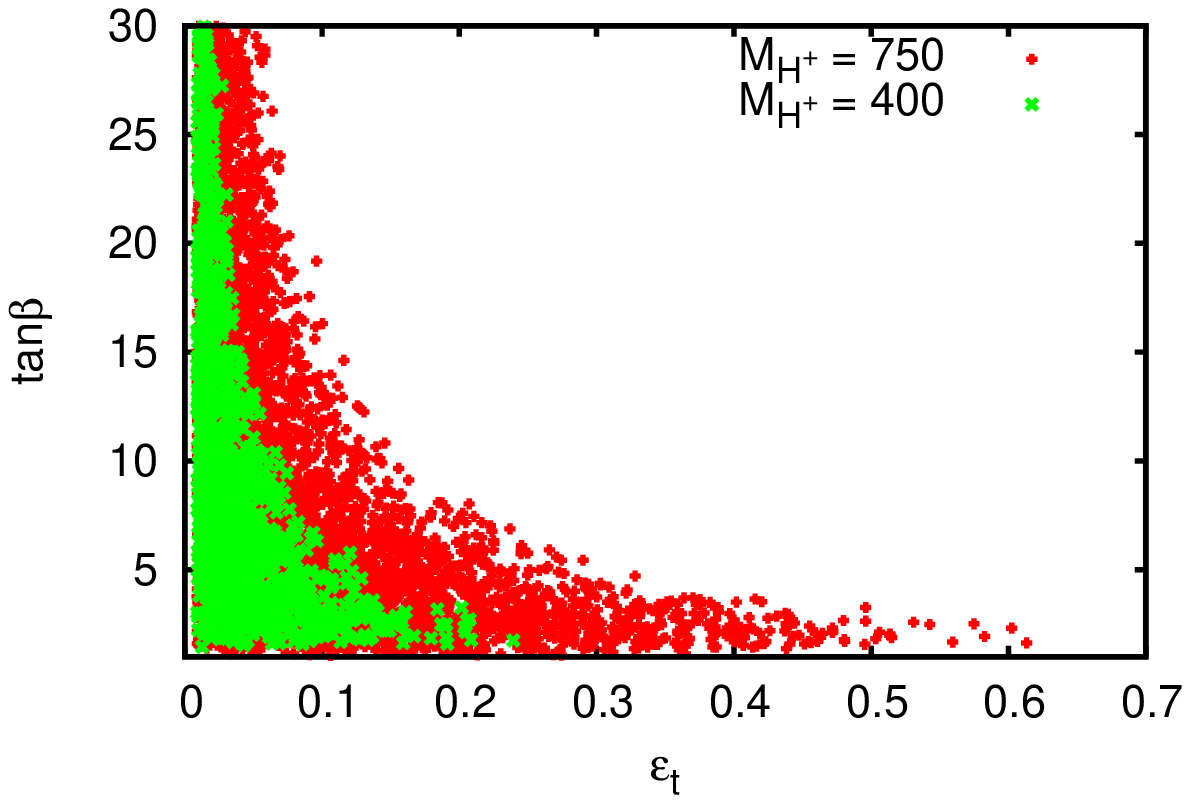,height=5cm,width=5cm,angle=0}
\caption{\emph{Same as Fig.~\ref{figI-vtb0} for the 4G2HDM-III.}}
\label{figIII-vtb0}
\end{center}
\end{figure}
\end{widetext}

In Figs.~\ref{figI-vtb0}, \ref{figII-vtb0} and \ref{figIII-vtb0}
we plot the
allowed ranges in the $m_{H^+} - \tan\beta$ (left plots) and the
$\tan\beta - \epsilon_t$ (right plots) planes,
in the 4G2HDM of types I, II and III, respectively,
using $|V_{t' b}| = 0.001$
(with $|\lambda^{t'}_{sb}| = 10^{-5}$ correspondingly).


We see that in the type-I 4G2HDM, the ``3+1" scenario typically imposes
$\tan\beta \sim 1$ with $\epsilon_t$ typically larger than about 0.4 when $m_{H^+} \lsim 500$ GeV.
In particular, for a fixed $\epsilon_t$ the upper bound on $\tan\beta$ is
reduced with the charged Higgs mass,
allowing $m_{H^+} \gsim 200$ GeV for $\tan\beta \sim 1$ and
restricting $m_{H^+} \gsim 500$ GeV for $\tan\beta \gsim 1.5$.
In the type II and type III 4G2HDMs we observe a similar correlation
between $\tan\beta$ and $m_{H^+}$, however, larger $\tan\beta$
are allowed for $\epsilon_t \lsim m_t/m_{t'}$
and a charged Higgs mass typically heavier than 400 GeV.

\begin{widetext}
\begin{figure}[htb]
\begin{center}
\epsfig{file=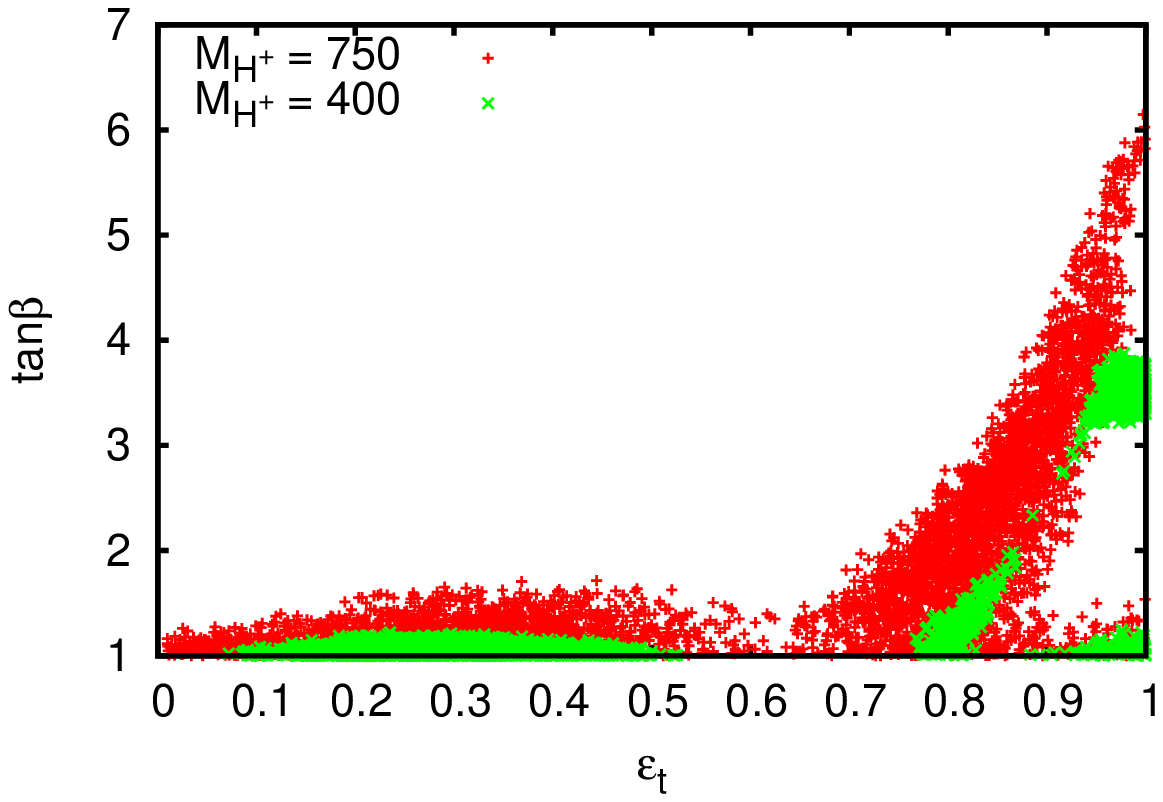,height=5cm,width=5cm,angle=0}
\epsfig{file=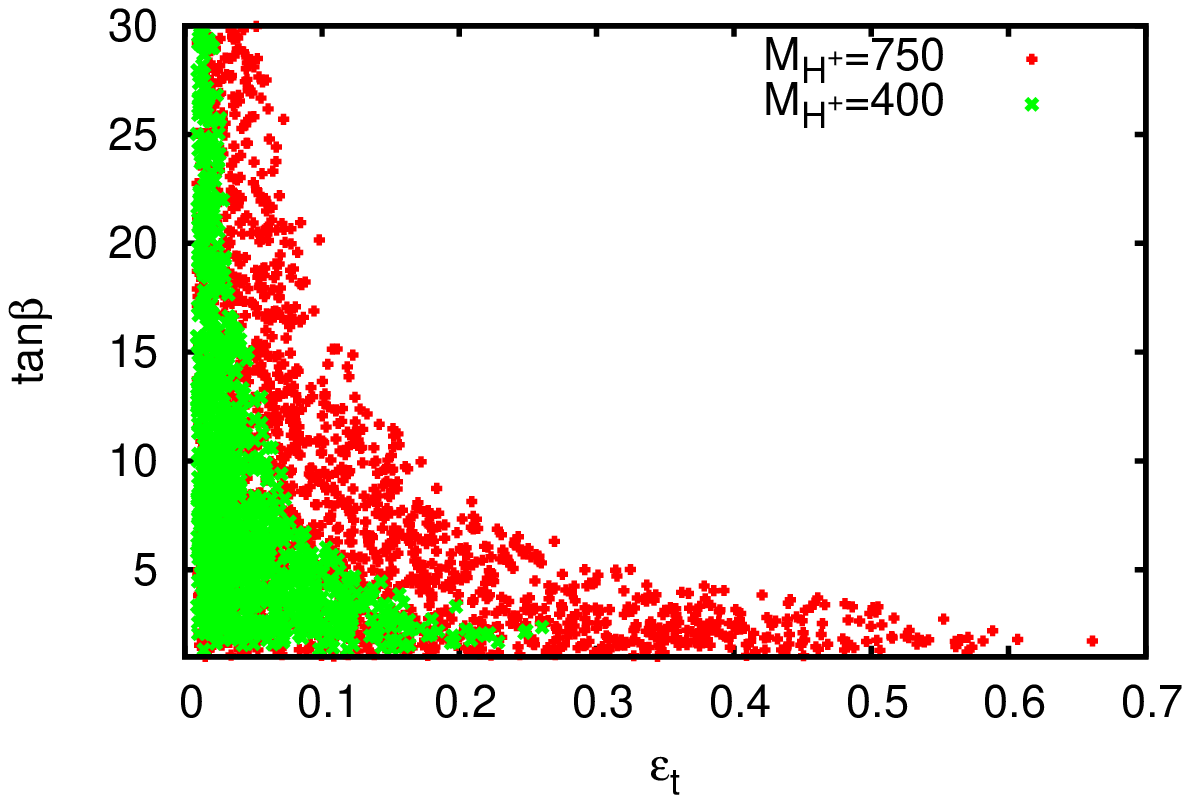,height=5cm,width=5cm,angle=0}
\epsfig{file=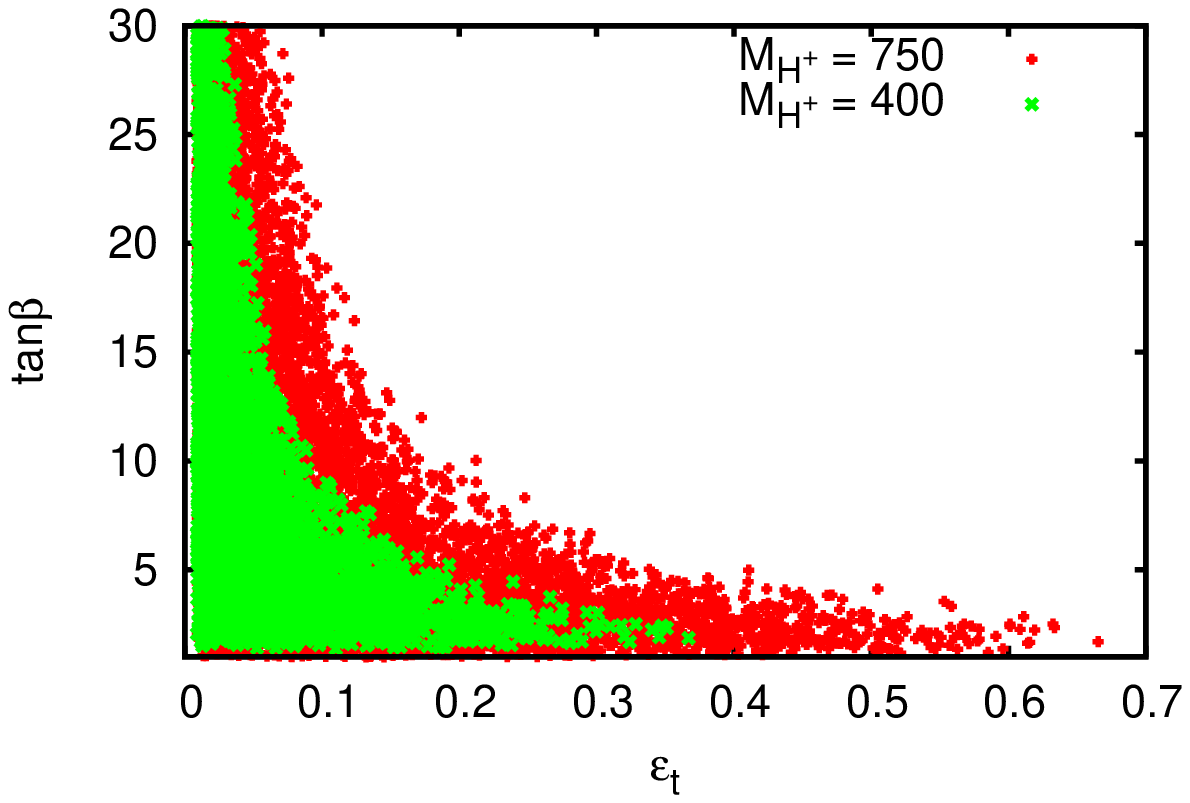,height=5cm,width=5cm,angle=0}
\caption{\emph{Allowed parameter space in the
$\tan\beta - \epsilon_t$ plane in the 4G2HDM of type-I (left),
type-II (middle) and type-III (right), for
$|V_{t' b}| = 0.2$, $m_{t'}=500$ GeV, $m_{b'}=450$ GeV,
$\epsilon_b = m_b/m_{b'}$, $|\lambda^{t'}_{sb}|=0.004$ and
with $m_{H^+}=400$ and 750 GeV.}}
\label{figI-vtb}
\end{center}
\end{figure}
\end{widetext}

Let us now turn to the case of a Cabbibo size
mixing between the 4th and 3rd generation quarks,
setting $|V_{t' b}| = |V_{t b'}| =0.2$.
In Fig.~\ref{figI-vtb} we show the allowed parameter space in the $\tan\beta - \epsilon_t$ plane in
the 4G2HDM-I, II and III with $|V_{t' b}| = 0.2$, $m_{t'}=500$ GeV, $m_{b'}=450$ GeV and
$\epsilon_b = m_b/m_{b'}$. In addition, we take $|\lambda^{t'}_{sb}|=0.004$ for Type-I and 0.001 for Type-II and III models and depict these correlations for two different values of the charged Higgs mass: $M_{H^+} = 400$ and 750 GeV.
In the type II and type III 4G2HDMs we see a similar behavior as in the no mixing case
($V_{t' b} \to 0$), while in the 4G2HDM-I we see that ``turning on''
$V_{t' b}$ allows for a slightly larger $\tan\beta$, i.e., up to $\tan\beta \sim 5$ for $\epsilon_t \gsim 0.9$.

\begin{widetext}
\begin{figure}[htb]
\begin{center}
\epsfig{file=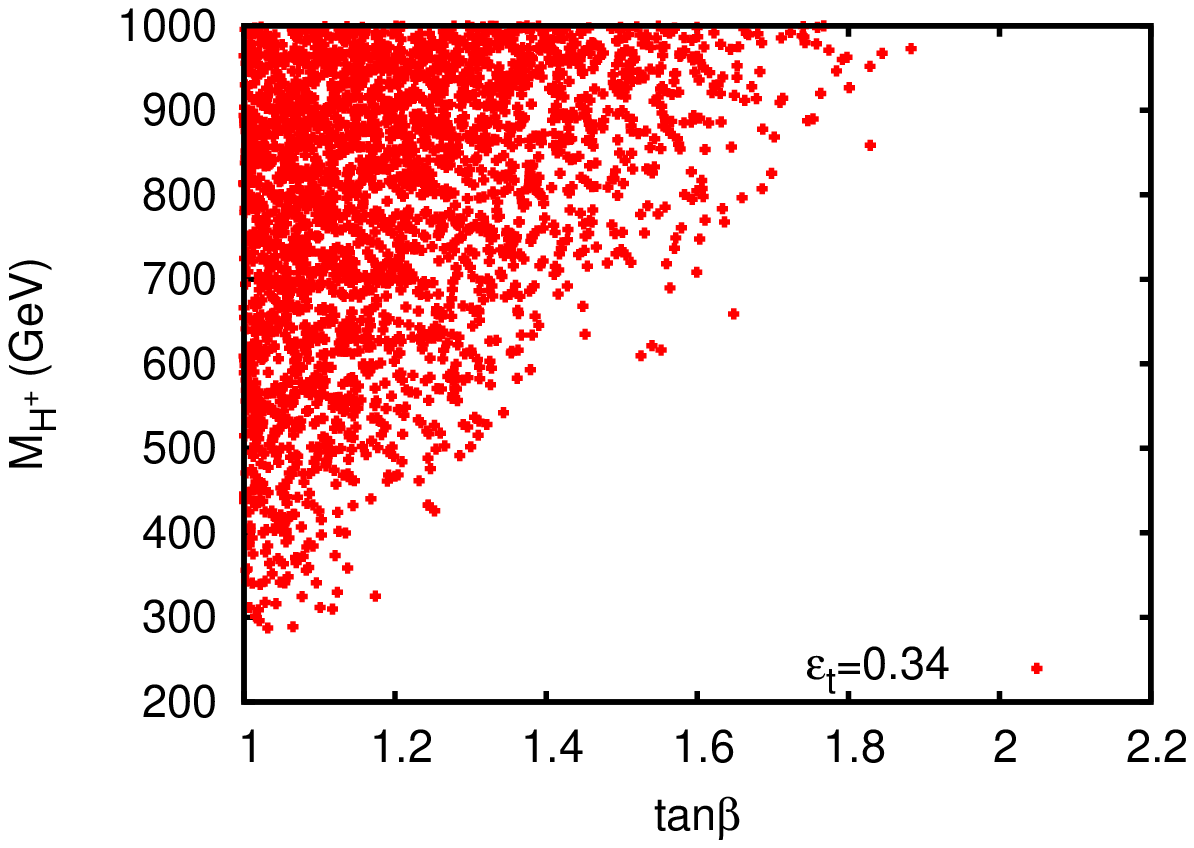,height=5cm,width=5cm,angle=0}
\epsfig{file=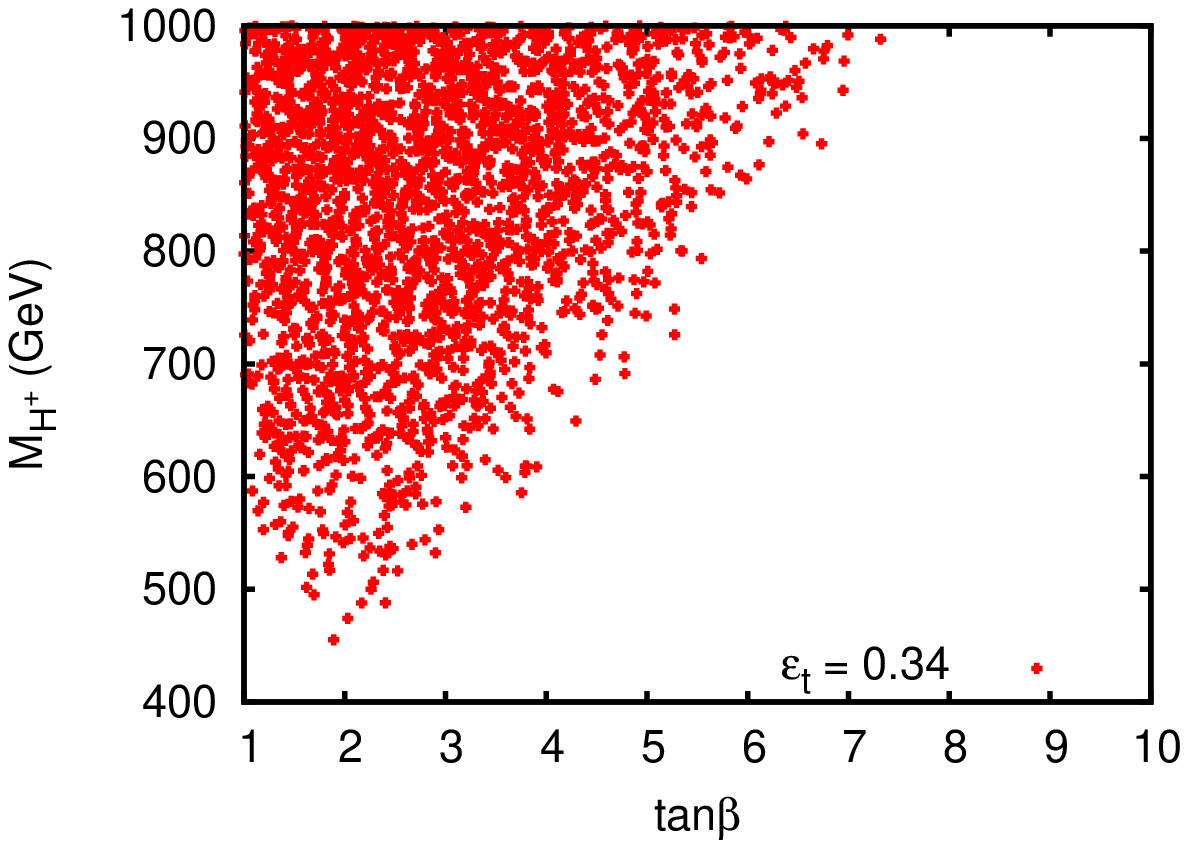,height=5cm,width=5cm,angle=0}
\epsfig{file=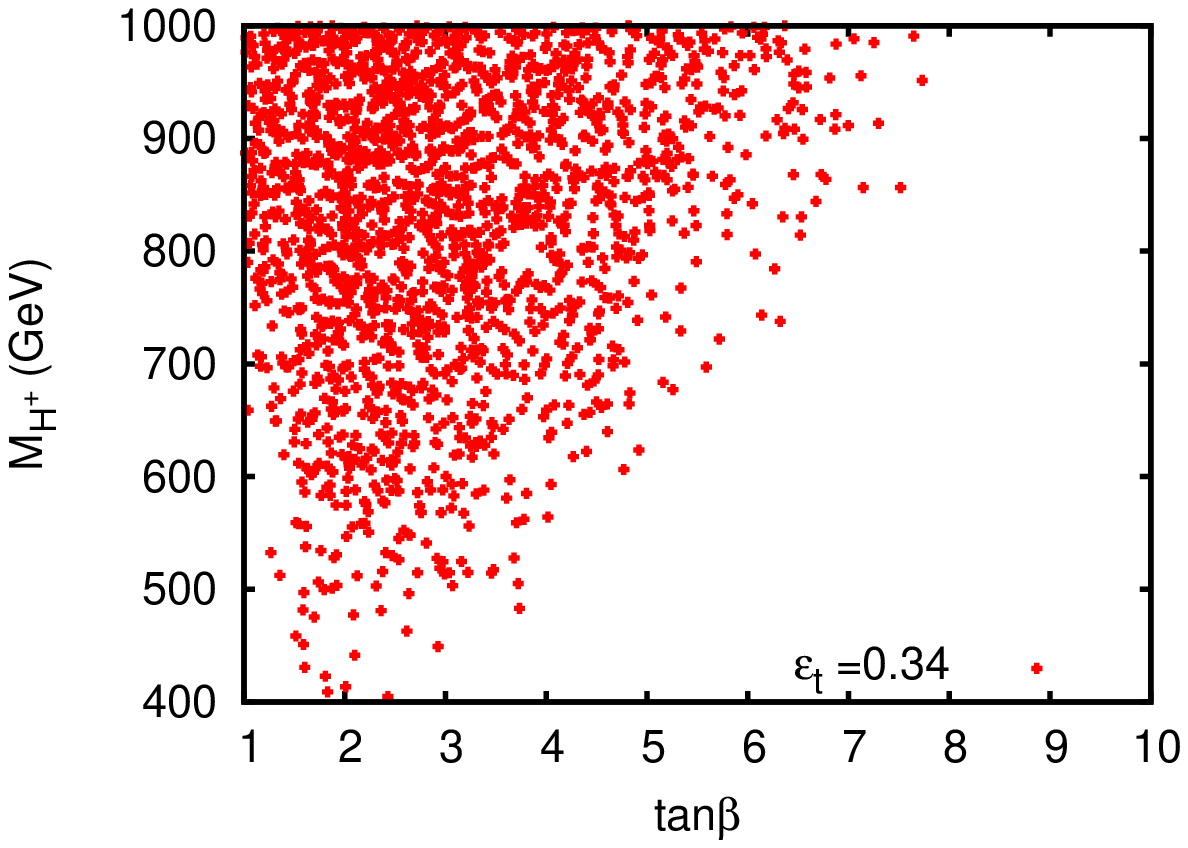,height=5cm,width=5cm,angle=0}
\caption{\emph{Allowed parameter space in the
$m_{H^+} - \tan\beta$ plane in the 4G2HDM of type-I (left),
type-II (middle) and type-III (right), for
$|V_{t' b}| = 0.2$, $m_{t'}=500$ GeV, $m_{b'}=450$ GeV,
$\epsilon_b = m_b/m_{b'}$, $|\lambda^{t'}_{sb}|=0.004$ and
$\epsilon_t = 0.34 (\sim m_t/m_{t'})$.}}
\label{figII-vtb}
\end{center}
\end{figure}
\end{widetext}

\begin{widetext}
\begin{figure}[htb]
\begin{center}
\epsfig{file=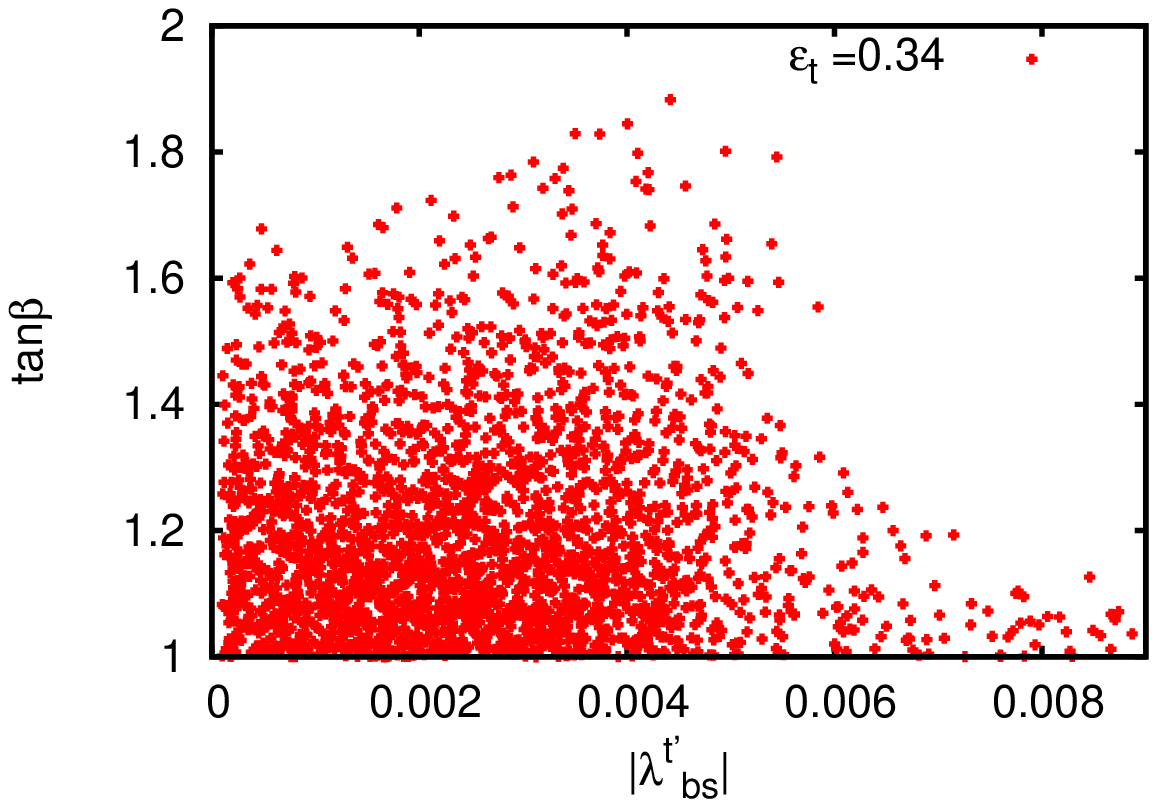,height=5cm,width=5cm,angle=0}
\epsfig{file=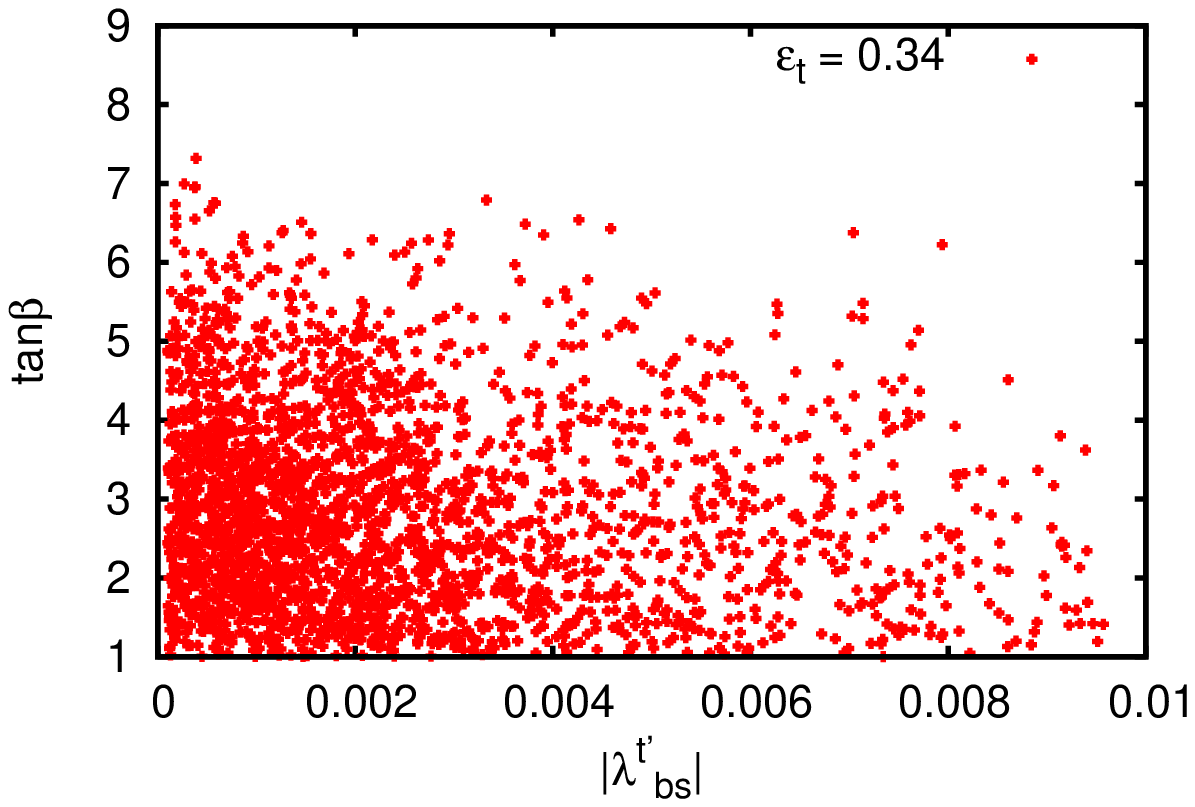,height=5cm,width=5cm,angle=0}
\epsfig{file=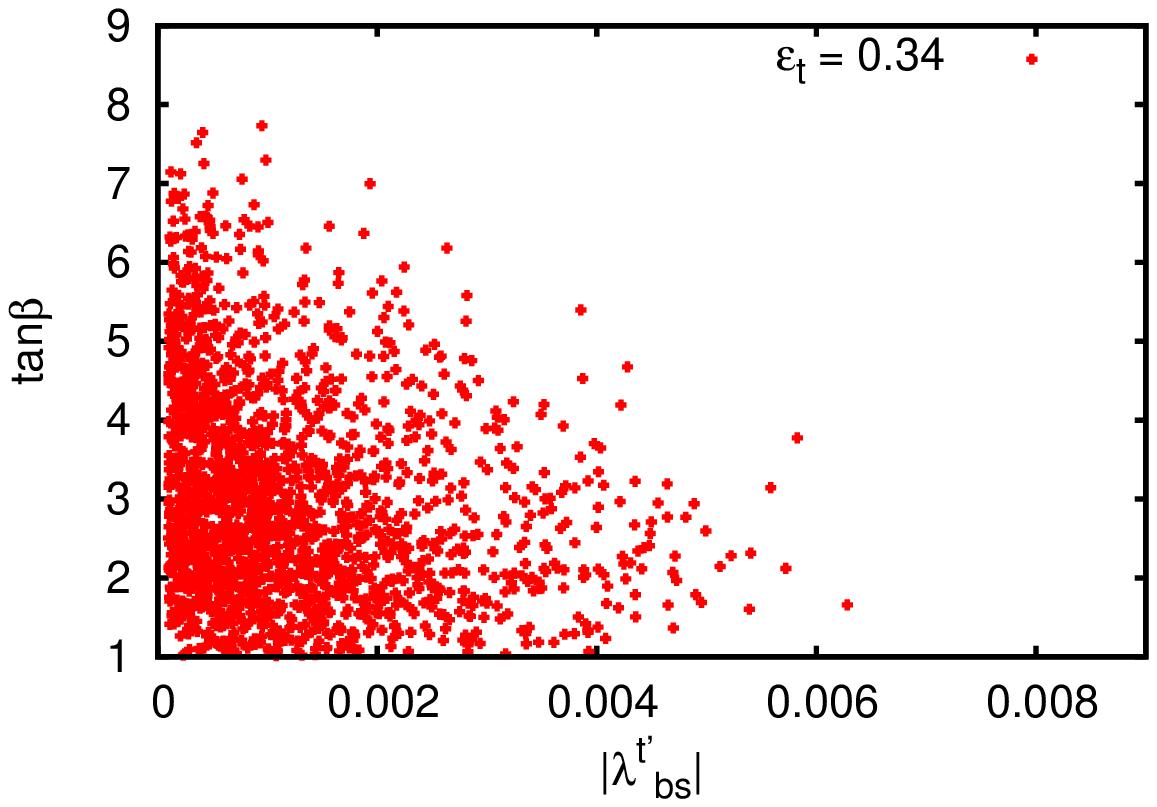,height=5cm,width=5cm,angle=0}
\caption{\emph{Allowed parameter space in the
$\tan\beta - |\lambda^{t'}_{sb}|$ plane in the 4G2HDM of type-I (left),
type-II (middle) and type-III (right), for a fixed $|V_{t' b}|=0.2$,
$\epsilon_t \sim m_t/m_{t'}$ and for
$m_{t'}=500$ GeV, $m_{b'}=450$ GeV and
$\epsilon_b = m_b/m_{b'}$.}}
\label{figIII-vtb}
\end{center}
\end{figure}
\end{widetext}

With a similar set of inputs, setting now $\epsilon_t \sim {m_t}/{m_{t'}}$,
in Figs.~\ref{figII-vtb} and \ref{figIII-vtb} we plot $\tan\beta$ as
a function of $M_{H^+}$ (where $|\lambda^{t'}_{sb}|$ is kept free)
and of $\lambda^{t'}_{bs}$ (where $M_{H^+}$ is kept free), respectively, in the
three different types of our 4G2HDMs.
We note that, similar to the no mixing case, larger values of $\tan\beta$
are allowed in the 4G2HDM of types II and III. Furthermore,
$m_{H^+} \sim 300$ GeV and $\tan\beta \sim 1$ are allowed in the 4G2HDM-I, and
from Fig.~\ref{figII-vtb} we see that $|\lambda^{t'}_{bs}|$ up to 0.01 is allowed in the case
of the 4G2HDM-I and II, while in
4G2HDM-III $|\lambda^{t'}_{sb}| \lsim 0.005$ is typically required.


To summarize this section, we find that the parameter space of our 4G2HDMs, when subject
to constraints from $Br(B\to X_s \gamma)$ and $B_q - {\bar B_q}$ mixing,
can be characterized by the following features:
\begin{itemize}
\item In the type II and III 4G2HDMs large
$\tan\beta \gsim 20$ are allowed for $\epsilon_t \lsim 0.1$.
\item In the 4G2HDM-I $\tan\beta$ is typically
restricted
to be $\tan\beta \sim {\cal O}(1)$ with $\epsilon_t \sim m_t/m_{t'}$,
reaching at most $\tan\beta \sim 5$ if $\epsilon_t \sim 1$
and $V_{t' b} \sim {\cal O}(0.1)$, i.e., of the
size of the Cabbibo angle.
\item The charged Higgs mass is typically heavier
than about 400 GeV in the type II and III 4G2HDM
and is allowed to be as light as 200-300 GeV (depending on $V_{t' b}$)
in the 4G2HDM-I. In all three models the lower bound on
$m_{H^+}$ increases (typically linearly)
with $\tan\beta$; reaching $m_{H^+} \sim 1$ TeV already
for $\tan\beta \sim 2$ in the 4G2HDM-I and
$\tan\beta \sim 7$ in the type II and III 4G2HDMs
if $\epsilon_t \sim m_t/m_{t'}$.
\item In the 4G2HDM-III, $|\lambda^{t'}_{sb}| \lsim 0.005$
is required if $\epsilon_t \sim m_t/m_{t'}$, but
values up to $|\lambda^{t'}_{sb}| \sim 0.01$ are still allowed
in the 4G2HDMs of types I and II.
\end{itemize}

\subsection{Constraints from $Z \to b \bar b$}

It has been long known that the decay $Z \to b \bar b$ is very sensitive to
effects of new heavy particles, in particular, to the dynamics of multi-Higgs models
through loop exchanges of both neutral and charged Higgs particles (see e.g., \cite{ARS,Haber}).
The $Zb \bar b$ vertex can be parameterized as follows:
\begin{eqnarray}
V_{qqZ} \equiv -i \frac{g}{c_W} \bar q \gamma_\mu \left( \bar g_{qL} L + \bar g_{qR} R \right) q Z^\mu ~,
\end{eqnarray}
where $s_W(c_W)=\sin\theta_W(\cos\theta_W)$, $L(R) = \left( 1-(+) \gamma_5 \right)/2$ and
\begin{eqnarray}
\bar g_{qL,R} = g_{qL,R}^{SM} + g_{qL,R}^{new}~,
\end{eqnarray}
so that $g_{qL,R}^{SM}$ are the SM (1-loop) quantities and $g_{qL,R}^{new}$ are
the new physics 1-loop corrections.

The effects of the new physics, $g_{qL,R}^{new}$,
is best studied via the well measured quantity $R_b$:
\begin{eqnarray}
R_b \equiv \frac{\Gamma(Z \to b \bar b)}{\Gamma(Z \to {\rm hadrons})}~,
\end{eqnarray}
which is a rather clean test of the SM. In particular,
being a ratio between two hadronic rates,
most of the electroweak, oblique and QCD corrections cancel between numerator and denumerator.

Following the analysis in \cite{ARS}, we parameterize the effects of new physics in $R_b$
in terms of the corrections $\delta_b$ and $\delta_c$ to
the decays $Z \to b \bar b$ and $Z \to c \bar c$, respectively:
\begin{eqnarray}
R_b = R_b^{SM} \frac{1+\delta_b}{1+ R_b^{SM} \delta_b + R_c^{SM} \delta_c} \label{Rb}~,
\end{eqnarray}
where $R_b^{SM}=0.21578 \pm 0.00005$ and $R_c^{SM}= 0.17224 \pm 0.00003$ \cite{EL2009}
are the corresponding 1-loop quantities calculated
in the SM, and $\delta_q$ are the new physics corrections defined in terms of the
$Z q \bar q$ couplings as:
\begin{eqnarray}
\delta_q = 2 \frac{ g_{qL}^{SM} g_{qL}^{new} + g_{qR}^{SM} g_{qR}^{new} }
{ \left( g_{qL}^{SM} \right)^2 + \left( g_{qR}^{SM} \right)^2 }~,
\end{eqnarray}
\begin{figure}[htb]
\begin{center}
\epsfig{file=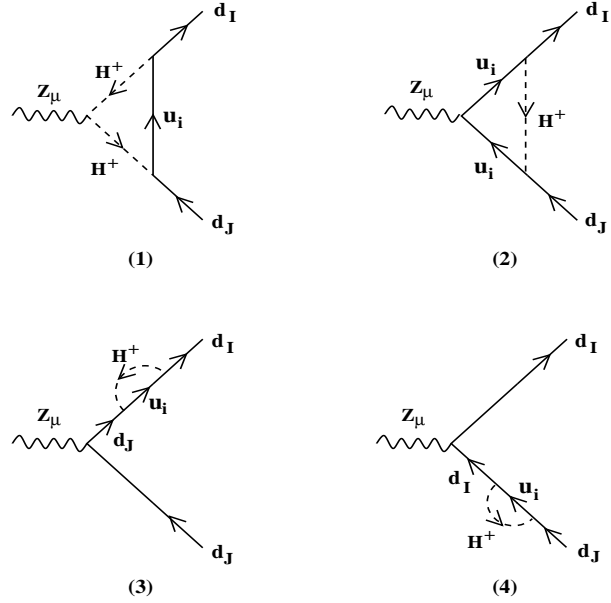,height=8cm,width=8cm,angle=0}
\caption{\emph{One-loop diagrams for corrections to $Z \to d_I \bar d_J$ from charged Higgs loops.}}
\label{diagrams}
\end{center}
\end{figure}
\begin{figure}[htb]
\begin{center}
\epsfig{file=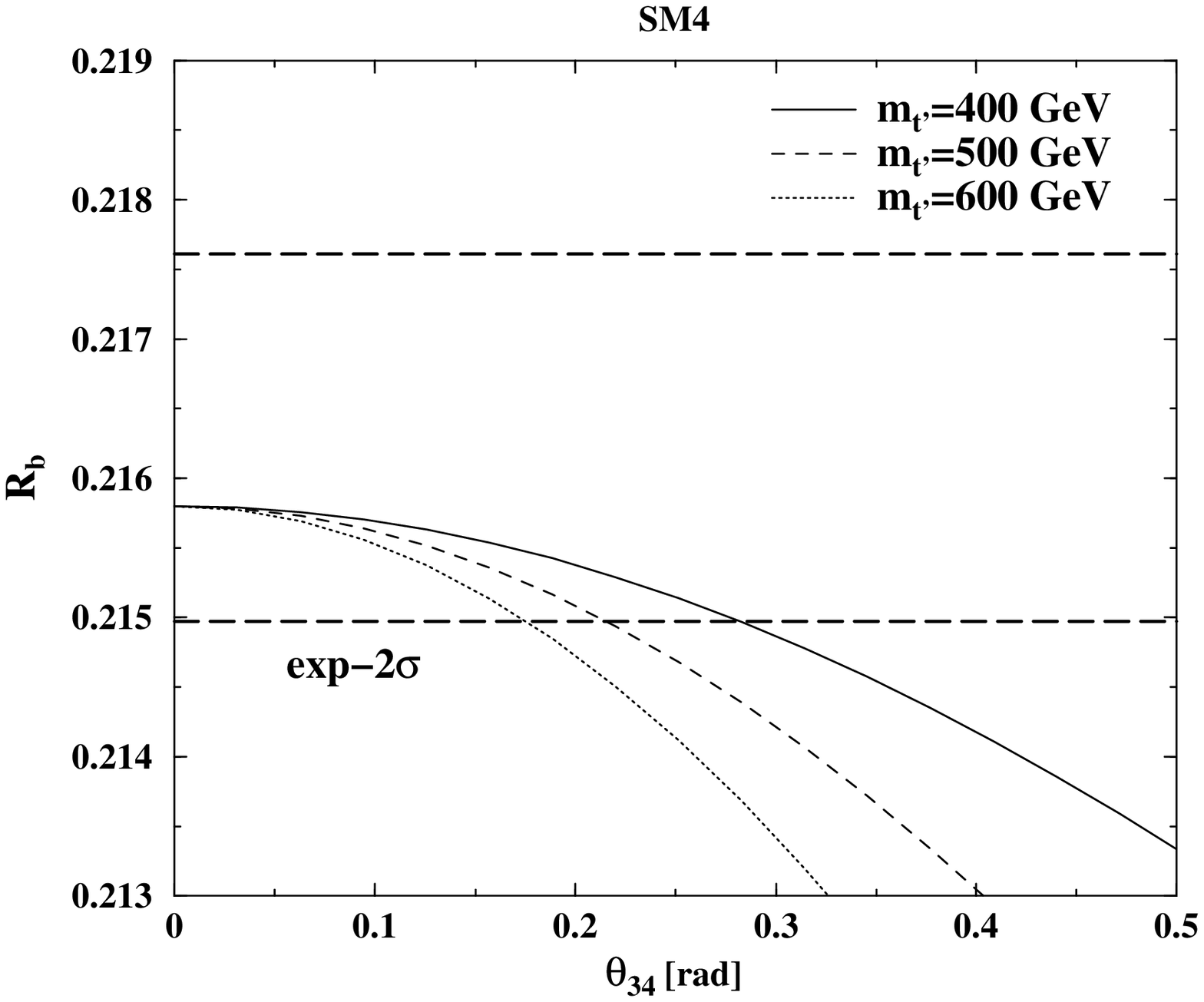,height=8cm,width=8cm,angle=0}
\epsfig{file=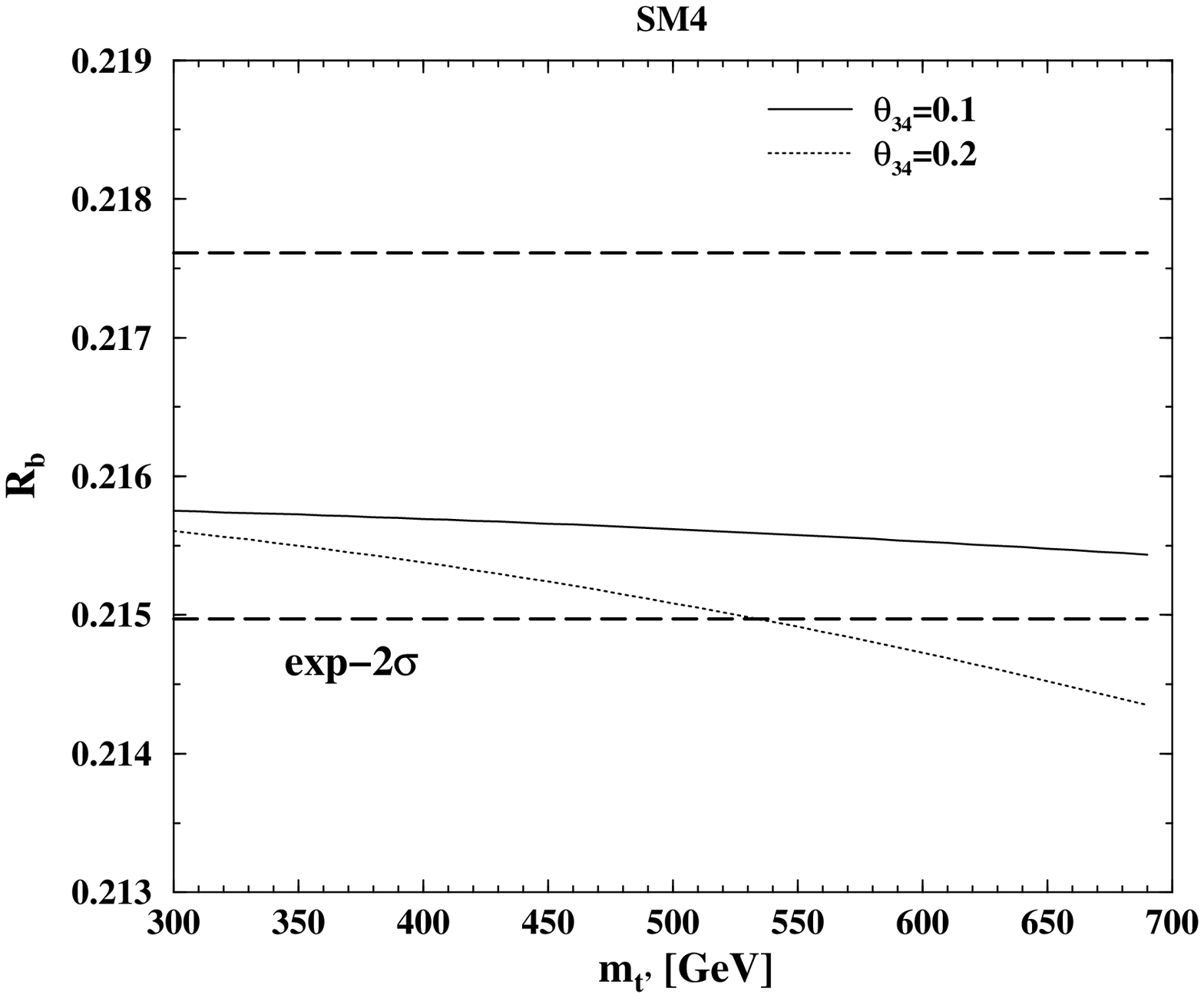,height=8cm,width=8cm,angle=0}\\
\epsfig{file=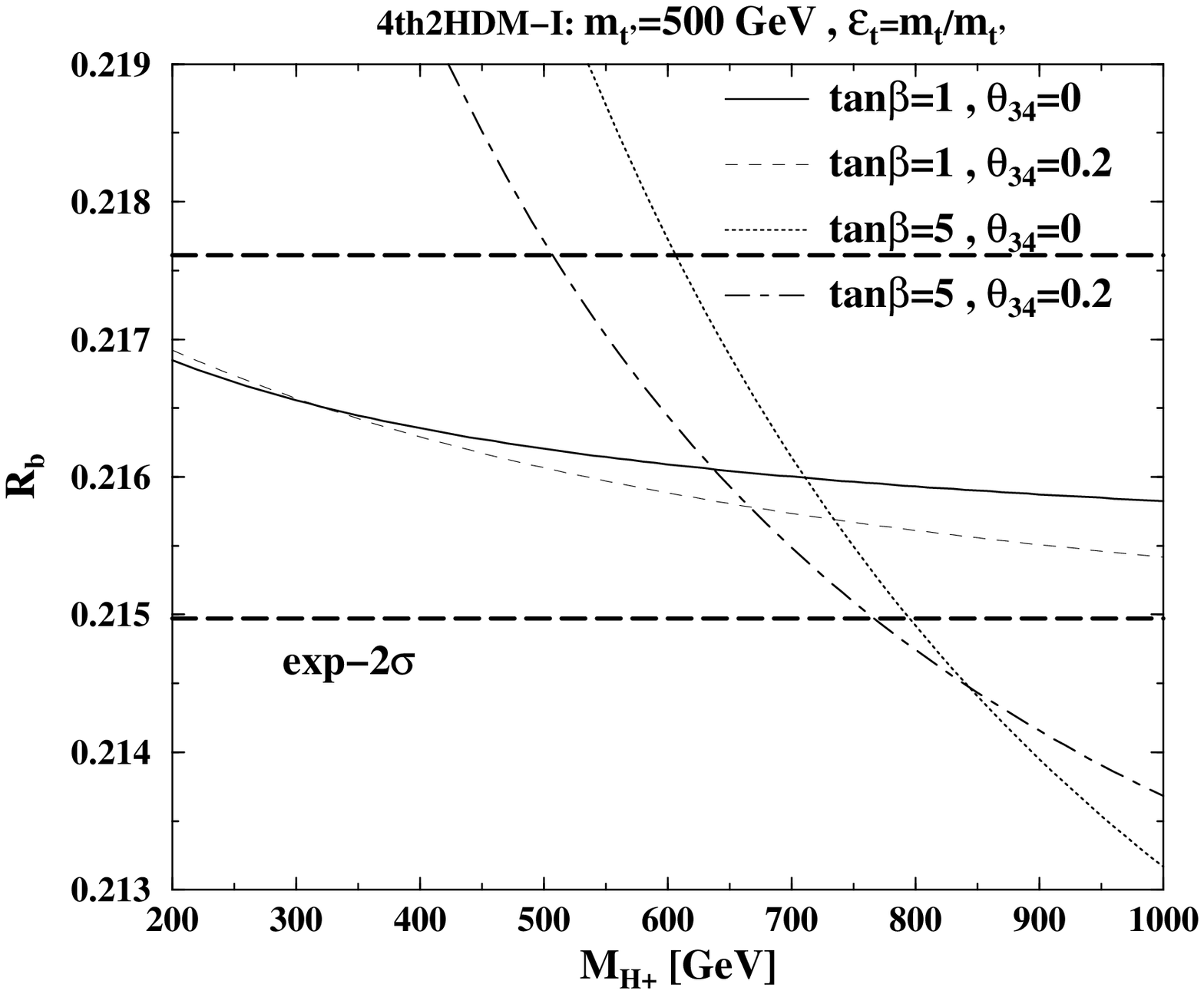,height=8cm,width=8cm,angle=0}
\epsfig{file=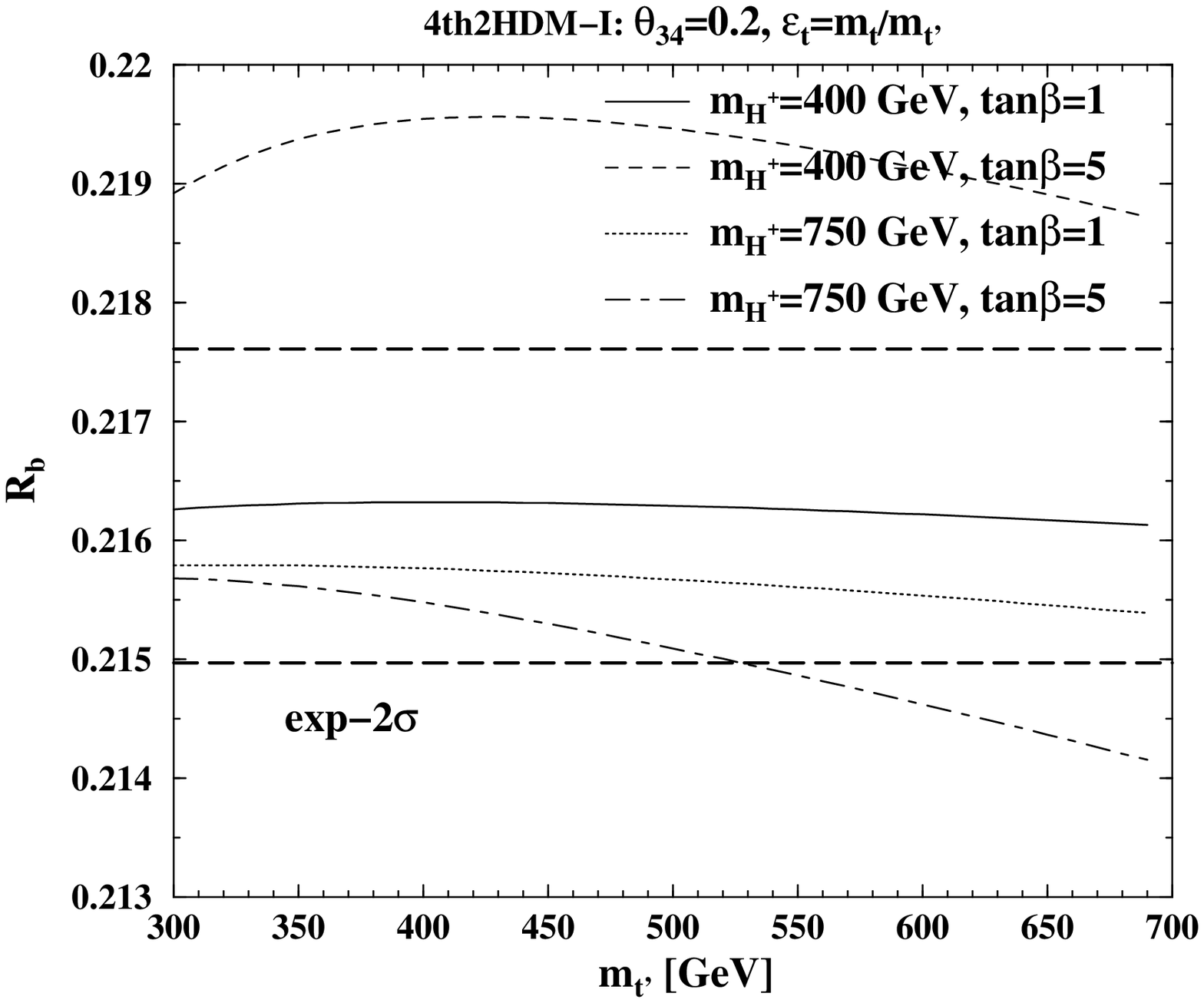,height=8cm,width=8cm,angle=0}
\caption{\emph{Upper plots: $R_b$ in the SM4,
as a function of $\theta_{34}$ for several values
of the $t^\prime$ mass (left) and as a function
of $m_{t^\prime}$ for $\theta_{34}=0.1$ and $0.2$ (right).
Lower plots: $R_b$ in the
4G2HDM-I, as a function of
the charged Higgs mass for
$m_{t^\prime}=500$ GeV, $\epsilon_t = m_t/m_{t^\prime}$
and for $(\tan\beta,\theta_{34})=(1,0),(1,0.2),(5,0),(5,0.2)$ (left),
and as
a function of $m_{t^\prime}$ for $\theta_{34}=0.2$,
$\epsilon_t = m_t/m_{t^\prime}$ and
for $(m_{H^+}~[{\rm GeV}],\tan\beta)=(400,1),(400,5),(750,1),(750,5)$ (right).}}
\label{figRb}
\end{center}
\end{figure}
\begin{widetext}
\begin{figure}[htb]
\begin{center}
\epsfig{file=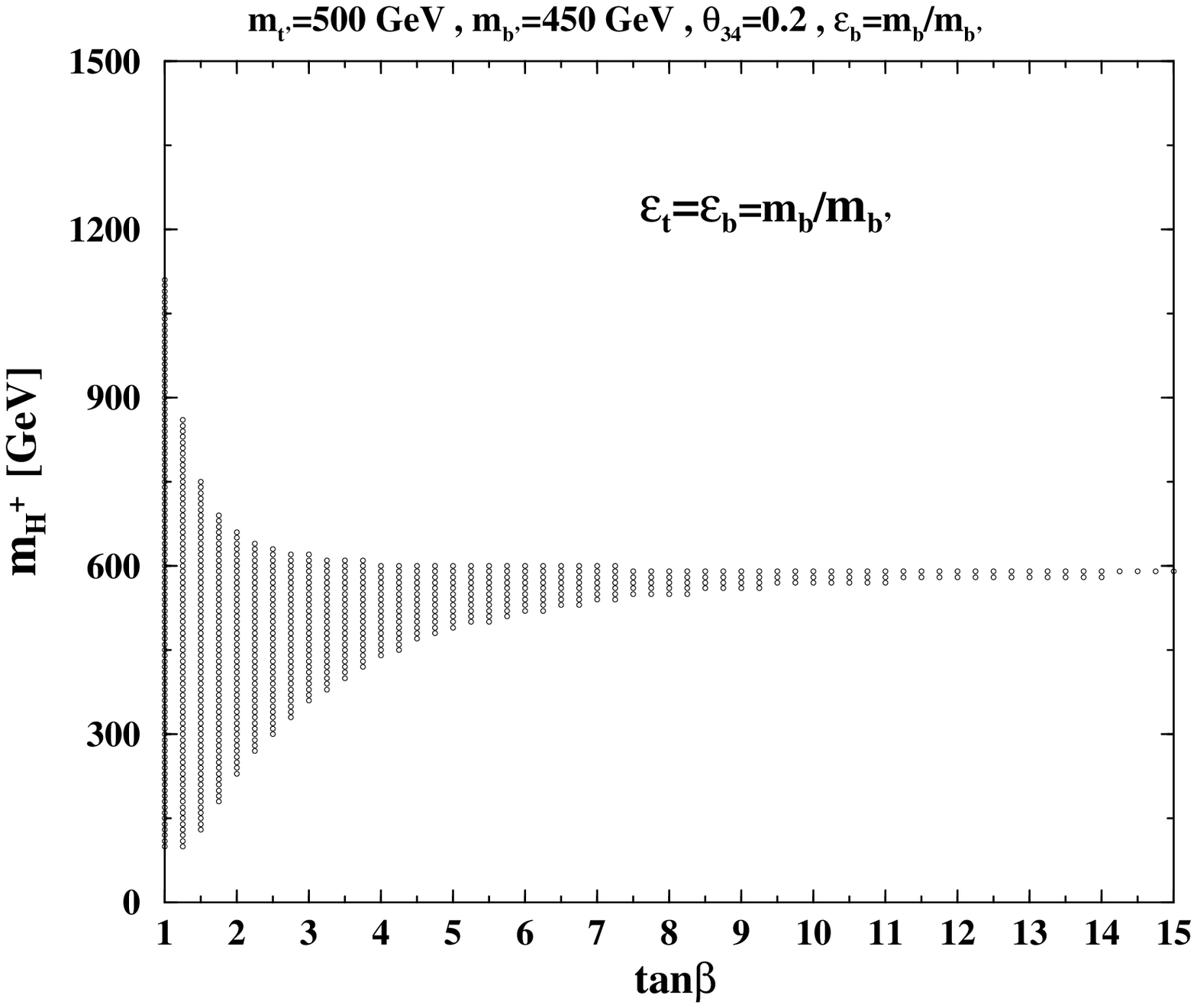,height=5cm,width=5cm,angle=0}
\epsfig{file=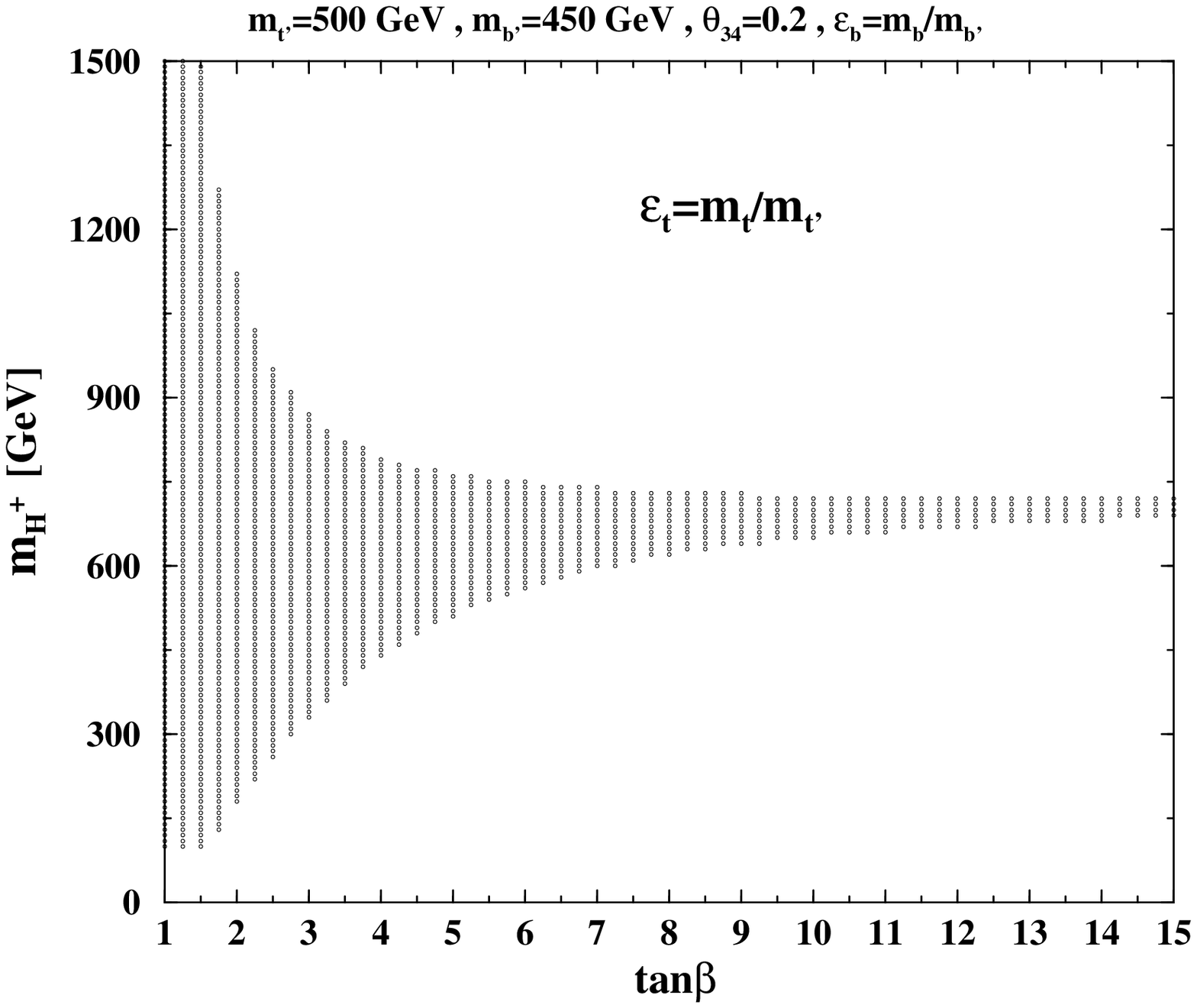,height=5cm,width=5cm,angle=0}
\epsfig{file=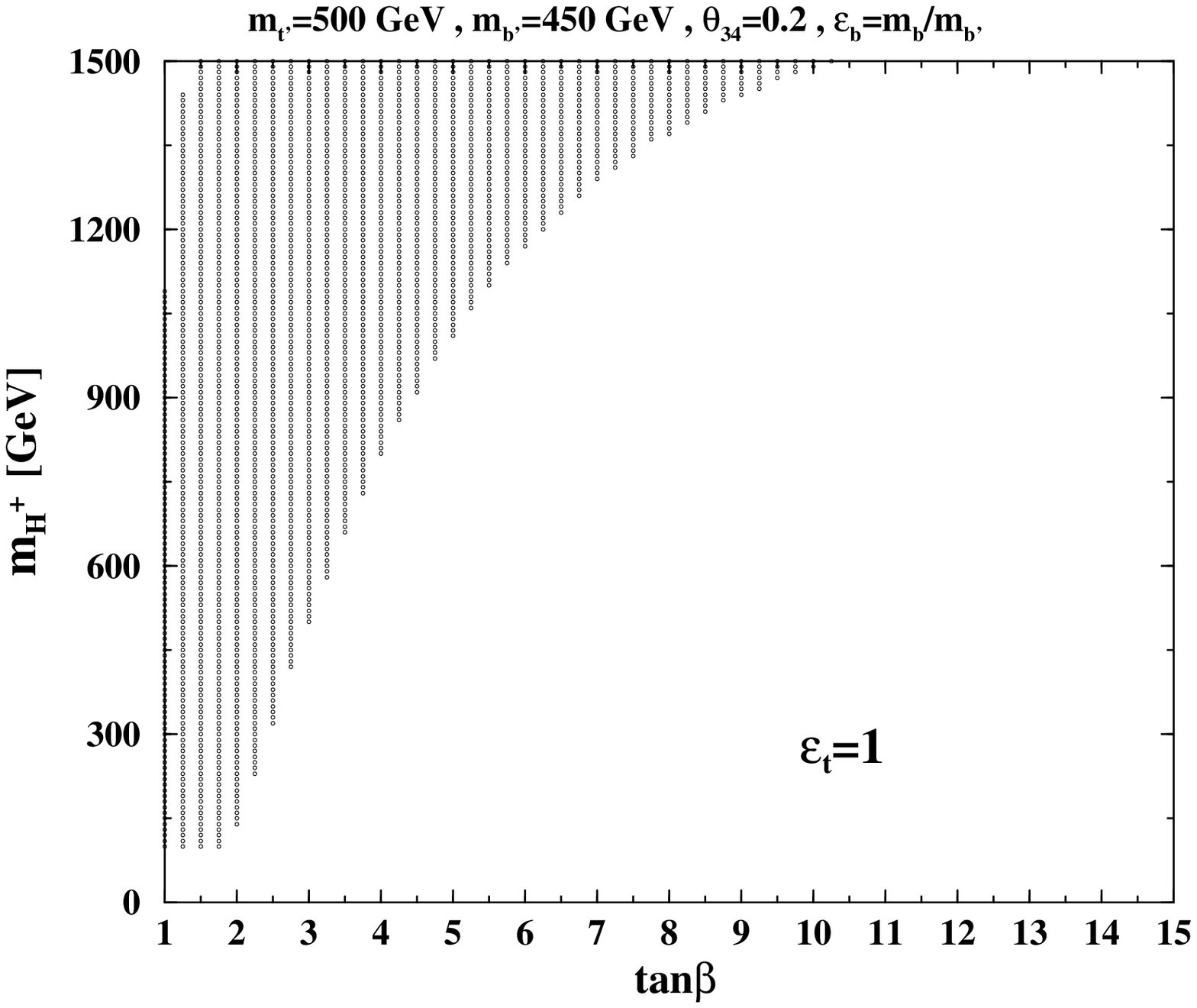,height=5cm,width=5cm,angle=0}
\caption{\emph{Allowed
area in the $m_{H^+} - \tan\beta$ in the 4G2HDM-I, subject to the
$R_b$ measurement (within $2\sigma$),
for $m_{t^\prime}=500$ GeV, $m_{b^\prime}=450$ GeV, $\theta_{34}=0.2$, $\epsilon_b = m_b/m_{b^\prime}$ and for three values of
the $t-t^\prime$ mixing parameter: $\epsilon_t = \epsilon_b \sim 0.01$ (left plot), $\epsilon_t = m_t/m_{t^\prime} \sim 0.35$ (middle plot) and $\epsilon_t = 1$ (right plot).}}
\label{figmhctb}
\end{center}
\end{figure}
\end{widetext}

With the new scalar-fermion interactions in Eqs.~\ref{Sff1}-\ref{Sff2},
the corrections to $R_b$ from a 4th generation quarks in our 4G2HDMs
are of three types: (i) the SM4-like corrections due to the 1-loop
$W-t^\prime$ exchanges (see also \cite{SAGMN10,chenowitz,RbSM4}),
(ii) the 1-loop diagrams in Fig.~\ref{diagrams}
involving the $H^+ - t^\prime$ exchanges and (iii) the 1-loop
corrections involving the FC ${\cal H}^0 b b^\prime$
interactions (coming from the non-diagonal 34 and 43 elements in $\Sigma^d$),
where ${\cal H}^0 = h,H$ or $A$.

Let us first consider the SM4-like (non-decoupling) correction to $R_b$, i.e.,
$g_{qL}^{SM4}$ from the 1-loop diagrams involving the $W-t^\prime$ exchanges
(which are also present in our 4G2HDMs). It is given by \cite{SAGMN10,chenowitz}:
\begin{eqnarray}
g_{qL}^{SM4} = \frac{g^2}{64 \pi^2 c_W^2}
\left( \frac{m_{t^\prime}^2}{m_Z^2} - \frac{m_{t}^2}{m_Z^2} \right) \sin^2\theta_{34} \label{SM4cor}~,
\end{eqnarray}

\noindent where $\theta_{34}$ is the mixing angle between the 3rd and 4th generation
quarks, i.e., defining $|V_{t^\prime b}| = |V_{t b^\prime}| \equiv \sin\theta_{34}$.
This SM4-like effect on $R_b$ is plotted in Fig.~\ref{figRb}. We see
that $R_b$ puts rather stringent constraints on the
$m_{t^\prime} - \theta_{34}$ plane which is the SM4 subspace of the
parameter space of our 4G2HDMs. In particular,
increasing the $t^\prime$ mass would tighten the constraints on $\theta_{34}$;
e.g., for $m_{t^\prime} \sim 500$ GeV
the $t^\prime - b$ mixing angle is restricted to $\theta_{34} \lsim 0.2$.
The upper bound on $\theta_{34}$ stays roughly the same in our 4G2HDMs
where the effects from the charged Higgs loops are included.
For concreteness, for the rest of this section we will fix $\theta_{34}$ to
either $\theta_{34} = 0$
or $\theta_{34} =0.1,~0.2$, representing the no-mixing or mixing
cases.

Using the generic formula given in \cite{ourZbs},
we calculated the 1-loop corrections to $R_b$ from the
charged-Higgs and from the FC neutral-Higgs exchanges
and found that:

\begin{itemize}

\item In all three models, i.e., 4G2HDM-I,II,III,
$\delta_c \ll \delta_b$, so that we can safely neglect the new effects in
$Z \to c \bar c$.

\item The 1-loop FC neutral-Higgs contributions are much smaller than
the 1-loop charged-Higgs contributions
shown in Fig.~\ref{diagrams}, in particular for $\epsilon_b \ll 1$. We, therefore,
focus below only on the leading effects coming
from the charged-Higgs sector.

\item The charged-Higgs interactions in models 4G2HDM-II and 4G2HDM-III
have negligible effects on $R_b$ and are,
therefore, not constrained by this quantity. On the other hand,
$R_b$ is rather sensitive to the
charged Higgs loop exchanges within our type I 4G2HDM.
\end{itemize}

In light of the above findings, we plot in Fig.~\ref{figRb} the quantity
$R_b$ in the 4G2HDM-I (calculated from Eq.~\ref{Rb}),
as a function of the charged Higgs and $t^\prime$
masses, fixing $\epsilon_t = m_t/m_{t^\prime}$ and focusing
on the values $\tan\beta=1,5$, $\theta_{34}=0,0.2$ and
$m_{H^+}=400,750$ GeV.
We see that, while there are no constraints from $R_b$
on the charged Higgs and $t^\prime$
masses if $\tan\beta=1$, for higher values of $\tan\beta$
a more restricted region of the charged Higgs mass is allowed which again
depends on $\theta_{34}$, e.g.,
for $\tan\beta=5$,
$ 550  ~{\rm GeV} \lsim m_{H^+} \lsim 800 ~{\rm GeV} $, and
$m_{t^\prime} \lsim 500$ GeV is required in order for
$R_b$ to be within its $2\sigma$ measured value ($R_b^{exp}=0.21629 \pm 0.00066$ \cite{EL2009}).

In Fig.~\ref{figmhctb} we show the allowed ranges in the $m_{H^+} - \tan\beta$ plane
in the 4G2HDM-I, subject to the $R_b$ constraint ($2\sigma$), for $\tan\beta$ in the range
1-15, fixing $m_{t^\prime} =500$ GeV, $m_{b^\prime} =450$ GeV, $\theta_{34}=0.2$, $\epsilon_b = m_b/m_{b^\prime}$ (which also enters the $t^\prime b H^+$ vertex)
and for three representative values of
the $t-t^\prime$ mixing parameter: $\epsilon_t = \epsilon_b \sim 0.01$, $\epsilon_t = m_t/m_{t^\prime} \sim 0.35$
and $\epsilon_t = 1$.
As expected, when $\tan\beta$ is lowered,
the constraints on the charged Higgs mass are weakened.
We see e.g., that for $\epsilon_t = m_t/m_{t^\prime} \sim 0.35$,
$\tan\beta \sim 1$ is compatible with $m_{H^+}$ values ranging from 200 GeV up to the TeV scale, while for $\tan\beta \sim 5$ the charged Higgs mass is restricted to be
within the range $ 450 ~{\rm GeV} \lsim m_{H^+} \lsim 750 ~{\rm GeV} $.
Note however, that in the 4G2HDM-I, $\tan\beta =5$ with $\epsilon_t = m_t/m_{t^\prime}$ is not
allowed by constraints from B-physics flavor data (see previous section).

\subsection{Constraints from the Oblique parameters}

The sensitivity of 4th generation fermions to PEWD within the minimal SM4 framework was extensively analyzed
in the past decade \cite{PDG,polonsky,novikov,Kribs_EWPT,langacker,chenowitz,chenowitz1}.
One of the immediate interesting consequences of the presence of the
4th generation fermion doublet
(with respect to the PEWD constraints) is that it allows for a considerably heavier
Higgs, thus removing the slight tension between the
LEPII bound on the mass of the SM Higgs $m_H \gsim 115$ GeV and
the corresponding theoretical best fitted value (to PEWD) $m_H = 87^{+35}_{-26}$ GeV \cite{PDG}.
In fact, a Higgs with $m_H \gsim 300$ GeV becomes favored
in the SM4 when $m_{t^\prime} - m_{b^\prime} \sim 50$ GeV and $\theta_{34}$ is
of the size of the Cabbibo angle, see e.g., \cite{chenowitz,Kribs_EWPT}.
On the other hand, if, as in our case, the 4th generation fermions are embedded in a
2HDM framework,
then there is a wider range of parameter space for which a lighter Higgs with a
mass of ${\cal O}(100)$ GeV is allowed
(see \cite{hashimoto2} and our analysis below). In addition, in the 2HDM case,
the LEPII lower bound $m_H \gsim 115$ GeV can be relaxed, depending on the value
of $\sin(\alpha-\beta)$ [$\sin(\alpha-\beta)=1$ corresponds to the current
SM bound] which controls the $ZZH$ coupling responsible for the Higgs
production mechanism at LEP.

In general, the contributions to the oblique parameters ($S$,$T$,$U$)
of 4th generation fermions ($\Delta S_f,\Delta T_f,\Delta U_f$) and
of extra scalars ($\Delta S_s,\Delta T_s,\Delta U_s$) are calculated with respect
to the SM values and are bounded by a fit to PEWD \cite{gfitter}:
\begin{eqnarray}
\Delta S &=& S - S_{SM} = 0.02 \pm 0.11 \nonumber \\
\Delta T &=& T - T_{SM} = 0.05 \pm 0.12 \nonumber \\
\Delta U &=& U - U_{SM} = 0.07 \pm 0.12 \label{stpar}~,
\end{eqnarray}
where, following the fit made in \cite{gfitter},
the SM values are defined for a Higgs mass reference value of $M_h^{ref}=120$ GeV
and for $m_t=173.2$ GeV. The effects of our models (and in general
of any heavy new physics) on the parameter U can be neglected. We, therefore,
consider below the constraints from the 2-dimensional ellipse in the $S-T$ plane
which, for a given confidence level (CL), is defined by (see e.g., \cite{lenz_fourth12}):
\begin{widetext}
\begin{eqnarray}
\left(
\begin{array}{c} S - S_{exp} \\ T - T_{exp}  \end{array}
\right)^T
\left(
\begin{array}{cc} \sigma_S^2 & \sigma_S \sigma_T \rho \\
\sigma_S \sigma_T \rho & \sigma_T^2 \end{array}
\right)
\left(
\begin{array}{c} S - S_{exp} \\ T - T_{exp}  \end{array}
\right)
= - 2 {\rm ln}\left( 1 - CL \right) ~,
\end{eqnarray}
\end{widetext}
where $S_{exp} = 0.02$ and $T_{exp} = 0.05$ are the best fitted (central)
values in Eq.~\ref{stpar}, $\sigma_S = 0.11, \sigma_T = 0.12$
are the corresponding standard deviations and $\rho=0.879$ \cite{gfitter}
is the (strong) correlation factor between S and T.

Note that the contribution of the Higgs spectrum of our 4G2HDMs to
$S$ and $T$ are identical to that of any general 2HDM.
We thus use the analytical expressions given in \cite{polonsky}, where we
also include in $\Delta T_f$ the new contributions from the
$W t^\prime b$ and $W t b^\prime$ off-diagonal CKM mixing angles
(see e.g., \cite{chenowitz}):
\begin{widetext}
\begin{eqnarray}
\Delta T_f = \frac{3}{8 \pi s_W^2 c_W^2}
\left( |V_{t^\prime b^\prime}|^2 F_{t^\prime b^\prime} +
|V_{t^\prime b}|^2 F_{t^\prime b} +
|V_{t b^\prime}|^2 F_{t b^\prime}  -
|V_{t b}|^2 F_{t b} +
\frac{1}{3} F_{\ell_4 \nu_4} \right) ~,
\end{eqnarray}
\end{widetext}
with
\begin{eqnarray}
F_{ij} = \frac{x_i+x_j}{2} - \frac{x_i x_j}{x_i-x_j} {\rm log}\frac{x_i}{x_j}~,
\end{eqnarray}
and $x_k \equiv (m_k/m_Z)^2$.

We first ``blindly" (randomly) scan our parameter space,
varying them in the ranges:
$\tan\beta \leq 30$, $\theta_{34} \leq 0.3$,
$100~{\rm GeV} \leq m_{h} \leq 1~{\rm TeV}$,
$m_h \leq m_{H} \leq 1.5~{\rm TeV}$,
$100~{\rm GeV} \leq m_{A} \leq 1~{\rm TeV}$,
$400~{\rm GeV} \leq m_{t^\prime},m_{b^\prime} \leq 600~{\rm GeV}$,
$100~{\rm GeV} \leq m_{\nu^\prime},m_{\tau^\prime} \leq 1.2~{\rm TeV}$,$^{\footnotemark[6]}$\footnotetext[6]{Note
that the perturbative unitarity upper bounds on the lepton masses
are about twice larger than those on the quark masses \cite{PU}; thus allowing
4th generation lepton masses around 1 TeV.}
and the CP-even neutral Higgs mixing angle in the range
$0 \leq \alpha \leq 2 \pi$.

\begin{figure}[htb]
\begin{center}
\epsfig{file=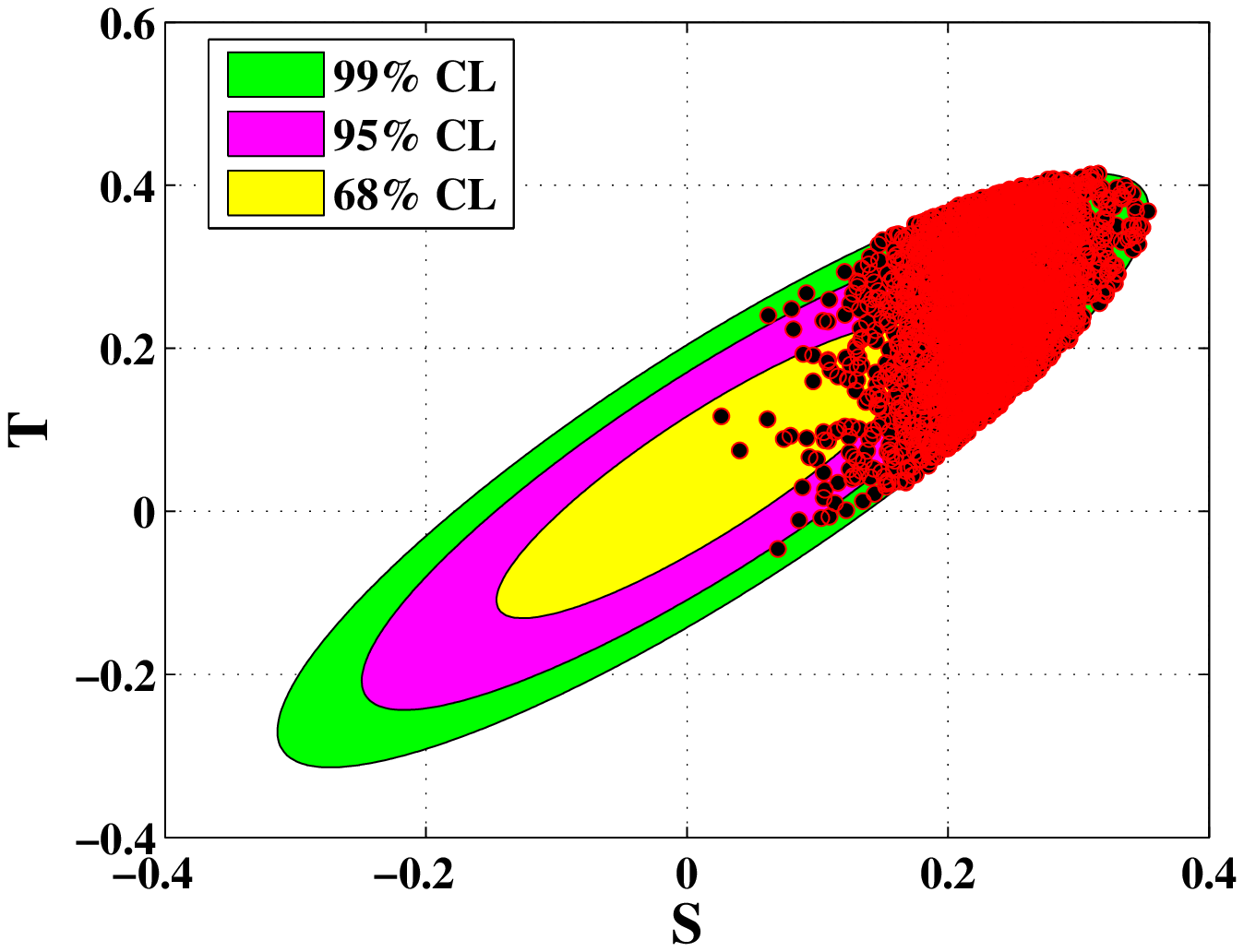,height=8cm,width=8cm,angle=0}
\epsfig{file=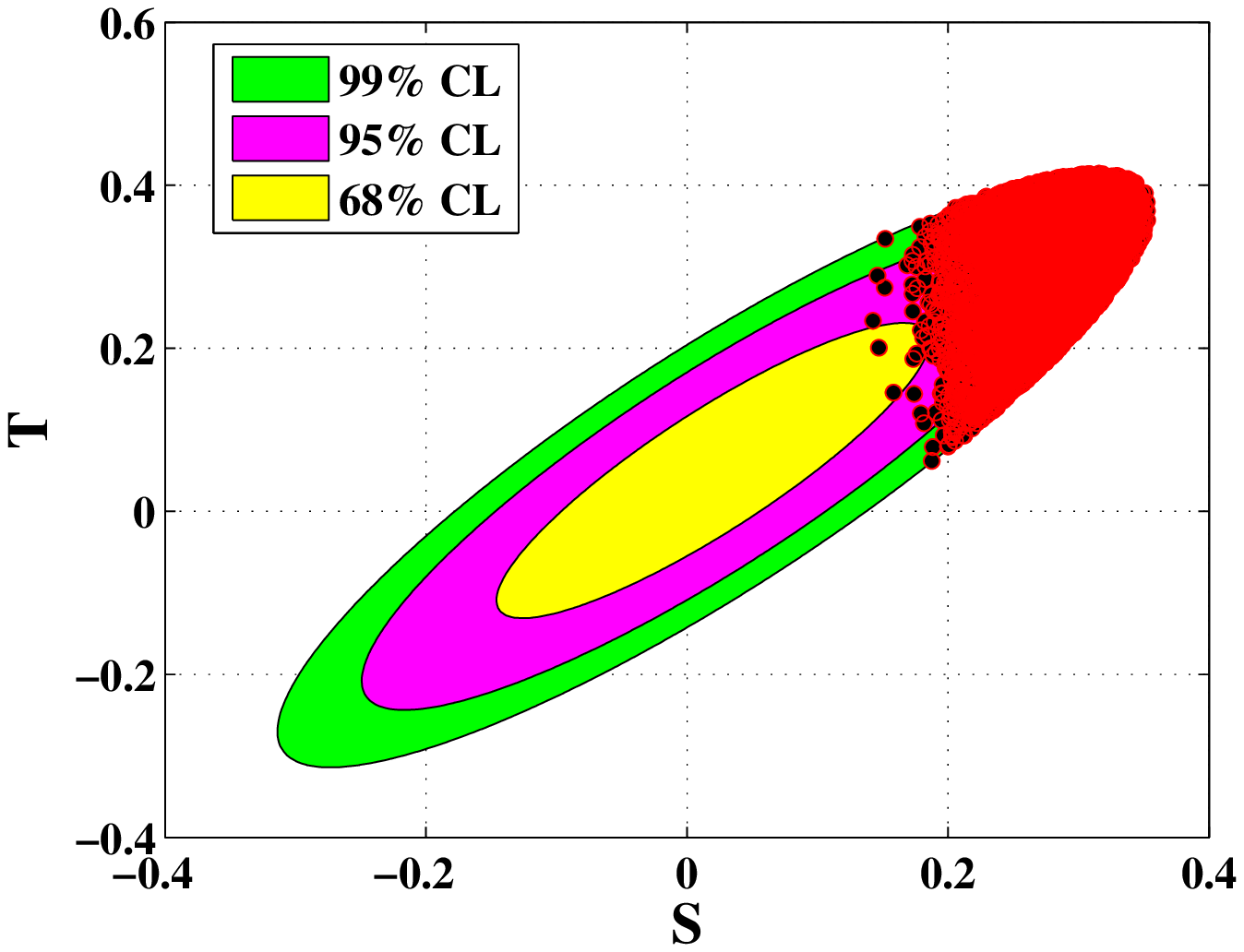,height=8cm,width=8cm,angle=0}
\caption{\emph{The
allowed points in parameter space projected onto the
68\%, 95\% and 99\% allowed contours in the S-T plane, in
the 4G2HDMs (left) and in the SM4 (right). The data points are
varied in the ranges: $\tan\beta \leq 30$, $\theta_{34} \leq 0.3$,
$100~{\rm GeV} \leq m_{h} \leq 1~{\rm TeV}$,
$m_h \leq m_{H} \leq 1.5~{\rm TeV}$,
$100~{\rm GeV} \leq m_{A} \leq 1~{\rm TeV}$,
$400~{\rm GeV} \leq m_{t^\prime},m_{b^\prime} \leq 600~{\rm GeV}$,
$100~{\rm GeV} \leq m_{\nu^\prime},m_{\tau^\prime} \leq 1.2~{\rm TeV}$
and the CP-even neutral Higgs mixing angle in the range
$0 \lsim \alpha \lsim 2 \pi$.}}
\label{blindscan}
\end{center}
\end{figure}

We use a sample of 100000 models (i.e., points
in parameter space varied in the above specified ranges)
and plot the result in Fig.~\ref{blindscan}.
We find that out of the 100000 models
about 3000 are within the 99\%CL contour, 1500 within
the 95\%CL contour and 100 within
the 68\%CL contour. We compare these results to the SM4
case also shown in Fig.~\ref{blindscan} (again using a sample of 100000 models),
where the 4th generation quark and lepton masses as well as the (single) neutral
Higgs mass are varied in the same ranges as specified above.
We find that in the SM4 case only a few points (out of the 100000)
are within the 68\%CL S-T contour, while the number
of SM4 points within the 95\%CL and 99\%CL allowed contours are comparable
to the 2HDM case.
This quantifies the slight preferability of the 2HDM
(with respect to the amount of fine tuning required for compatibility
with the available precision data)
as an underlying framework for a 4th generation model.

\begin{figure}[htb]
\begin{center}
\epsfig{file=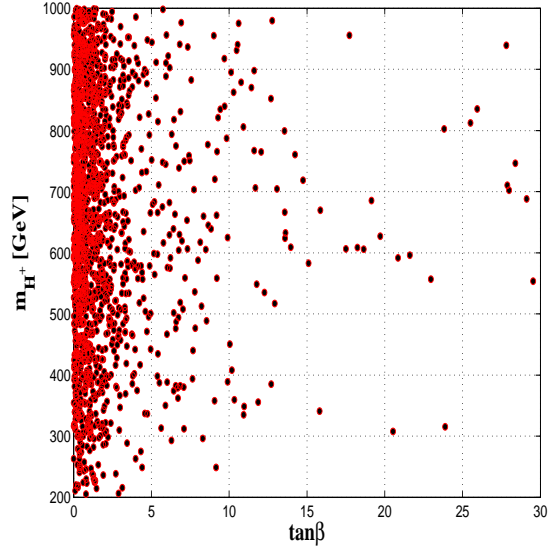,height=8cm,width=8cm,angle=0}
\caption{\emph{95\% CL allowed range in the $m_{H^+} - \tan\beta$ plane.
The data points are varied as specified in Fig.~\ref{blindscan}.}}
\label{mhtb}
\end{center}
\end{figure}

We also examined the correlation in the $m_{H^+} - \tan\beta$ plane, when
subject to the PEWD S-T constraint.
This is shown in Fig.~\ref{mhtb}, where the data points
are taken from the same 100000 sample used in Fig.~\ref{blindscan} (i.e.,
the rest of the parameter space was varied in the ranges specified above).
We see that compatibility with PEWD mostly requires
$\tan\beta \sim {\cal O}(1)$ with a small number of points in parameter space
having $\tan\beta \gsim 5$.


\begin{figure}[htb]
\begin{center}
\epsfig{file=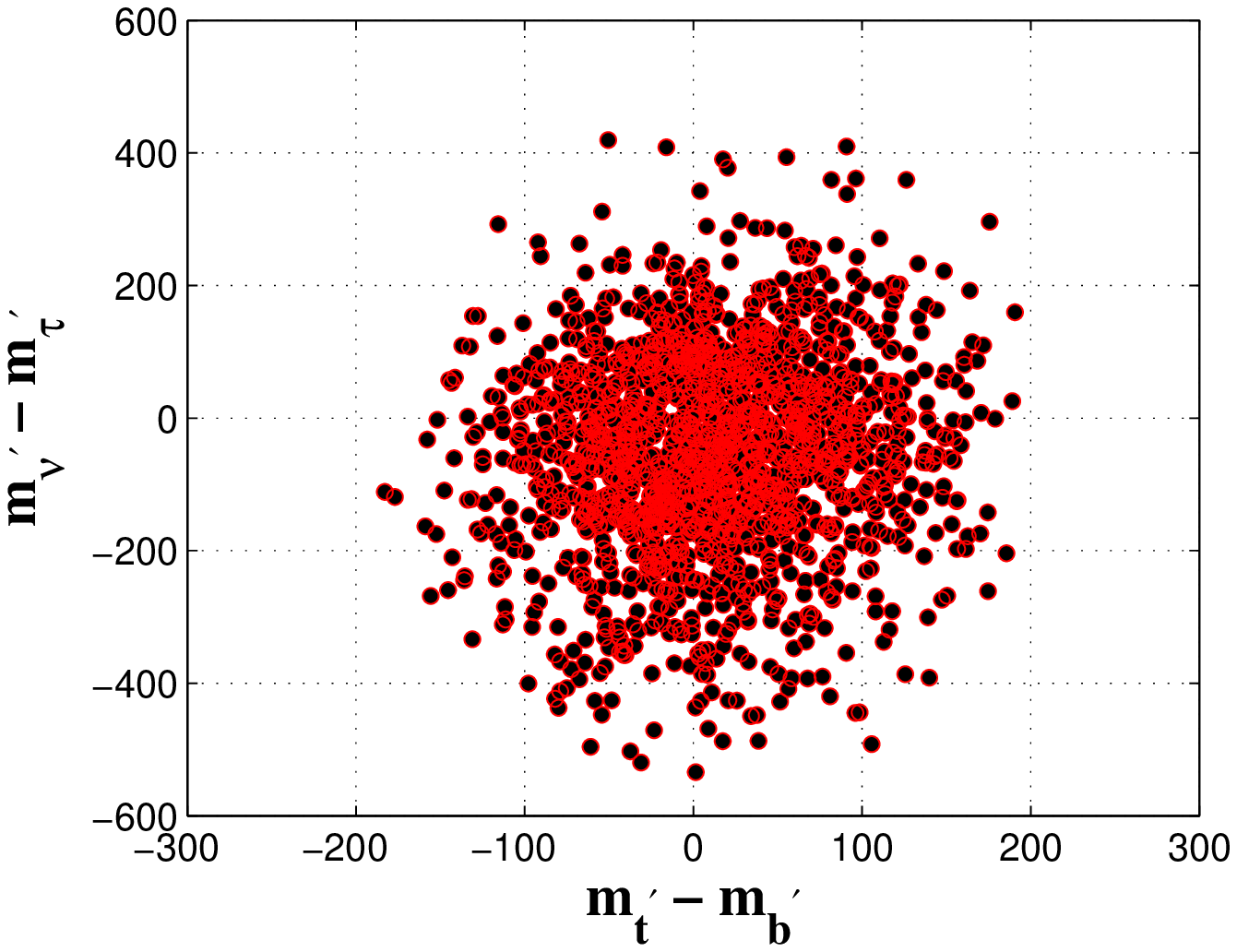,height=8cm,width=8cm,angle=0}
\epsfig{file=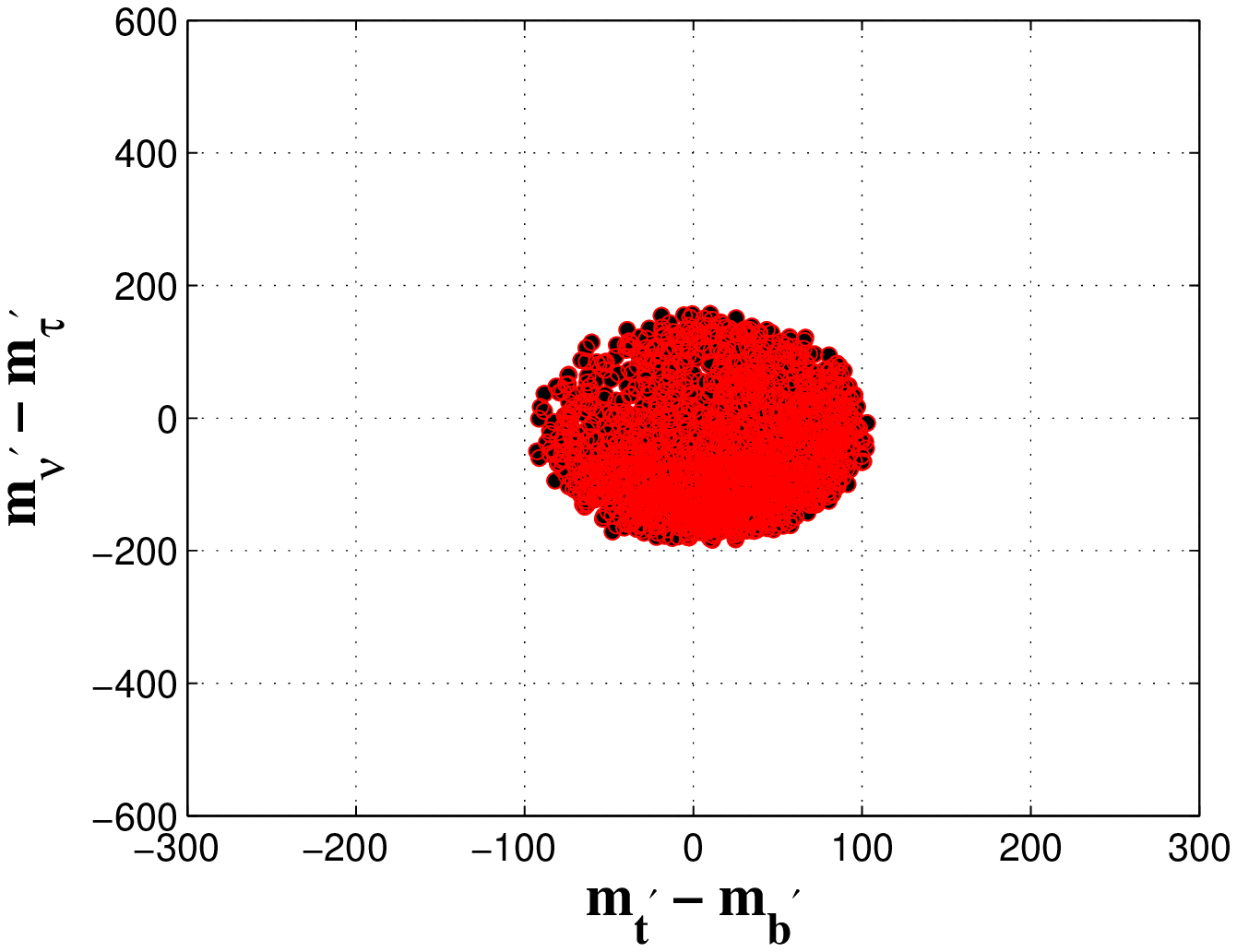,height=8cm,width=8cm,angle=0}
\caption{\emph{Allowed regions in the
$\Delta m_{q^\prime} - \Delta m_{\ell^\prime}$ plane
within the $95\%$CL contour in the S-T plane,
for the 4G2HDMs (left) and
for the SM4 (right).
The data points are
varied as in Fig.~\ref{blindscan}.}}
\label{deltamblind}
\end{center}
\end{figure}

Next we consider the correlations between the mass splitting among
the 4th generation quark masses,
$\Delta m_{q^\prime} \equiv m_{t^\prime} - m_{b^\prime}$,
and the lepton masses $\Delta m_{\ell^\prime} \equiv m_{\nu_4} - m_{\ell_4}$.
In Fig.~\ref{deltamblind} we plot the $95\%$CL allowed regions
(i.e., subject to the measured $95\%$CL contour in the S-T plane)
for both the 4G2HDMs and the SM4 in the
$\Delta m_{q^\prime} - \Delta m_{\ell^\prime}$ plane, again
using the same data set of 100000 models used in Fig.~\ref{blindscan}.
We see that, while in the SM4 case the allowed mass splittings are restricted to
$-100~{\rm GeV} < \Delta m_{q^\prime} < 100~{\rm GeV}$
and $-200~{\rm GeV} < \Delta m_{\ell^\prime} < 200~{\rm GeV}$,
in the 4G2HDMs these
mass splitting ranges are significantly extended
to: $-200~{\rm GeV} < \Delta m_{q^\prime} < 200~{\rm GeV}$
and $-500~{\rm GeV} < \Delta m_{\ell^\prime} < 400~{\rm GeV}$.

\begin{figure}[htb]
\begin{center}
\epsfig{file=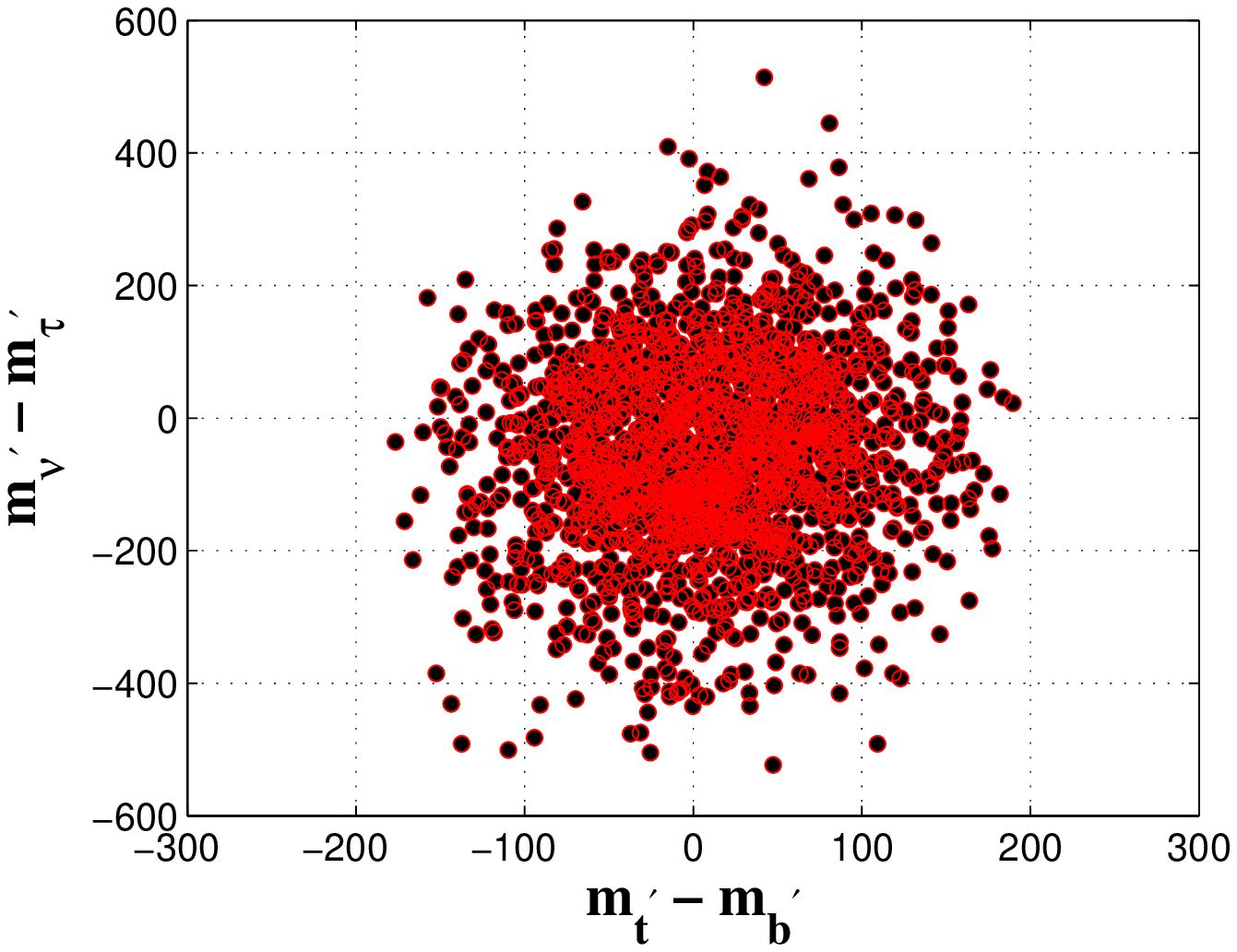,height=8cm,width=8cm,angle=0}
\epsfig{file=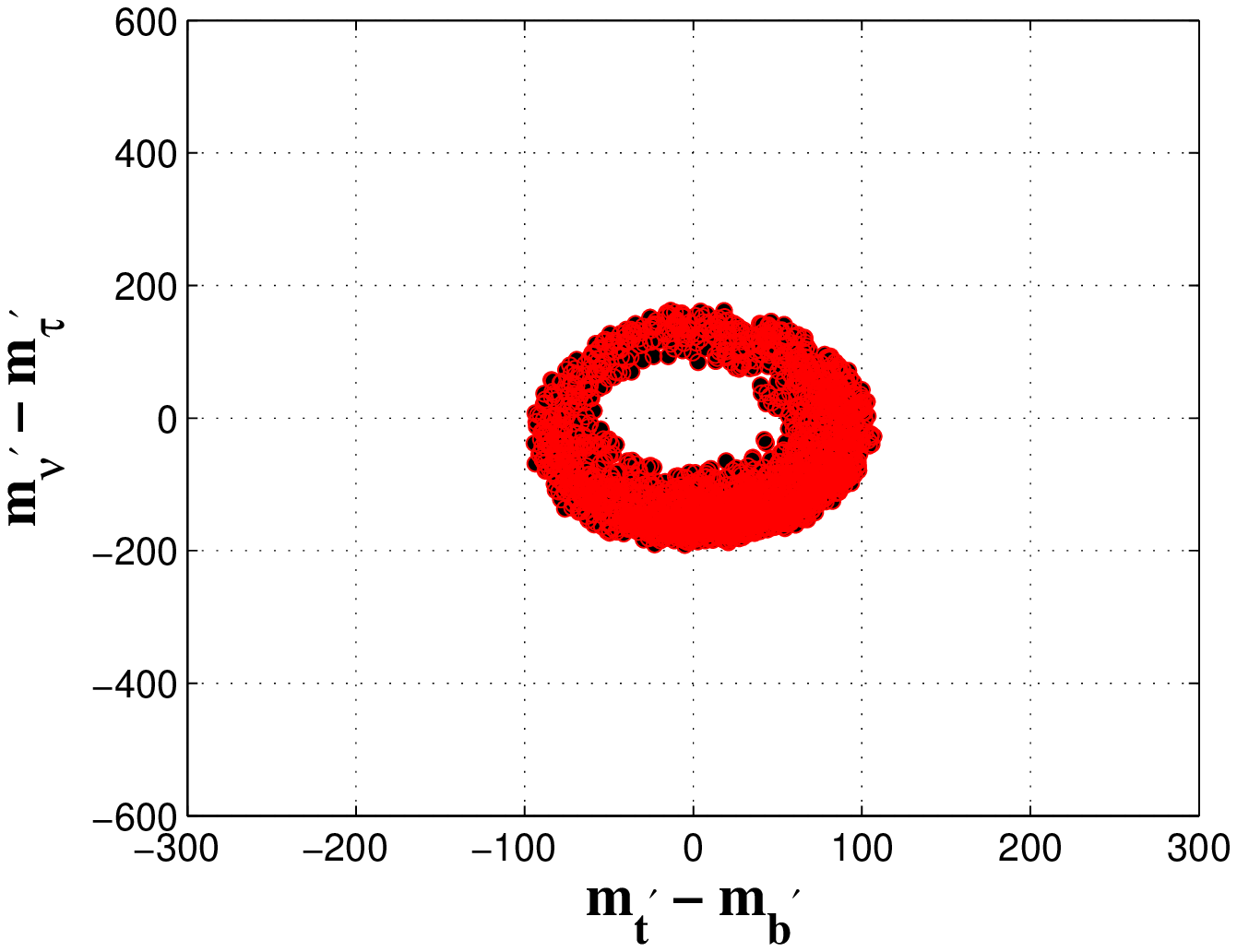,height=8cm,width=8cm,angle=0}
\caption{\emph{Same as Fig.~\ref{deltamblind} but for $\theta_{34}=0$;
the rest of the parameter space is varied as in Fig.~\ref{deltamblind}.}}
\label{deltam34zero}
\end{center}
\end{figure}

In Figs.~\ref{deltam34zero} we again plot
the $95\%$CL allowed regions in the
$\Delta m_{q^\prime} - \Delta m_{\ell^\prime}$ plane,
for both the 4G2HDMs and the SM4, considering now
the ``3+1" scenario, i.e., with a vanishing mixing
between the 4th generation quarks and the lighter three generations;
$\theta_{34}=0$. The rest of the parameter space is varied
as in Fig.~\ref{deltamblind}.
We see that in the SM4 with $\theta_{34} \to 0$
there are no solutions where both the
quark and lepton 4th generation doublets are degenerate, in particular,
no solutions where both $|\Delta m_{q^\prime}| \lsim 50$ GeV and
$|\Delta m_{\ell^\prime}| \lsim 100$ GeV.
On the other hand, the implications of the no-mixing case on the 4G2HDMs
are mild as there are still points/models for which the 4th generation quarks
and leptons
are both almost degenerate. For such small (or no) 4th generation fermion mass
splitting the amount of isospin breaking required to compensate for the effect of the
extra fermions and Higgs particles on S and T is provided by a mass splitting among the
Higgs particles, as is shown below.

In order to demonstrate the interplay between the
mass splittings in the Higgs and fermion sectors, we choose
a more specific framework - partly motivated by our theoretical prejudice
towards the possibility of dynamical EWSB,
driven by the condensation(s) of the 4th generation fermions.
In particular, we set $\alpha \sim \pi/2$,
for which case $H \sim {\rm Re}(\Phi_h^0)$ and
$h \sim {\rm Re}(\Phi_\ell)$; the heavier Higgs may be thus identified
as a possible $\bar Q^\prime Q^\prime$ ($Q^\prime = t^\prime,b^\prime$) condensate,
with a typical mass of $m_H \lsim 2 m_{Q^\prime}$ \cite{mHest}.
We thus set $m_H=1$ TeV and take a nearly degenerate
4th generation quark doublet with
$m_{t^\prime}=500$ GeV and $m_{b^\prime}=490$ GeV.
We further study two representative values for $\tan\beta$:
$\tan\beta=1$ and $\tan\beta=5$,
recalling that for $\tan\beta \sim {\cal O}(1)$, $H^+$ and $A$ are
roughly equal admixtures of $\Phi_\ell$ and $\Phi_h$, while if
$\tan^2\beta >> 1$, one has $H^+ \sim \Phi_\ell^+$ and $A \sim {\rm Im}(\Phi_h^0)$.
The charged Higgs mass is set to $m_{H^+} = 600$ GeV, so that
it is within the $R_b$ constraints for both $\tan\beta=1$ and $\tan\beta=5$
when $ \frac{m_b}{m_b'} \lsim \epsilon_t \lsim \frac{m_t}{m_t'}$ (see previous
section).
For simplicity we furthermore set $\theta_{34} =0$
and vary the 4th generation lepton masses in the range
$100~{\rm GeV} \lsim m_{\nu^\prime},m_{\tau^\prime} \lsim 1.2~{\rm TeV}$
and the masses of the neutral Higgs particles, $h$ and $A$, in the range
$100~{\rm GeV} \lsim m_{h},m_A \lsim 1000~{\rm GeV}$.

\begin{figure}[]
\begin{center}
\epsfig{file=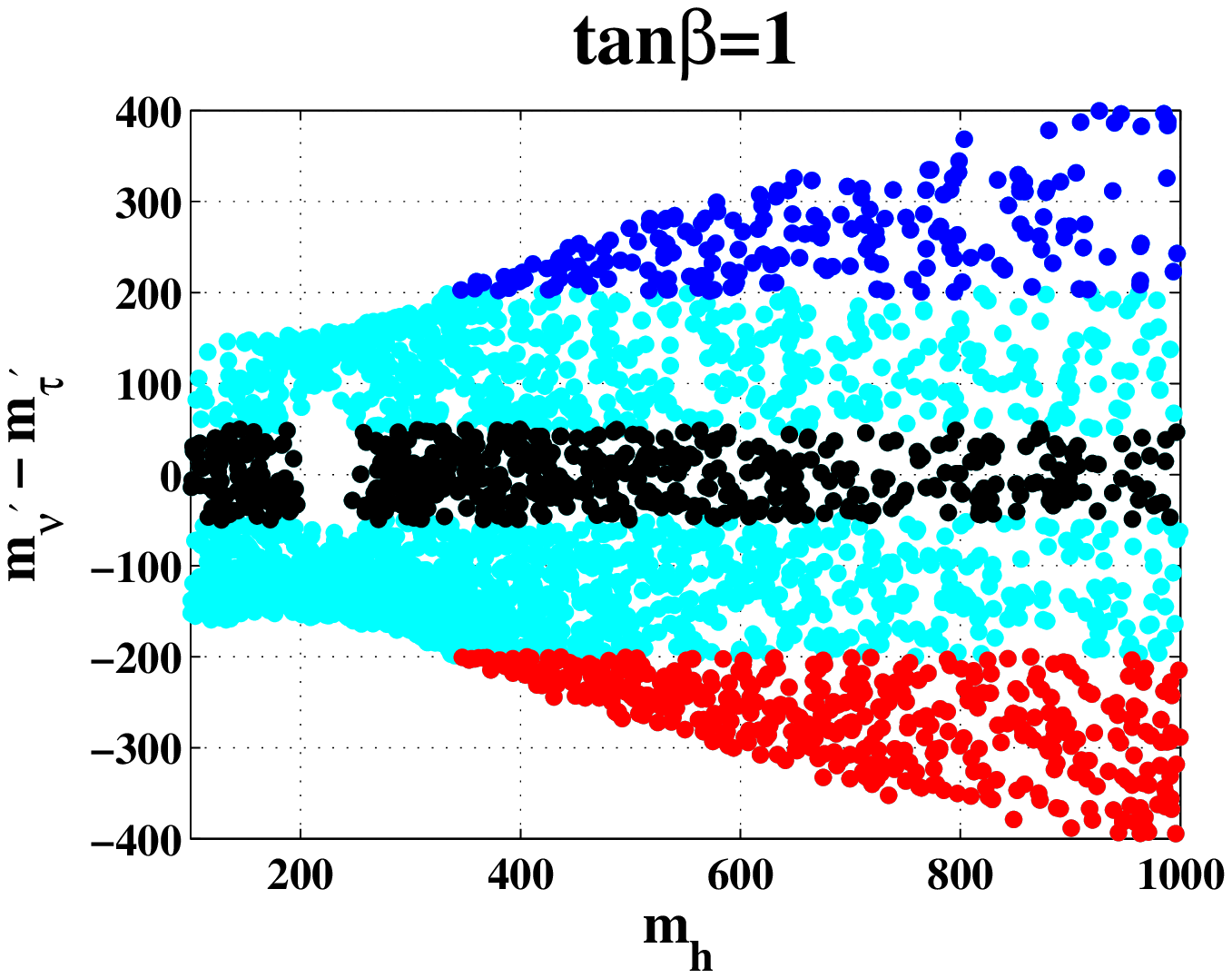,height=5cm,width=5cm,angle=0}
\epsfig{file=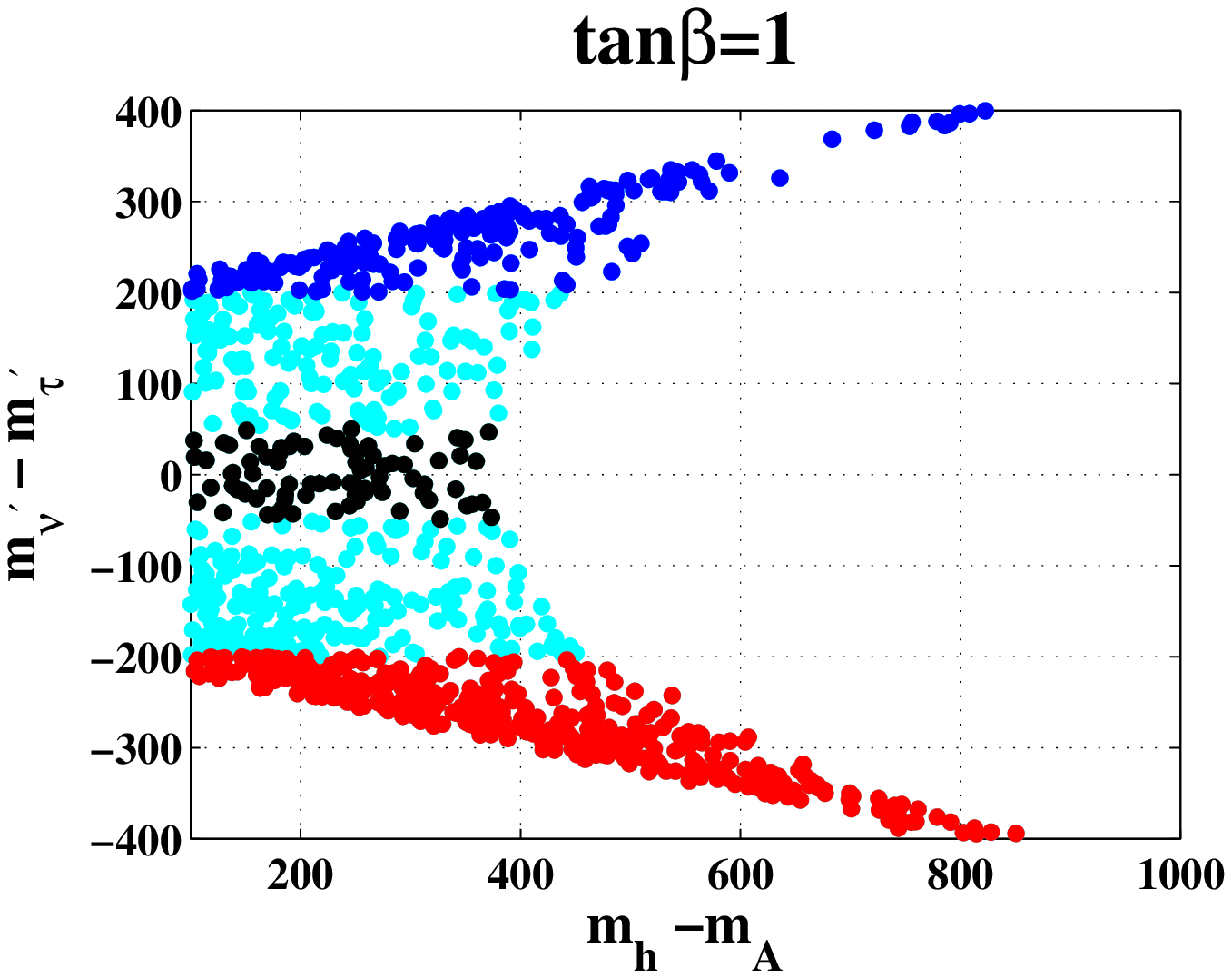,height=5cm,width=5cm,angle=0}
\epsfig{file=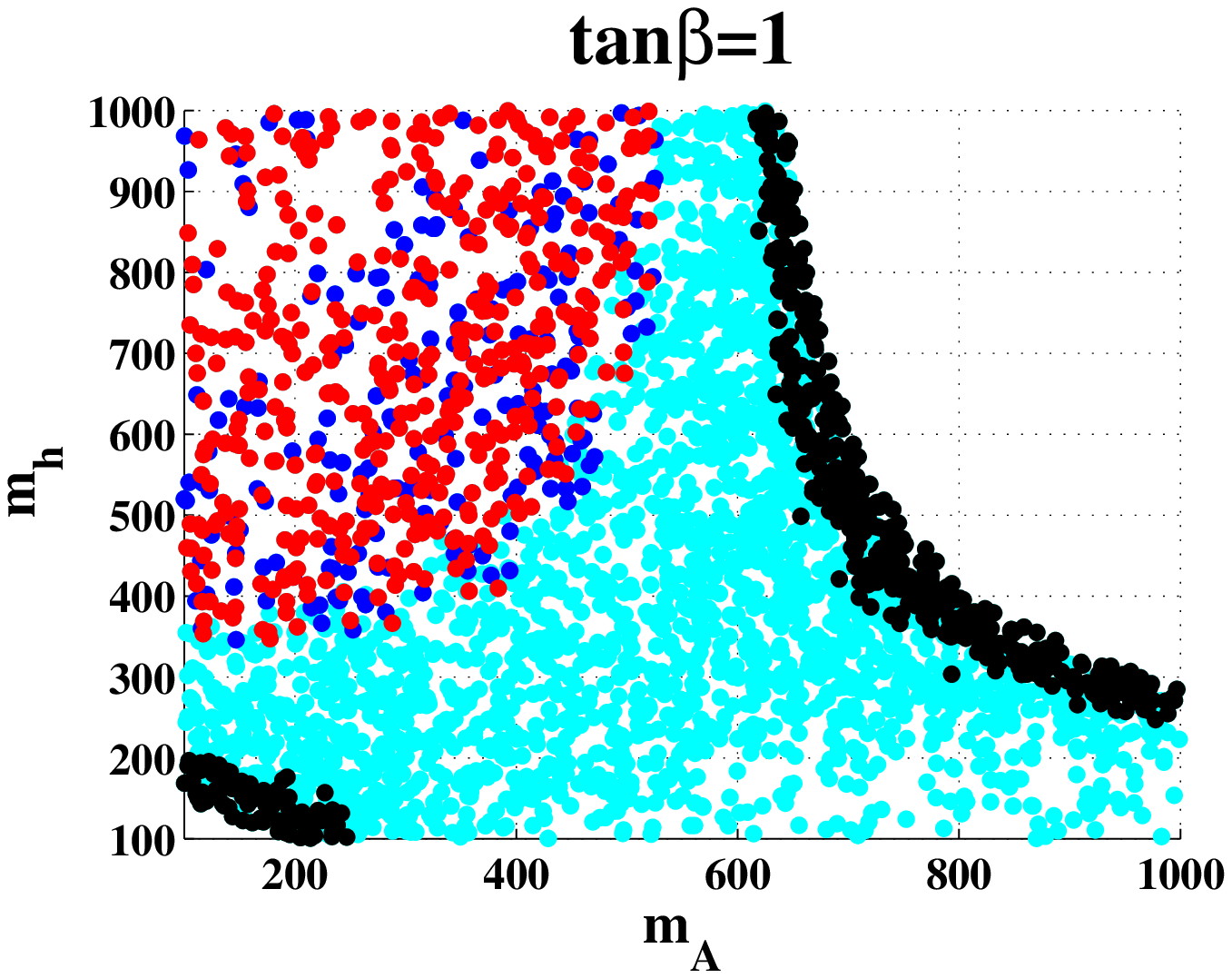,height=5cm,width=5cm,angle=0}
\caption{\emph{$95\%$ CL allowed regions in the
$\Delta m_{\ell^\prime} - m_h$ plane (left), in
the $\Delta m_{\ell^\prime} - (m_h - m_A)$ plane (middle) and
in the $m_h - m_A$ plane (right),
for $\tan\beta=1$, $m_{H^+}=600$ GeV, $\theta_{34}=0$,
$\alpha \sim \pi/2$,
$m_{t^\prime}=500$ GeV and $m_{b^\prime}=490$ GeV.
The lepton masses and Higgs masses are varied in the ranges:
$100~{\rm GeV} \lsim m_{\nu^\prime},m_{\tau^\prime} \lsim 1.2~{\rm TeV}$
and $100~{\rm GeV} \lsim m_{h},m_A \lsim 1000~{\rm GeV}$.
The black dots correspond to solutions with $|\Delta m_{\ell^\prime}|<50$ GeV,
the red dot to solutions with $m_{\tau^\prime} - m_{\nu^\prime} > 200$ GeV and
the blue dots to solutions with $m_{\nu^\prime} - m_{\tau^\prime} > 200$ GeV.
}}
\label{deltamnu1}
\end{center}
\end{figure}

\begin{figure}[]
\begin{center}
\epsfig{file=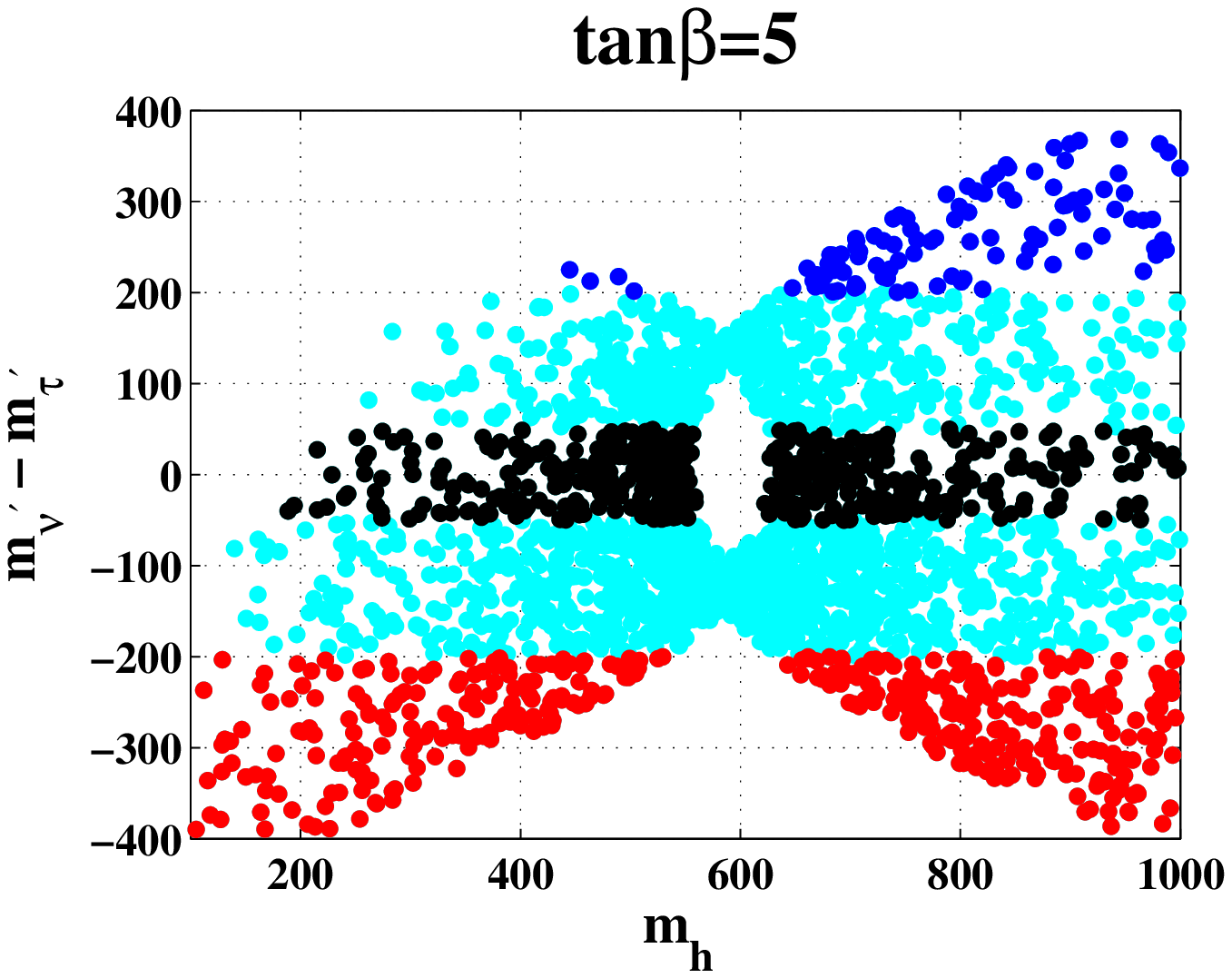,height=5cm,width=5cm,angle=0}
\epsfig{file=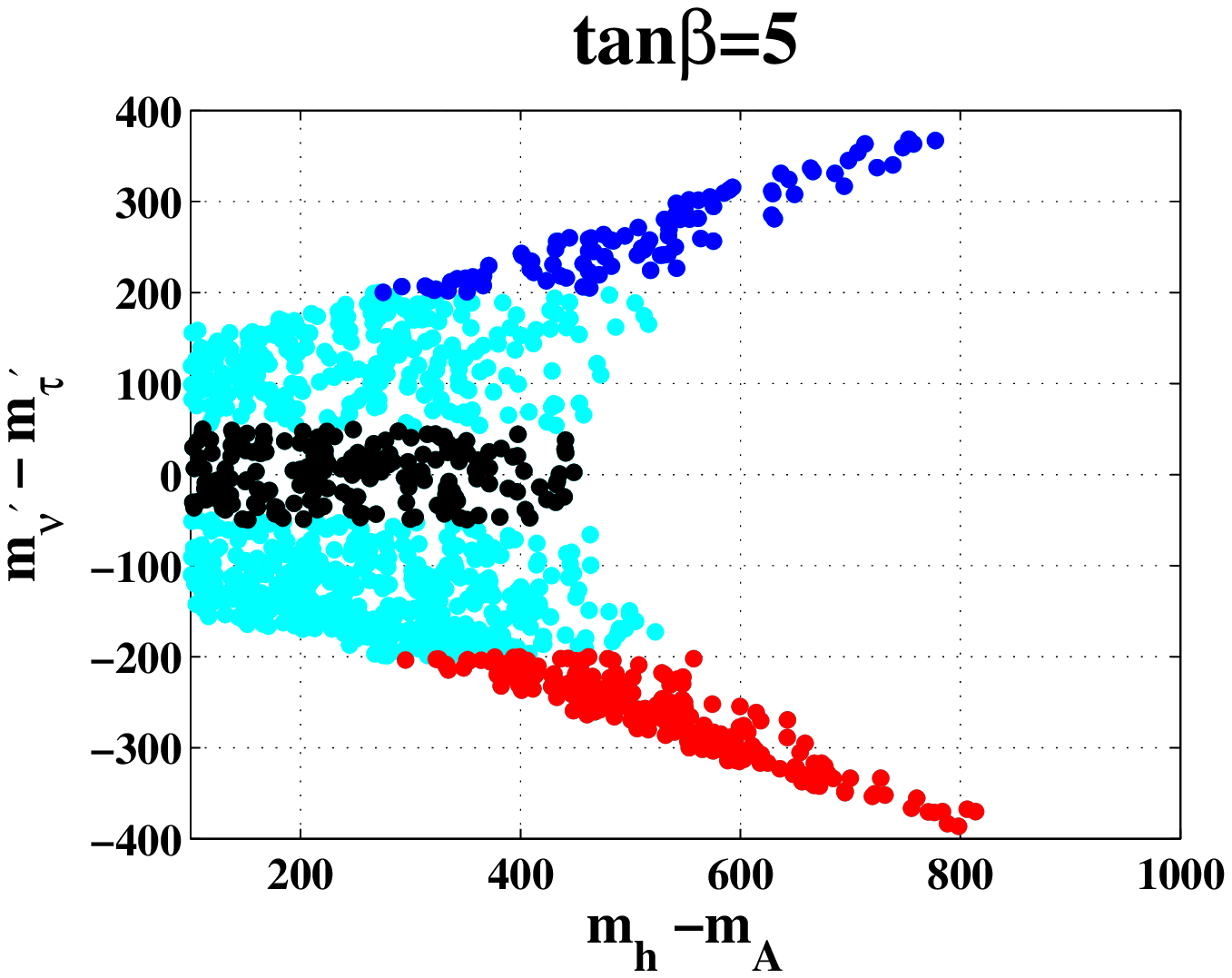,height=5cm,width=5cm,angle=0}
\epsfig{file=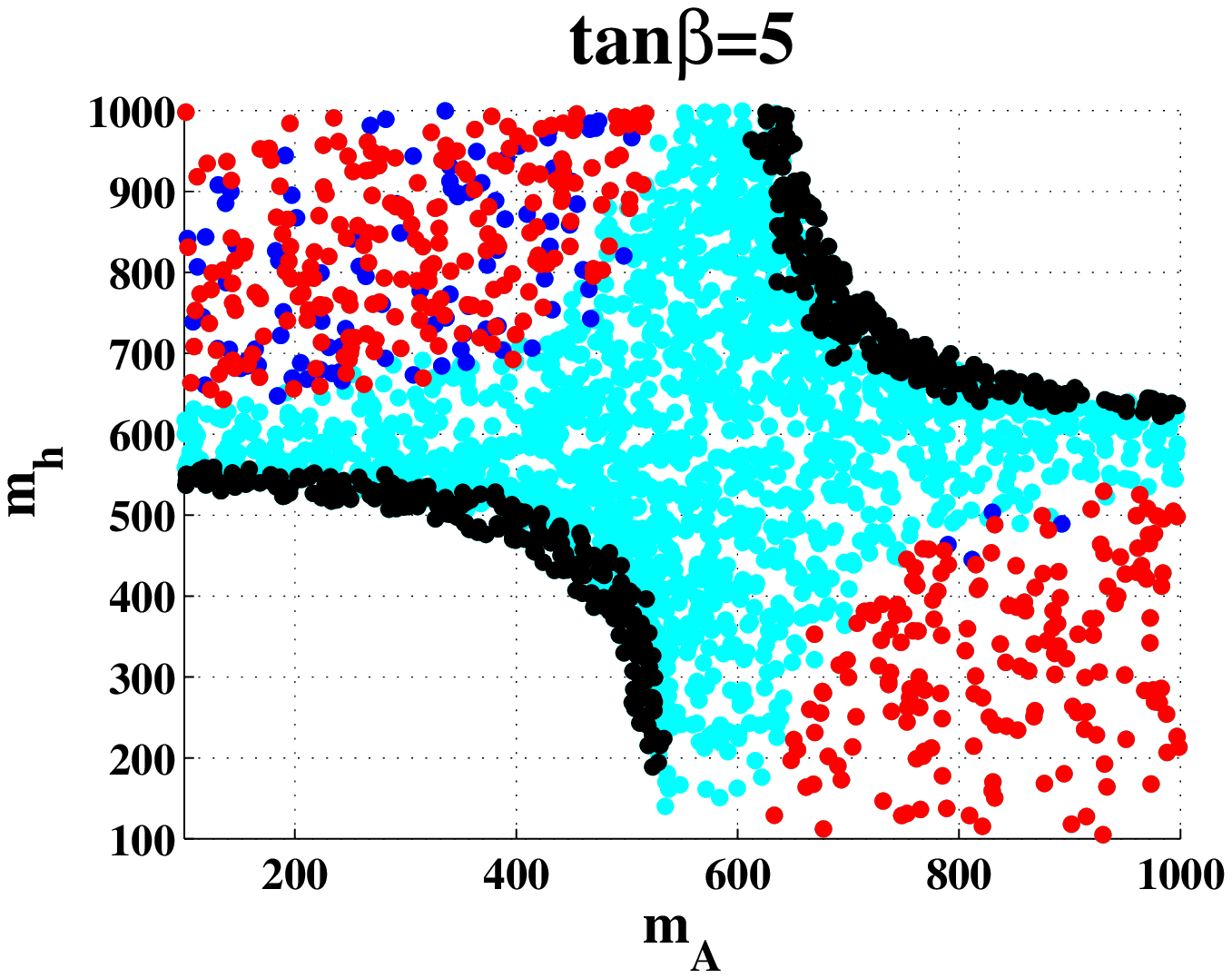,height=5cm,width=5cm,angle=0}
\caption{\emph{Same as Fig.~\ref{deltamnu1} for $\tan\beta=5$.
}}
\label{deltamnu2}
\end{center}
\end{figure}

Using the above set of assumptions on our parameter space,
we plot in Figs.~\ref{deltamnu1} and \ref{deltamnu2} the 95\%CL allowed region
in the $\Delta m_{\ell^\prime} - m_h$, the $m_h - m_A$
and the $\Delta m_{\ell^\prime} - (m_h-m_A)$ planes, using again
a sample of 100000 models with
$\tan\beta=1$ and $\tan\beta=5$, respectively.
Under the above set of inputs, we find the following noticeable features:
\begin{itemize}
\item There are allowed sets of points in parameter space (i.e., models)
where both the 4th generation quarks and
leptons are nearly degenerate with a mass splitting smaller than 50 GeV.
These solutions require $m_h$ and $m_A$ to have a mass splitting
smaller than about 400 GeV and to be within
the narrow black bands in the $m_h - m_A$ plane, as seen in
Figs.~\ref{deltamnu1} and \ref{deltamnu2}.
\item There are allowed sets of points with a large splitting between
the 4th generation leptons,
$|m_{\nu^\prime} - m_{\tau^\prime}| > 300$ GeV. These cases require
a large splitting also among $m_h$ and $m_A$;
$  400 ~{\rm GeV} \lsim |m_h - m_A| \lsim 800 ~{\rm GeV}$ if $\tan\beta=1$ and
$  600 ~{\rm GeV} \lsim |m_h - m_A| \lsim 800 ~{\rm GeV}$  if $\tan\beta=5$.
\item For $\tan\beta=1$, a splitting in the 4th generation lepton sector
of $|m_{\nu^\prime} - m_{\tau^\prime}| > 200$ GeV requires
$m_h$ to be larger than about 400 GeV.
\end{itemize}

Finally, in Table~\ref{tab2} we give a list of interesting points (models) in parameter space (of our
4G2HDM of types I, II and III) that pass all the constraints considered in this chapter, i.e., from the S and T
parameters, from $R_b$ and from B-physics flavor data. In particular, the list includes models with
mass splittings between the up and down partners of both the 4th family quarks and leptons larger than 150 GeV,
models with a light $100 - 200$ GeV neutral Higgs, models with degenerate 4th generation doublets,
models with a large inverted mass hierarchy in the quark doublet, i.e.,
$m_{b^\prime} - m_{t^\prime} > 150$ GeV, models with a light charged Higgs with a mass smaller than 500 GeV,
models with a Cabbibo size as well as an ${\cal O}(0.01)$ size $t^\prime - b$/$t - b^\prime$
mixing angle (i.e., $\theta_{34}$).

\begin{table}[]
\begin{center}
\begin{tabular}{c|c|c|c|c|c|c|c|c|c|c|c}
\hline \hline
~ & \multicolumn{11}{c}{$\tan\beta=1$, $\epsilon_t = m_t/m_{t^\prime}$} \\
\hline \hline
  Point \# & Model & $m_{t^\prime}$ & $m_{b^\prime}$  & $m_{\nu^\prime}$  & $m_{\tau^\prime}$ &
  $m_h$ & $m_A$ & $m_H$ & $m_{H^+}$ & $\sin\theta_{34}$ & $\alpha$ \\
\hline
1 & 4G2HDM-I,II,III & 542 & 358 & 144 & 462 &  260 & 296 & 1357 & 654 & 0.153 & $0.74 \pi$ \\
2 & 4G2HDM-I & 511 & 353 & 426 & 455 &  261 & 296 & 1075 & 428 & 0.09 & $0.705 \pi$ \\
3 & 4G2HDM-I,II,III & 548 & 372 & 413 & 434 &  199 & 272 & 1088 & 707 & 0.063 & $1.88 \pi$ \\
4 & 4G2HDM-I & 367 & 525 & 829 & 993 &  347 & 491 & 1227 & 681 & 0.011 & $1.82 \pi$ \\
5 & 4G2HDM-I & 356 & 537 & 121 & 310 &  675 & 238 & 1306 & 542 & 0.056 & $0.97 \pi$ \\
6 & 4G2HDM-I & 440 & 456 & 619 & 634 &  169 & 332 & 405  & 479 & 0.082 & $0.82 \pi$ \\
7\footnote{point requires $|\lambda^{t^\prime}_{sb}| \lsim 10^{-5}$.} & 4G2HDM-I,II,III & 526 & 534 & 403 & 420 &  152 & 875 & 550 & 461 & 0.007 & $0.87 \pi$ \\
8 & 4G2HDM-I & 416 & 510 & 370 & 536 &  216 & 153 & 1032  & 333 & 0.14 & $0.96 \pi$ \\
9 & 4G2HDM-II,III & 520 & 369 & 738 & 744 &  102 & 882 & 238  & 781 & 0.129 & $1.28 \pi$ \\
10 & 4G2HDM-I & 500 & 450 & 302 & 414 &  220 & 793 & 1001  & 750 & 0.05 & $\pi/2$ \\
11 & 4G2HDM-I & 500 & 450 & 424 & 410 &  120 & 597 & 1479  & 750 & 0.2 & $\pi/2$ \\
12 & 4G2HDM-I & 500 & 450 & 147 & 127 &  350 & 716 & 506  & 400 & 0.05 & $\pi/2$ \\
13\footnote{point requires $\epsilon_b \sim m_b/m_{b^\prime}$ in order to have
${\rm BR}(b^\prime \to t H^+) \sim {\cal O}(1)$ (see Fig.~\ref{fig3BR}.}) & 4G2HDM-I & 450 & 500 & 225 & 235 &  220 & 782 & 303  & 300 & 0.2 & $\pi/2$ \\
\hline \hline
~ & \multicolumn{11}{c}{$\tan\beta=5$, $\epsilon_t = m_t/m_{t^\prime}$} \\
\hline \hline
14 & 4G2HDM-II,III & 542 & 386 & 938 & 740 &  126 & 458 & 1141 & 738 & 0.094 & $0.9 \pi$ \\
15 & 4G2HDM-II,III & 544 & 367 & 305 & 310 &  179 & 417 & 1255 & 706 & 0.117 & $1.09 \pi$ \\
16 & 4G2HDM-II,III & 517 & 366 & 393 & 211 &  295 & 130 & 1347 & 801 & 0.188 & $0.12 \pi$ \\
17 & 4G2HDM-II,III & 430 & 412 & 193 & 175 &  246 & 568 & 904 & 617 & 0.18 & $0.12 \pi$ \\
18 & 4G2HDM-II,III & 463 & 451 & 398 & 418 &  170 & 593 & 1218 & 715 & 0.026 & $0.25 \pi$ \\
19 & 4G2HDM-II,III & 381 & 465 & 545 & 622 &  135 & 145 & 1084 & 803 & 0.051 & $0.77 \pi$ \\
20 & 4G2HDM-II,III & 514 & 371 & 106 & 610 &  122 & 295 & 1495 & 819 & 0.031 & $1.89 \pi$ \\
21 & 4G2HDM-II,III & 496 & 399 & 541 & 617 &  148 & 343 & 1054 & 780 & 0.03 & $1.77 \pi$ \\
22 & 4G2HDM-II,III & 463 & 481 & 959 & 784 &  105 & 319 & 918 & 760 & 0.188 & $0.03 \pi$ \\
23 & 4G2HDM-II,III & 504 & 508 & 497 & 545 &  140 & 118 & 1175 & 748 & 0.166 & $0.14 \pi$ \\
\hline \hline
~ & \multicolumn{11}{c}{$\tan\beta=20$, $\epsilon_t < 0.1$} \\
\hline \hline
24 & 4G2HDM-II,III & 521 & 362 & 178 & 191 &  177 & 231 & 775 & 525 & 0.03 & $1.96 \pi$ \\
25 & 4G2HDM-II,III & 535 & 381 & 568 & 399 &  435 & 573 & 1500 & 954 & 0.073 & $0.81 \pi$ \\
26 & 4G2HDM-II,III & 542 & 372 & 106 & 314 &  510 & 268 & 1382 & 450 & 0.158 & $0.94 \pi$ \\
27 & 4G2HDM-II,III & 369 & 527 & 212 & 565 &  571 & 233 & 1335 & 669 & 0.175 & $1.88 \pi$ \\
28 & 4G2HDM-II,III & 459 & 440 & 684 & 702 &  142 & 455 & 631 & 400 & 0.101 & $0.12 \pi$ \\
29 & 4G2HDM-II,III & 546 & 517 & 260 & 661 &  111 & 216 & 1347 & 940 & 0.186 & $0.08 \pi$ \\
30 & 4G2HDM-II,III & 411 & 456 & 126 & 423 &  140 & 163 & 1261 & 940 & 0.1843 & $0.11 \pi$ \\
\hline \hline
\end{tabular}
\caption{List of points (models) in parameter space for our 4G2HDMs of types I, II and III,
allowed at 95\%CL by PEWD and B-physics flavor data. The 2nd column denotes the model(s)
for which the point is applicable. Points 1-3,14-16 and 24 have
$m_{t^\prime} - m_{b^\prime} > 150 $ GeV with a light CP-even Higgs of mass $m_h \lsim 300$ GeV,
while points 4,5,27 have a large inverted splitting
$m_{b^\prime} - m_{t^\prime} > 150 $ GeV with a heavier h.
Points 6,7 and 17,18,28 have nearly degenerate 4th generation quark
and lepton doublets, while points 22,23 have a nearly degenerate 4th generation quark doublet with
a lepton doublet heavier than the quark doublet. Points 8,19 have
$m_{b^\prime} - m_{t^\prime} > m_W $ and a light charged Higgs,
while points 9,16 have $m_{t^\prime} - m_{b^\prime} \sim 150 $ GeV with a light Higgs mass of
$m_h \sim 100$ GeV. Points 1,8,16,17,22,23,26,27,29,30 all have a large $t^\prime - b/t-b^\prime$
mixing angle: $\theta_{34} \gsim 0.15$. Finally,
points 10,11 give ${\rm BR}(t^\prime \to t h) \sim {\cal O}(1)$ (see Fig.~\ref{fig1BR} in the next section),
point 12 gives ${\rm BR}(t^\prime \to b H^+) \sim {\cal O}(1)$ (see Fig.~\ref{fig2BR} in the next section) and
point 13 gives ${\rm BR}(b^\prime \to t H^+) \sim {\cal O}(1)$ (see Fig.~\ref{fig3BR} in the next section).}
\label{tab2}
\end{center}
\end{table}

\section{\label{sec:phen2hdmI} Phenomenology of the Yukawa sector in the 4G2HDM-I}

Although this paper is not aimed to explore in detail
the phenomenological consequences of the modifications to the
Higgs Yukawa interactions involving the 4th generation quarks in
our 4G2HDMs,
in order to give a feel for their importance for collider searches of
the 4th generation fermions, we consider below
some phenomenological aspects of the 4G2HDM-I which is defined by $\left(\alpha_d,\beta_d,\alpha_u,\beta_u\right)=\left(0,1,0,1\right)$.
Recall that in this case, the $\Sigma$ mixing matrices simplify to (keeping terms up to ${\cal O}(\epsilon_q^2)$,
$q=b,t$):
\begin{eqnarray}
\Sigma^d \simeq \left(\begin{array}{cccc}
0 & 0 & 0 & 0 \\
0 & 0 & 0 & 0 \\
0 & 0 &  |\epsilon_b|^2 & \epsilon_b^\star \\
0 & 0 &  \epsilon_b  & \left( 1- \frac{|\epsilon_b|^2}{2} \right)
\end{array}\right)~,~
\Sigma^u \simeq \left(\begin{array}{cccc}
0 & 0 & 0 & 0 \\
0 & 0 & 0 & 0 \\
0 & 0 &  |\epsilon_t|^2 & \epsilon_t^\star \\
0 & 0 &  \epsilon_t  & \left( 1- \frac{|\epsilon_t|^2}{2} \right)
\end{array}\right)
 \label{sigmaI}~,
\end{eqnarray}
\begin{figure}[htb]
\begin{center}
\epsfig{file=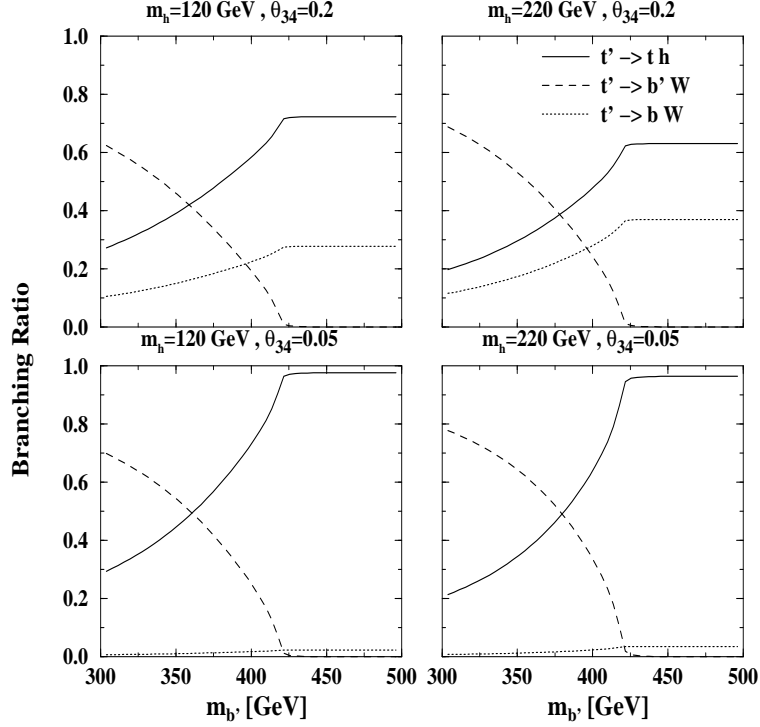,height=10cm,width=10cm,angle=0}
\caption{\emph{The branching ratio for the $t^\prime$ decay channels
$t^\prime \to t h$, $t^\prime \to bW$ and $t^\prime \to b^\prime W^{(\star)}$ ($W^{(\star)}$
is either on-shell or off-shell depending on the $b^\prime$ mass),
as a function of $m_{b^\prime}$ for $m_{t^\prime}=500$ GeV, $\epsilon_t =m_t/m_{t^\prime}$,
$\tan\beta=1$ and $(m_h~{\rm [GeV]},\theta_{34})=(120,0.05),(120,0.2),(220,0.05),(220,0.2)$, as indicated.
Also, $\alpha=\pi/2$ and $m_{H^+} > m_{t^\prime}$, $m_{A} > m_{t^\prime}$ is assumed.}}
\label{fig1BR}
\end{center}
\end{figure}
\begin{figure}[htb]
\begin{center}
\vspace{1.0cm}
\epsfig{file=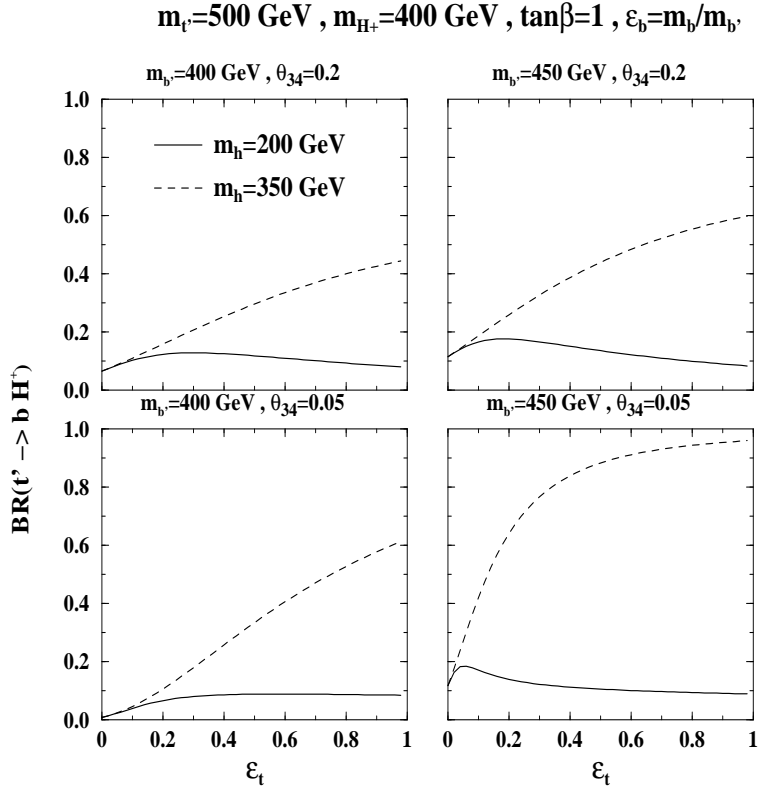,height=10cm,width=10cm,angle=0}
\caption{\emph{The branching ratios for the decay
$t^\prime \to b H^+$
as a function of $\epsilon_t$ for $m_{t^\prime}=500$ GeV, $m_{H^+}=400$ GeV,
$\tan\beta=1$, $\epsilon_b=m_b/m_{b^\prime}$,
$m_h=220$ and 350 GeV
and $(m_{b^\prime}~{\rm [GeV]},\theta_{34})=(400,0.05),(400,0.2),(450,0.05),(450,0.2)$, as indicated.
Also, $\alpha=\pi/2$ and $m_{A} > m_{t^\prime}$ is assumed.}}
\label{fig2BR}
\end{center}
\end{figure}
\begin{figure}[htb]
\begin{center}
\vspace{1.0cm}
\epsfig{file=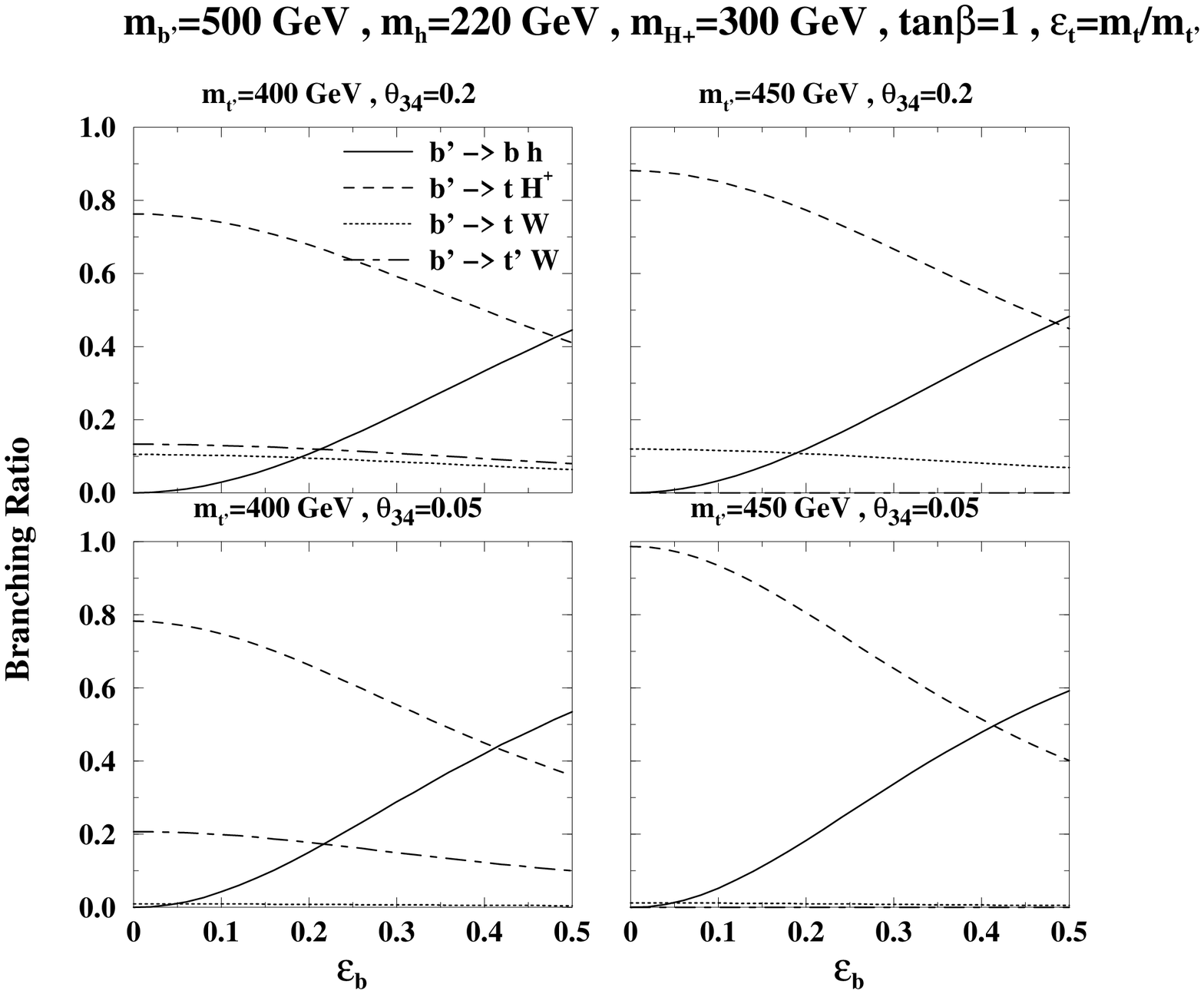,height=10cm,width=10cm,angle=0}
\caption{\emph{The branching ratios for the decay channels
$b^\prime \to b h$, $b^\prime \to t H^+$, $b^\prime \to tW$ and $b^\prime \to t^\prime W$,
as a function of $\epsilon_b$ for $m_{b^\prime}=500$ GeV,
$m_h=220$ GeV, $m_{H^+}=300$ GeV, $\tan\beta=1$, $\epsilon_t=m_t/m_{t^\prime}$
and $(m_{t^\prime}~{\rm [GeV]},\theta_{34})=(400,0.05),(400,0.2),(450,0.05),(450,0.2)$, as indicated.
Also, $\alpha=\pi/2$ and $m_{A} > m_{b^\prime}$ is assumed.}}
\label{fig3BR}
\end{center}
\end{figure}
so that $\Sigma^{u,d} = 0$ if $i ~ {\rm or} ~ j \neq 3,4$. This
leads to new interesting patterns (in flavor space) of the ${\cal H}^0q_iq_j$ Yukawa
interactions in Eqs.~\ref{Sff1}-\ref{Sff2} (${\cal H}^0=h,H,A$).
In particular, the most notable new features of the
4G2HDM-I
are:
\begin{enumerate}
\item There are no tree-level FC neutral currents (FCNC) among
the quarks of the 1st, 2nd and 3rd generations. That is,
no tree-level $c \to u$, $s \to d$ transitions,
as well as no $t \to u$, $t \to c$,
$b \to d$ and $b \to s$ ones.
\item There are no tree-level FCNC effects involving transitions between
the quarks of the 4th generation and the 1st and 2nd generations, i.e, no
$t^\prime \to u$, $t^\prime \to c$, $b^\prime \to d$ and $b^\prime \to s$ transitions.
This, makes the 4G2HDM-I compatible with all FCNC constraints coming
from light meson mixings and decays, i.e., in the K and D systems.
\item There are new potentially large tree-level FCNC effects in the ${\cal H}^0q_iq_j$ couplings involving the 3rd and 4th generation quarks (i.e., $i,j=3,4$),
which can have drastic phenomenological consequences for high-energy collider searches of the 4th generation fermions, as we will further discuss below.
In particular, the
FC ${\cal H^0}t^\prime t$ and ${\cal H^0} b^\prime b$
interactions are (taking $\alpha \to \pi/2$):
\begin{widetext}
\begin{eqnarray}
{\cal L}(h t^\prime t) &=&
-\frac{g}{2}\frac{m_{t^\prime}}{m_W} \epsilon_t \sqrt{1+ t_\beta^2}
~ \bar t^\prime \left( R + \frac{m_t}{m_{t^\prime}} L \right) t h
\label{htpt}~, \\
{\cal L}(H t^\prime t) &=&
-\frac{g}{2}\frac{m_{t^\prime}}{m_W} \epsilon_t \frac{\sqrt{1+ t_\beta^2}}{t_\beta}
~ \bar t^\prime \left( R + \frac{m_t}{m_{t^\prime}} L \right) t H
\label{hhtpt}~, \\
{\cal L}(A t^\prime t) &=&
i \frac{g}{2}\frac{m_{t^\prime}}{m_W} \epsilon_t \frac{1+ t_\beta^2}{t_\beta}
~ \bar t^\prime \left( R - \frac{m_t}{m_{t^\prime}} L \right) t A
\label{Atpt}~,
\end{eqnarray}
\end{widetext}
and similarly for the ${\cal H}^0 b^\prime b$ interactions
by changing $\epsilon_t \to \epsilon_b$ (and an extra minus sign
in the $A b^\prime b$ coupling).

We thus see that, if $\epsilon_t \sim m_t/m_{t^\prime}$, then the above
couplings can become sizable, e.g., to the level that it might
dominate the decay pattern of the $t^\prime$ (see below).
In fact, we also expect large FC effects in $b^\prime \to b$ transitions
since, even for a very small $\epsilon_b \sim m_b / m_{b^\prime}$,
the FC $h b^\prime b$ and $A b^\prime b$ Yukawa couplings can become sizable
if e.g., $\tan\beta \sim 5 $, i.e., in which case
they are $\propto \frac{5 m_b}{m_W}$.
\item The flavor diagonal interactions
of the Higgs species with the up-quarks, ${\cal H}^0 uu$,
are proportional to $\tan\beta$, thus being a factor of $\tan^2\beta$ larger than
the corresponding ``conventional" 2HDMs couplings, for which
these couplings are
$\propto \cot\beta$ (e.g., as in the 2HDM of type II
which also underlies the supersymmetric Higgs sector).
In particular, the ${\cal H}^0 tt$ couplings in our 4G2HDM-I are given by:
\begin{widetext}
\begin{eqnarray}
{\cal L}(h t t) & \approx & \frac{g}{2}\frac{m_t}{m_W} \sqrt{1+ t_\beta^2} \left(1 -|\epsilon_t|^2 \right)
~ \bar t t h
\stackrel{|\epsilon_t|^2 \ll 1}{\longrightarrow}
\frac{g}{2}\frac{m_t}{m_W} \sqrt{1+ t_\beta^2}
~ \bar t t h
\label{htt}~, \\
{\cal L}(H t t) &\approx& - \frac{g}{2}\frac{m_t}{m_W} \frac{\sqrt{1+ t_\beta^2}}{t_\beta} |\epsilon_t|^2
~ \bar t t H
\label{hhtt}~, \\
{\cal L}(A tt) &\approx& - i \frac{g}{2}\frac{m_t}{m_W} t_\beta
\left[ 1 - \left(1 + t_\beta^{-2} \right) |\epsilon_t|^2 \right]
~ \bar t \gamma_5 t A
\stackrel{|\epsilon_t|^2 \ll 1}{\longrightarrow}
-i \frac{g}{2}\frac{m_t}{m_W} t_\beta
~ \bar t \gamma_5 t A
 \label{Att}~.
\end{eqnarray}
\end{widetext}

We see that the $htt$ and $Att$ Yukawa interactions are indeed enhanced
by a factor of $t_\beta^2$ relative to the conventional $htt$ and $Att$ couplings
in multi-Higgs models (with no suppression from $t -t^\prime$ mixing parameter $\epsilon_t$).
On the other hand, the $ht^\prime t^\prime$ and $At^\prime t^\prime$ couplings
are suppressed by the $t -t^\prime$ mixing parameter and by $t_\beta$, respectively:
%
\begin{eqnarray}
{\cal L}(h t^\prime t^\prime) &\approx& \frac{g}{4}\frac{m_{t^\prime}}{m_W}
\sqrt{1+ t_\beta^2} |\epsilon_t|^2
~ \bar t^\prime t^\prime h \label{htptp}~, \\
{\cal L}(H t^\prime t^\prime) &\approx& - \frac{g}{2}\frac{m_{t^\prime}}{m_W}
\frac{\sqrt{1+ t_\beta^2}}{t_\beta} \left( 1 - \frac{|\epsilon_t|^2}{2} \right)
~ \bar t^\prime t^\prime H \label{hhtptp}~, \\
{\cal L}(A t^\prime t^\prime) &\approx&  - i \frac{g}{2}\frac{m_{t^\prime}}{m_W}
t_\beta \left[ 1 - \left(1 + t_\beta^{-2} \right) \left(1- \frac{|\epsilon_t|^2}{2} \right) \right]
~ \bar t^\prime \gamma_5 t^\prime A
\stackrel{|\epsilon_t|^2 \ll 1}{\longrightarrow}
i \frac{g}{2}\frac{m_t}{m_W} \frac{1}{t_\beta}
~ \bar t \gamma_5 t A
\label{Atptp}~.
\end{eqnarray}
%
\item The charged Higgs couplings involving the 3rd and 4th generation quarks
are completely altered by the presence of the $\Sigma$ matrix in Eq.~\ref{Sff2}.
For instance, the $H^+ t^\prime b$ and $H^+ t b^\prime$ couplings have new terms
proportional to $V_{tb}$ and $V_{t^\prime b^\prime}$. In particular, in the ``3+1" scenario where
$V_{t^\prime d_i},V_{u_i b^\prime} \to 0$ for $i=1,2,3$, we have:
\begin{eqnarray}
&&{\cal L}(H^+ t^\prime b) \approx \frac{g}{\sqrt{2}m_W} t_\beta
\left(1 + t_\beta^{-2} \right)
\bar t^\prime \left( m_t \epsilon_t V_{tb} L - m_{b^\prime} \epsilon_b V_{t^\prime b^\prime}
R \right) b H^+ ~, \\
&&{\cal L}(H^+ t b^\prime) \approx
 \frac{g}{\sqrt{2}m_W} t_\beta \left(1 + t_\beta^{-2} \right)
\bar t \left( m_t^\prime \epsilon_t^\star V_{t^\prime b^\prime}  L - m_{b} \epsilon_b^\star V_{tb}
R \right) b^\prime H^+
\label{hpcoup}~.
\end{eqnarray}

Recall that in the standard 2HDM of type II that also underlies supersymmetry (assuming
four generations of fermions)
the $\bar t^\prime_R b_L H^+$ would be $\propto m_{t^\prime} V_{t^\prime b} / t_\beta$. We thus see
that in our 4G2HDM-I the $\bar t^\prime_R b_L H^+$ coupling is potentially
enhanced by a factor of
$t_\beta^2 \cdot \epsilon_t \cdot (m_t/m_{t^\prime}) \cdot (V_{tb}/V_{t^\prime b})$. For example,
if $t_\beta=3$, $m_{t^\prime} \sim 500$ GeV and $\epsilon_t \sim m_t/m_{t^\prime}$
we get a factor of $V_{tb}/V_{t^\prime b}$ enhancement to the
$\bar t^\prime_R b_L H^+$ interaction.
\end{enumerate}

The implications of the above new Yukawa interactions can be far reaching with regard
to the decay patterns of the
$t^\prime$ and the $b^\prime$ and the search strategies for these heavy quarks.
In particular, in Fig.~\ref{fig1BR} we plot the branching ratios of the
leading $t^\prime$ decay channels (assuming $m_{H^+},m_A > m_{t^\prime}$):
$t^\prime \to t h, ~ bW ,~b^\prime W^{(\star)}$
[$W^{(\star)}$ stands for either on-shell or off-shell $W$ depending
on the $m_{b^\prime}$], as a function of
the $b^\prime$ mass.
We use
$m_{t^\prime}=500$ GeV, $\tan\beta=1$, $\epsilon_t = m_t / m_{t'}$ and
the following values for $m_h$ and $\theta_{34}$: $(m_h~{\rm [GeV]},\theta_{34})=(120,0.05),(120,0.2),(220,0.05),(220,0.2)$.
We see that the $BR(t^\prime \to t h)$ can easily reach
${\cal O}(1)$, in particular when $m_{t^\prime} - m_{b^\prime} < m_W$ and
even for a rather large $\theta_{34} \sim 0.2$; see e.g., points 10 and 11 in Table \ref{tab2}
for which $BR(t^\prime \to t h) \sim {\cal O}(1)$.

In Fig.~\ref{fig2BR} we take $m_{H^+}=400$ GeV (again assuming
$m_A > m_{t^\prime}$ so that $t^\prime \to t A$ is
still kinematically closed) and
 plot $BR(t^\prime \to b H^+)$
as a function
of $\epsilon_t$, for $m_{t^\prime}=500$ GeV, $\tan\beta=1$,
$\epsilon_b = m_b / m_{b'} \sim 0.01$
$m_h= 200$ and 350 GeV
and
the following values for $m_{b'}$ and $\theta_{34}$
$(m_{b^\prime}~{\rm [GeV]},\theta_{34})=(400,0.05),(400,0.2),(450,0.05),(450,0.2)$.
We see that the decay channel $t^\prime \to b H^+$ can become
important and even dominate if $\epsilon_t \gsim m_t/m_{t^\prime}$,
in particular, when $m_{t^\prime} - m_{b^\prime} < m_W$ and
a small mixing angle of $\theta_{34} \sim {\cal O}(0.05)$; see e.g.,
point 12 in Table \ref{tab2}
for which $BR(t^\prime \to b H^+) \sim {\cal O}(1)$.

In Fig.~\ref{fig3BR} we plot the branching ratios of the leading
$b^\prime$ decay channels, as a function
of $\epsilon_b$ for $m_{b^\prime}=500$ GeV, $\tan\beta=1$, $m_{H^+}=300$ GeV,
$m_h =220$ GeV,
$\epsilon_t = m_t / m_{t'}$
and
the following values for $m_{t'}$ and $\theta_{34}$
$(m_{t^\prime}~{\rm [GeV]},\theta_{34})=(400,0.05),(400,0.2),(450,0.05),(450,0.2)$.
We see that in the $b^\prime$ case the dominance of $b^\prime \to t H^-$ (if kinematically allowed) is much more pronounced due to the expected smallness of the $b - b^\prime$ mixing parameter, $\epsilon_b$, which controls the FC decay $b^\prime \to b h$; see e.g.,
point 13 in Table \ref{tab2}
for which $BR(b^\prime \to b H^-) \sim {\cal O}(1)$.

This change in the decay pattern of the 4th generation quarks can
have important consequences for collider searches of these heavy fermions. For example, as was already noticed in \cite{sher1}, if
$t^\prime \to t h$ dominates then $t^\prime$ production at the LHC
via $gg \to t^\prime \bar t^\prime$ will lead to the dramatic
signature of $t \bar t hh$. If $m_h < 2 m_W$ (so that
$h$ decays to $b \bar b$) this will
give a $6b+2W$ signature (i.e., after the top decays via $t \to b W$), while
if $m_h > 2 m_W, 2m_Z$ the $t \bar t hh$ final state can lead
to either $t \bar t hh \to t \bar t W^+ W^-$ and/or
 $t \bar t hh \to t \bar t Z Z$. In particular, notice that
 the former $t \bar t W^+ W^-$ is the one conventionally used
 for $b^\prime$ searches \cite{limits}, while the latter will lead to e.g.,
 a $2b+2W+4\ell$ signature which is expected to have a rather small
 irreducible SM-like background (e.g., coming from $gg \to t \bar t h$)
 that can be further controlled using the kinematic features of the
    process $gg \to t^\prime \bar t^\prime \to t \bar t hh \to t \bar t Z Z$. If, on the other hand, $t^\prime \to b H^+$ dominates,
   then the signature
   $b \bar b H^+ H^-$ should be focused on. In this case
   the $t^\prime$ searches will depend on the $H^+$ decays,
   e.g., $H^+ \to t b$ or $H^+ \to \tau \nu$, which will lead
   to $gg \to t^\prime \bar t^\prime \to 6b + 2W$ or
   $gg \to t^\prime \bar t^\prime \to 2b+2\tau+\missET$, respectively.

   For the $b^\prime$ the situation is similar, i.e,
   the new decays $b^\prime \to b h$ and/or $b^\prime \to t H^-$ can
also alter the search strategies for $b^\prime$. For example,
if $b^\prime \to t H^-$ dominates the $b^\prime$ decays, then
$gg \to b^\prime \bar b^\prime$ will lead to, e.g., a
$t \bar t H^- H^+ \to 4t + 2b$ signature as opposed to the
``standard" $2t+2W$ one when the $b^\prime$ decays via
$b^\prime \to t W$ \cite{dsg2011}.

Clearly, these new 4th generation quark signatures deserve
a detailed investigation which is beyond the scope of this paper
and will be considered elsewhere \cite{inprogress}.

\section{\label{sec:summary} Summary}

We have introduced a class of 2HDMs, which we named the 4G2HDM of types I, II and III.
Our models are
``designed" to give an effective low-energy
description for the apparent heaviness of the 4th generation fermions and to address
the possibility of dynamical EWSB which is driven by the condensates of these new heavy
fermionic states. This is done by giving a special status to the 4th family fermions which are
coupled to the scalar doublet that has the heavier VEV. Such setups give rise to very distinct
Yukawa textures which can have drastic implications on the phenomenology of
4th generation fermions systems. We studied the constraints
from PEWD and from flavor physics
in B-systems and outlined the allowed parameter space of our 4G2HDMs, which we find to
have various different features than the simpler SM4 version with a single Higgs boson and
a 4th family of fermions. For example, we find that the mass splitting $m_{t^\prime} - m_{b^\prime}$
and the inverted mass splitting $m_{b^\prime} - m_{t^\prime}$ can be as large as 200 GeV,
and that the mass splitting in the 4th generation lepton doublet can be as large as 400 GeV.

We focused on the 4G2HDM-I, where the Higgs doublet with the heavier VEV is coupled only to
the 4th generation doublet while the ``lighter" Higgs doublet is coupled to all other
quarks. This model is, in our view, somewhat better motivated as it
provides a more natural setup in the leptonic sector, i.e., addressing the
existence of a 4th generation EW-scale neutrino.
In addition, it has very distinctive features in flavor space: there are
no tree-level FCNC among the 1st three generation of fermions
as well as no FCNC among the 4th generation fermions and the light fermions of the
1st and 2nd generations.
On the other hand, the 4G2HDM-I does give rise to potentially large tree-level
FCNC $t^\prime \to t$ and $b^\prime \to b$ transitions, which, as we briefly
explored in the paper, can have significant implications
on the search for the 4th generation quarks at high-energy colliders.
For example, the FC decay $t^\prime \to t h$ can become the dominant
$t^\prime$ decay channel and should therefore effect the search strategy
for the $t^\prime$.

Finally, we note that the 4G2HDM setups can also alter the
production and decay patterns of the Higgs particles at hadron colliders.
For example, the di-photon Higgs channel
$gg \to h \to \gamma \gamma$ can be dramatically enhanced
or suppressed (to the level of being unobservable at the LHC)
compared to the SM4 case. The phenomenology of the production and decay
channels of the Higgs particles in the 4G2HDMs will be
considered elsewhere.

\bigskip

{\bf Acknowledgments:} SBS acknowledge research support from the Technion.
SN would like to thank Paolo Gambino for useful discussions and the
NSERC of Canada for financial support.
The work of AS was supported in part by the U.S. DOE contract
\#DE-AC02-98CH10886(BNL).


\begin{thebibliography}{99}

\newcommand{\np}[3]{Nucl. Phys. {\bf B#1} (#2) #3}
\newcommand{\pr}[3]{Phys. Rev.  {\bf D#1} (#2) #3}
\newcommand{\prp}[3]{Phys. Rept. {\bf #1} (#2) #3}
\newcommand{\zpc}[3]{Z. Phys. {\bf C#1} (#2) #3}
\newcommand{\PL}[3]{Phys. Lett. {\bf B#1} (#2) #3}

\bibitem{sl07} E.~Lunghi and A.~Soni, Phys.\ Lett.\  {\bf B666}, 162 (2008).

\bibitem{sl08} E.~Lunghi and A.~Soni, JHEP {\bf 0908}, 051 (2009).

\bibitem{lenz1} A.~Lenz {\it et al.} [CKMfitter Group], arXiv:1008.1593 [hep-ph].

\bibitem{bona} M.~Bona {\it et al.}  [UTfit Collaboration], Phys.\ Lett.\  B {\bf 687}, 61 (2010).

\bibitem{slprl10} E.~Lunghi and A.~Soni, Phys.\ Rev.\ Lett.\ {\bf 104}, 251802 (2010).

\bibitem{sl10} E.~Lunghi and A.~Soni, arXiv:1010.6069 [hep-ph].

\bibitem{hfag10} The Heavy Flavor Averaging Group {\it et al.}, arXiv:1010.1589 [hep-ex].

\bibitem{d0dimuonprd} V.~M.~Abazov {\it et al.}  [D0 Collaboration], Phys.\ Rev.\ {\bf D82}, 032001 (2010).

\bibitem{d0dimuonprl} V.~M.~Abazov {\it et al.}  [D0 Collaboration],
Phys.\ Rev.\ Lett.\ {\bf 105}, 081801 (2010).

\bibitem{sher0} P.H. Frampton, P.Q. Hung and M. Sher, Phys. Rept. {\bf 330}, 263 (2000).

\bibitem{hou2009-rev}
B. Holdom, W.S. Hou, T. Hurth, M.L. Mangano, S. Sultansoy, G. Unel, talk presented
at {\it Beyond the 3rd SM generation at the LHC era workshop}, Geneva, Switzerland, Sep 2008,
arXiv:0904.4698 [hep-ph], published in PMC Phys. {\bf A3}, 4 (2009).

\bibitem{SM4proc} For older literature on the 4th generation SM, see:
Proceedings of the First (February 1987) and the Second (February 1989) International
Symposiums on the {\it fourth family of quarks and leptons}, Santa Monica, CA,
published in
Annals of the New York Academy of Sciences, 517 (1987) \& 578 (1989),
edited by D. Cline and A. Soni.

\bibitem{DEWSB} B. Holdom, Phys. Rev. Lett. {\bf 57}, 2496 (1986)
[Erratum-ibid. {\bf 58}, 177 (1987);
W.A. Bardeen, C.T. Hill and M. Lindner, Phys. Rev. {\bf D41}, 1647 (1990);
S.F. King, Phys. Lett. {\bf B234}, 108 (1990);
C. Hill, M. Luty and E.A. Paschos, Phys. Rev. {\bf D43}, 3011 (1991);
P.Q. Hung and G. Isidori Phys. Lett. {\bf B402}, 122 (1997).

\bibitem{holdom-new} B. Holdom, JHEP {\bf 0608}, 76 (2006).

\bibitem{hung-new} P.Q. Hung and Chi Xiong, Nucl. Phys. {\bf B848}, 288 (2011).

\bibitem{SAGMN08}
A.~Soni, A.~K.~Alok, A.~Giri, R.~Mohanta and S.~Nandi,
Phys.\ Lett.\  {\bf B683}, 302 (2010).

\bibitem{SAGMN10} A.~Soni, A.~K.~Alok, A.~Giri, R.~Mohanta and S.~Nandi,
Phys.\ Rev.\ {\bf D82}, 033009 (2010).


\bibitem{ajb10B}
A.~J.~Buras, B.~Duling, T.~Feldmann, T.~Heidsieck, C.~Promberger and S.~Recksiegel,
JHEP {\bf 1009}, 106 (2010).

\bibitem{buras_charm}
A.~J.~Buras, B.~Duling, T.~Feldmann, T.~Heidsieck, C.~Promberger and S.~Recksiegel,
JHEP {\bf 1007}, 094 (2010).

\bibitem{gh10} W.~S.~Hou and C.~Y.~Ma, Phys.\ Rev.\ {\bf D82}, 036002 (2010).

\bibitem{NS3} S.~Nandi and A.~Soni, arXiv:1011.6091 [hep-ph].

\bibitem{lenz_fourth11}
M.~Bobrowski, A.~Lenz, J.~Riedl and J.~Rohrwild,
Phys.\ Rev.\  D {\bf 79}, 113006 (2009).

\bibitem{lenz_fourth12}
O.~Eberhardt, A.~Lenz and J.~Rohrwild,
Phys.\ Rev.\  D {\bf 82}, 095006 (2010).

\bibitem{Alok:2010zj} A.K. Alok, A. Dighe and D. London, arXiv:1011.2634 [hep-ph].

\bibitem{gustavo1} G. Burdman and L. Da Rold, JHEP {\bf 0712}, 86 (2007).

\bibitem{gustavo2} G. Burdman, L. Da Rold, O. Eboli and R. D'Elia Matheus, Phys. Rev. {\bf D79}, 075026 (2009);
G. Burdman, L. de Lima and R.D. Matheus, Phys. Rev. {\bf D83}, 035012 (2011).

\bibitem{hashimoto1} M. Hashimoto and V.A. Miransky, Phys. Rev. {\bf D81}, 055014 (2010).

\bibitem{gh08} W.S. Hou, Chin. J. Phys. {\bf 47}, 134 (2009),
arXiv:0803.1234 [hep-ph]; W.S. Hou, talk given at
{\it 34th International Conference on High Energy Physics (ICHEP 2008)},
Philadelphia, Pennsylvania, Jul 2008, arXiv:0810.3396 [hep-ph].

\bibitem{jarlskog}
C. Jarlskog and R. Stora, \PL(B208,288,1988);
F. del Aguila and J. A. Aguilar-Saavedra, \PL(B386,241,1996);
F. del Aguila and J. A. Aguilar-Saavedra and G. C. Branco, Nucl. Phys. {\bf B510}, 39 (1998).

\bibitem{fok} S.W. Ham, S.K. Oh, D. Son, Phys. Rev. {\bf D71}, 015001 (2005);
R. Fok, G.D. Kribs, Phys. Rev. {\bf D78}, 075023 (2008).

\bibitem{gh2011} G. W.S. Hou, arXiv:1101.2161 [hep-ph].

\bibitem{limits} V.M. Abazov {\it et al.} (D0 collaboration), arXiv:1104.4522 [hep-ex];
C.J. Flacco, D. Whiteson and M. Kelly, arXiv:1101.4976 [hep-ph];
C.J. Flacco, D. Whiteson, T.M.P. Tait and S. Bar-Shalom,
Phys. Rev. Lett. {\bf 105}, 111801 (2010);
T. Aaltonen {\it et al.} (CDF collaboration), Phys. Rev. Lett. {\bf 104}, 091801 (2010);
T. Aaltonen {\it et al.} (CDF collaboration), Phys. Rev. Lett. {\bf 100}, 161803 (2008);
A. Lister (CDF collaboration), arXiv:0810.3349 [hep-ex];
J. Conway {\it et al.} (CDF collaboration), public conference note
CDF/PUB/TOP/PUBLIC/10110; D. Whiteson {\it et al.} (CDF collaboration),
CDF public conference note CDF/PUB/TOP/PUBLICH/10243;
P.~Q.~Hung and M.~Sher, Phys. Rev. {\bf D77}, 037302 (2008).

\bibitem{luty} M.A. Luty, Phys. Rev. {\bf D41}, 2893 (1990).

\bibitem{sher1} E. De Pree, G. Marshall and M. Sher, Phys. Rev. {\bf D80}, 037301 (2009).

\bibitem{hung} P.Q. Hung and Chi Xiong, Nucl. Phys. {\bf B847}, 160 (2011);
{\it ibid.} Phys. Lett. {\bf B694}, 430 (2011).

\bibitem{hashimoto2} M. Hashimoto, Phys. Rev. {\bf D81}, 075023 (2010).

\bibitem{wise1} K. Ishiwata and M.B. Wise, Phys. Rev. {\bf D83}, 074015 (2011).

\bibitem{HHG}J. F. Gunion, H. E. Haber, G. Kane, S. Dawson, {}``The
Higgs Hunter's Guide'', Addison-Wesley (1990); see also: Errata,
SCIPP-92-58 (1992), arXiv:hep-ph/9302272.

\bibitem{bernr} W. Bernreuther, P. Gonzalez, M. Wiebusch, Eur. Phys. J. {\bf C69}, 31 (2010).

\bibitem{sher2HDM} Marc Sher, Phys. Rev. {\bf D61}, 057303 (2000).

\bibitem{shersusy} S. Litsey, M. Sher, Phys. Rev. {\bf D80}, 057701 (2009).

\bibitem{dawson} S. Dawson, P. Jaiswal, Phys. Rev. {\bf D82}, 073017 (2010).

\bibitem{rizzo} R.C. Cotta, J.L. Hewett, A. Ismail, M.-P. Le, T.G. Rizzo, arXiv:1105.0039 [hep-ph].

\bibitem{Das} A. Das, C. Kao, Phys. Lett. {\bf B372}, 106 (1996).

\bibitem{gustavo3} G. Burdman, L. Da Rold, R. D'Elia Matheus, Phys. Rev. {\bf D82}, 055015 (2010).

\bibitem{aps2005} K. Agashe, G. Perez, A. Soni, Phys. Rev. {\bf D71}, 016002 (2005).

\bibitem{neubert} S.~Casagrande, F.~Goertz, U.~Haisch, M.~Neubert and T.~Pfoh,
JHEP {\bf 0810}, 094 (2008);
M. Blanke, A.J. Buras, B. Duling, S. Gori and A. Weiler,
JHEP {\bf 0903}, 001 (2009).


\bibitem{king} For a minimal extension to the SM4 with only one Higgs doublet that can address
 the heavy 4th generation neutrino problem see, S.F. King, Phys. Lett. {\bf B281}, 295 (1992).

\bibitem{PDG} The Review of Particle Physics, K. Nakamura {\it et al.} (Particle Data Group),
J. Phys. {\bf G37}, 075021 (2010).

\bibitem{peskin} M.E. Peskin, T. Takeuchi, Phys. Rev. Lett. {\bf 65}, 964 (1990);
{\it ibid.}, Phys. Rev. {\bf D46}, 381 (1992).

\bibitem{CDGG98}  M.~Ciuchini, G.~Degrassi, P.~Gambino and G.~F.~Giudice,
Nucl.\ Phys.\  {\bf B527}, 21 (1998);
{\it ibid},  Nucl.\ Phys.\ {\bf B534}, 3 (1998).

\bibitem{DGG00} G.~Degrassi, P.~Gambino and G.~F.~Giudice,
JHEP {\bf 0012}, 009 (2000).

\bibitem{MPR98} M.~Misiak, S.~Pokorski and J.~Rosiek, hep-ph/9703442,
published in the Review Volume ``Heavy Flavors II'', eds. A.J.~Buras
and M.~Lindner, World Scientific Publishing Co., Singapore, 1998.

\bibitem{Chetyrkin:1996vx}
K.G.~Chetyrkin, M.~Misiak and M.~M\"unz,
Phys.\ Lett.\ {\bf B400}, 206 (1997),
[Erratum-ibid.\ Phys.\ Lett.\ {\bf B425}, 414 (1998)].

\bibitem{GHW96}   C.~Greub, T.~Hurth and D.~Wyler,
Phys.\ Rev.\  {\bf D54}, 3350 (1996);
A.J.~Buras, A.~Czarnecki, M.~Misiak and J.~Urban,
Nucl.\ Phys.\  {\bf B611}, 488 (2001).

\bibitem{CMM97} K.~G.~Chetyrkin, M.~Misiak and M.~Munz,
Phys.\ Lett.\ {\bf B400}, 206 (1997); [Erratum-ibid.\ {\bf B425}, 414 (1998)].

\bibitem{AG95}
A.~Ali and C.~Greub, Phys.\ Lett.\  {\bf B361}, 146 (1995).

\bibitem{P96} N.~Pott, Phys.\ Rev.\ {\bf D54}, 938 (1996).

\bibitem{MM95} M.~Misiak and M.~Munz,
Phys.\ Lett.\  {\bf B344}, 308 (1995).

\bibitem{match2} K.~Adel and Y.~P.~Yao, Phys.\ Rev.\  {\bf D49}, 4945 (1994);
C.~Greub and T.~Hurth, Phys.\ Rev.\  {\bf D56}, 2934 (1997);
A.J.~Buras, A.~Kwiatkowski and N.~Pott,
Nucl.\ Phys.\  {\bf B517}, 353 (1998).

\bibitem{Czakon:2006ss} M.~Czakon, U.~Haisch and M.~Misiak,
JHEP {\bf 0703}, 008 (2007).

\bibitem{Bobeth:1999mk} C.~Bobeth, M.~Misiak and J.~Urban,
Nucl.\ Phys.\ {\bf B574}, 291 (2000).

\bibitem{KN99} A.~L.~Kagan and M.~Neubert,
Eur.\ Phys.\ J.\  {\bf C7}, 5 (1999);
K. Kiers, A. Soni, G.-H. Wu, Phys. Rev. {\bf D62}, 116004 (2000).

\bibitem{CM98} A.~Czarnecki and W.~J.~Marciano,
Phys.\ Rev.\ Lett.\ {\bf 81}, 277 (1998).

\bibitem{BM00}
K.~Baranowski and M.~Misiak, Phys.\ Lett.\  {\bf B483}, 410 (2000).

\bibitem{GH00} P.~Gambino and U.~Haisch, JHEP {\bf 0009}, 001 (2000).

\bibitem{misiak08} M.~Misiak, arXiv:0808.3134 [hep-ph].

\bibitem{BMMP94}
A.~J.~Buras, M.~Misiak, M.~Munz and S.~Pokorski,
Nucl.\ Phys.\  {\bf B424}, 374 (1994).

\bibitem{Gambino:2001ew} P.~Gambino and M.~Misiak, Nucl.\ Phys.\ {\bf B611}, 338 (2001).

\bibitem{Hou:1987kf}
W.~S.~Hou and R.~S.~Willey,
Phys.\ Lett.\  {\bf B202}, 591 (1988).

\bibitem{Bobeth:1999ww} C.~Bobeth, M.~Misiak and J.~Urban,
Nucl.\ Phys.\  {\bf B567}, 153 (2000).

\bibitem{buras_2hdm}
A.~J.~Buras, P.~Krawczyk, M.~E.~Lautenbacher and C.~Salazar,
Nucl.\ Phys.\ {\bf B337}, 284 (1990).

\bibitem{PU} M.S. Chanowitz, M.A. Furman and I. Hinchliffe,
Phys. Lett. {\bf B78}, 285 (1978); {\it ibid.}, Nucl. Phys. {\bf B153}, 402 (1979).

\bibitem{Gamiz:2009ku}
E.~Gamiz, C.~T.~H.~Davies, G.~P.~Lepage, J.~Shigemitsu and M.~Wingate [HPQCD Collaboration],
Phys.\ Rev.\ {\bf D80}, 014503 (2009).

\bibitem{gamizp} E. Gamiz, private communication.

\bibitem{buras1} A.~J.~Buras, M.~Jamin and P.~H.~Weisz,
Nucl.\ Phys.\  {\bf B347}, 491 (1990).

\bibitem{ARS} D. Atwood, L. Reina, A. Soni, Phys. Rev. {\bf D54}, 3296 (1996);
{\it ibid.}, Phys. Rev. {\bf D55}, 3156 (1997).

\bibitem{Haber} H.E. Haber, H.E. Logan, Phys. Rev. {\bf D62}, 015011 (2000), and references
therein.

\bibitem{EL2009} J. Erler, P. Langacker, JHEP {\bf 0908}, 017 (2009).

\bibitem{chenowitz} M.S. Chanowitz, Phys. Rev. {\bf D79}, 113008 (2009).

\bibitem{RbSM4} See e.g., T. Yanir, JHEP {\bf 0206}, 044, (2002);
J. Alwall {\it et al.}, Eur. Phys. J. {\bf C49}, 791 (2007).

\bibitem{ourZbs} D. Atwood, S. Bar-Shalom, G. Eilam, A. Soni,
Phys. Rev. {\bf D66}, 093005 (2002).

\bibitem{polonsky} H.-J. He, N. Polonsky, S. Su, Phys. Rev. {\bf D64}, 053004 (2001).

\bibitem{novikov} V.A. Novikov, L.B. Okun, A.N. Rozanov, M.I. Vysotsky, 
Phys. Lett. {\bf B529}, 111 (2002).

\bibitem{Kribs_EWPT}
G.~D.~Kribs, T.~Plehn, M.~Spannowsky and T.~M.~P.~Tait,
Phys.\ Rev.\  {\bf D76}, 075016 (2007).

\bibitem{langacker}
J. Erler and P. Langacker, Phys. Rev. Lett. {\bf 105}, 031801 (2010).

\bibitem{chenowitz1} M.S. Chanowitz, Phys. Rev. {\bf D82}, 035018 (2010).

\bibitem{gfitter} G. Collaboration, http://gfitter.desy.de/.

\bibitem{mHest} See e.g.,
J. Carpenter, R. Norton, S. Siegemund-Broka and A. Soni, Phys. Rev. Lett. {\bf 65}, 153 (1990);
see also Ref.~\cite{sher0}.

\bibitem{dsg2011} B. Holdom and Q.-S. Yan, arXiv:1101.3844 [hep-ph];
D. Atwood, S.K. Gupta and A. Soni, arXiv:1104.3874 [hep-ph].


\bibitem{inprogress} M. Geller, S. Bar-Shalom and G. Eilam, work in progress.


\end{thebibliography}
\end{document}